\documentclass[prd,twocolumn,superscriptaddress,floatfix,showpacs,nofootinbib]{revtex4-2}
\usepackage{amsfonts}
\usepackage{amsmath}
\usepackage{amssymb}
\usepackage{xcolor}
\usepackage{graphicx}
\usepackage[normalem]{ulem}
\font\mybb=msbm10 at 10pt
\def\bb#1{\hbox{\mybb#1}}
\def\be{\begin{equation}}
\def\ee{\end{equation}}
\def\bea{\begin{eqnarray}}
\def\eea{\end{eqnarray}}
\renewcommand{\theequation}{\arabic{section}.\arabic{equation}}

\begin{document}
\begin{flushright}
%06/03/2026 IIB-quantization-etc.tex, \\
06/03/2026,  Printed \today
\end{flushright}
\bigskip
\title{Spinor moving frame, type II superparticle quantization, hidden $SU(8)$ symmetry of linearized 10D supergravity,
and superamplitudes
}

\author{Igor Bandos}
\email{igor.bandos@ehu.eus}
\affiliation{Department of Physics and EHU Quantum Center, University of the Basque Country UPV/EHU,
P.O. Box 644, 48080 Bilbao, Spain,}
\affiliation{IKERBASQUE, Basque Foundation for Science,
48011, Bilbao, Spain, }

\author{Mirian Tsulaia}
\email{mirian.tsulaia@oist.jp}
\affiliation{Okinawa Institute of Science and Technology,
1919-1 Tancha, Onna-son, Okinawa 904-0495, Japan}

\bigskip

\begin{abstract}
A covariant quantization of type IIB and type IIA superparticles in their spinor moving frame formulation reveals the hidden $SU(8)$ symmetry of the linearized type II supergravity. It acts on auxiliary variables parameterizing the possible choices of complex structures which is necessary to arrive at the realization of the quantum state vector as an analytic on-shell superfield, the one-particle counterpart of analytical superamplitudes for type IIB and type IIA supergravity multiplets.

The description of the type IIA supergraity multiplet in terms of an analytic on-shell superfield  is then identical to that for type IIB supermultiplet; the difference is in spacetime interpretation. However, the definition of suitable auxiliary variables in type IIA case requires introduction of covariantly constant $SO(8)$ vector which  can be related to  T-duality transform  between type IIA and type IIB superspaces.

The simplest analytic IIB superamplitudes discussed in the literature thus also describe type IIA processes. We
elaborate on these using the spinor moving frame (Lorentz harmonic) formalism and point out the restrictions on computation of many particle type IIA amplitudes in this approach. We also briefly discuss the initial steps towards  type IIA  superamplitudes involving, besides
supergravity multiplets, also D$0$-branes (Dirichlet superparticles), which was recently quantized in the similar formalism, and indicate some problems which appear are to be solved to proceed on this way.

\end{abstract}

\maketitle

\begin{widetext}

%\newpage

\tableofcontents

\section{Introduction}
Quantization of relativistic particle mechanics by operator methods can be used to obtain free relativistic field equations or their solutions
\cite{Berezin:1975md,Berezin:1976eg,Casalbuoni:1976tz,Brink:1976uf,Gershun:1979fb,Brink:1981nb,Shirafuji:1983zd,Gumenchuk:1990db,Bandos:1990ji,Galperin:1992pz,Bandos:2007mi,Bandos:2007wm} \footnote{
This list is not exhaustive; see also
\cite{Sorokin:1989jj,Bandos:1993ma,West:2012vka} for review and more references.}.  In the case of paricles with spin this implies the use of a model with worldline supersymmetry \cite{Berezin:1975md,Berezin:1976eg,Casalbuoni:1976tz,Brink:1976uf,Gershun:1979fb}, twistor approach \cite{Penrose:1972ia,Penrose:1986ca,Ferber:1977qx,Shirafuji:1983zd,Gumenchuk:1990db} or twistor like-methods \cite{Bandos:1990ji,Galperin:1992pz,Bandos:2006nr,Bandos:2007mi,Bandos:2007wm}. In the case of  quantization of the particle models with spacetime supersymmetry, the  result can be the set of superfield equations in target superspace or their solutions. Among such solutions an important role is played by D=4  on-shell superfields which depend on supertwistor variables \cite{Ferber:1977qx}. Their multiparticle   counterparts are $D=4$ on-shell superamplitudes the use of which, in particular in  ${\cal N}=4$ and  ${\cal N}=8$ cases,  was - and still is - the basis of the impressive  progress in calculation of amplitudes and superamplitudes of maximally supersymmetric theories \cite{Bianchi:2008pu,Brandhuber:2008pf,Arkani-Hamed:2008gz,Elvang:2009wd,Elvang:2015rqa}.
These superamplitudes are covariant under $SU({\cal N})$ symmetry, R-symmetry of ${\cal N}$-- extended supersymmetry algebra,  $SU(8)$ in the case of maximal ${\cal N}=8$ D=4 supergravity.

Notice that the bosonic variables of the (multiparticle) on-shell superspace are helicity spinors, the bosonic spinors which can be related to the twistor approach
\cite{Penrose:1972ia,Penrose:1986ca,Ferber:1977qx}. These  variables, as well as the  on-shell superfields
naturally appear in quantization of massless D=4 superparticle in its twistor--like formulation by Ferber and Shirafuji
\cite{Ferber:1977qx,Shirafuji:1983zd}.

Higher dimensional generalizations of superamplitudes were studied in \cite{Cheung:2009dc,Caron-Huot:2010nes,Boels:2012ie,Boels:2012zr,Boels:2012if,Wang:2015jna,Bandos:2016tsm,Bandos:2017eof,Bandos:2017zap,Cachazo:2018hqa,Albonico:2020mge, Kallosh:2024lsl}. The generalization of spinor helicity formalism to $10D$ was performed and used to study $10D$ SYM in
\cite{Caron-Huot:2010nes} and IIB superamplitudes in \cite{Boels:2012ie,Boels:2012zr,Wang:2015jna,Kallosh:2024lsl}. The clarification  of the relation of this formalism
with Lorenz harmonic approach \cite{Bandos:1990ji,Galperin:1991gk,Delduc:1991ir,Bandos:1991bh,Galperin:1992pz}, which is  also known under the name of spinor moving frame  approach \cite{Bandos:1991bh,Bandos:1992np,Bandos:1992ze,Bandos:1993yc,Bandos:1994eu,Bandos:1996ju} \footnote{The first name reflects the relation with the harmonic superspace approach to D=4 ${\cal N}=2,3$ SYM theory   \cite{Galperin:1984av,Galperin:1984bu,Galperin:2001seg}
(see \cite{Bandos:1988bn,Zupnik:2001hd,Howe:2001je,Eden:2001ec} for on-shell ${\cal N}\geq 4$ harmonic superspace studies.). Notice also the light-cone vector harmonic approach to superparticle
\cite{Sokatchev:1985tc,Sokatchev:1987nk}.}, allowed one of the authors to generalize the spinor helicity formalism to 11D case \cite{Bandos:2016tsm,Bandos:2017eof,Bandos:2017zap} (see also \cite{Geyer:2019ayz,Bandos:2019zqp} for related studies).

Notice that the $SU(8)$ R--symmetry of maximal ${\cal N}=8$ supergravity is not seen in the above mentioned descriptions of D=10 and D=11 superamplitudes. One of the aims of the present paper is to
reveal  hidden $SU(8)$ symmetries in the on-shell superfields and superamplitudes of $D=10$ type IIB and type IIA theories. The most natural way to this passes through a suitable covariant quantization of type IIB and type IIA superparticle models.

Surprisingly, to our best knowledge, the quantization of type IIB superparticle  has not been elaborated yet. In this paper we address this problem  in the frame of its spinor moving frame formulation. We  resolve some issues of covariant quantization of type IIB superparticle, quantize in a similar manner massless type IIA superparticle and discuss how the T-duality relation between these manifests itself in our approach.

Our quantization procedure results in a state vectors described by chiral (analytical) on--shell superfield defined in superspace with complex octuplet of fermionic coordinates $\Theta^{-A}$ carrying the index of fundamental representation of $SU(8)$. The description in terms of such a superfield is manifestly invariant under $SU(8)$ symmetry.
This is broken--or, better to say, realized dynamically-- when  the explicit spacetime interpretation of the components of this on--shell superfield is carried out.

The invariance of the quantum theory of type IIB superparticle under the $SU(8)$ symmetry occurs due to the presence of additional complex constrained variables $w_q^A$, carrying the index $A=1,...,8$ of $SU(8)$ fundamental representation and also $SO(8)$ s-spinor index $q=1,...,8$. They parametrize the possible choices of the complex structure in the quantization of type IIB model. The fact that they are actually St\"uckelberg  fields for $SU(8)$ gauge symmetry reflects the fact that the choice of complex structure in quantization of type IIB superparticle does not have
physical consequences. In the case of quantization of type IIA model introduction of complex structure requires also  the presence of a covariantly constant $8$-vector $k^i$ which is necessary to connect two sets of complex constrained matrix St\"uckelberg fields, $w_q^A$ and $w_{\dot{q}}^A$, the latter carrying  c-spinor index of $SO(8)$, $\dot{q}=1,\ldots, 8$.

Notice that the St\"uckelberg fields were known to be inevitable for construction of the formulations of 11D supergravity with manifest
$E7$ and $SU(8)$ invariance  \cite{deWit:1985iy,deWit:1986mz}, as well as for its $E8$ and $SO(16)$ symmetric formulation  \cite{Nicolai:1986jk,Drabant:1988bk}. This was the reason of some critics and doubts in importance of such formulations. However, the 11D supergrvaity formulations of  \cite{deWit:1985iy,deWit:1986mz,Nicolai:1986jk,Drabant:1988bk} were proved to be useful first in studying of gaugings of lower dimensional supergravities (see e.g.  \cite{Hull:1988jw}) and then by discovery of $E7$ and $E8$ Exceptional field theories \cite{Hohm:2013pua,Hohm:2013uia,Hohm:2014fxa}, which can be considered as development of this line. The same applies to   very recent findings of  unusual $\beta$--symmetry of type IIA and $\gamma$--symmetry of 11D supergravity \cite{Baron:2022but,Baron:2025bli}.

Our study clarifies the appearance of the hidden $SU(8)$ symmetry in perturbative description of $10D$ type II supergravity theories and reveals its relation with introduction of complex structure in these theories. Clearly, as $E7$ symmetry is realized nonlinearly on the  scalar fields of supergravity multiplet, when the theory is linearized over flat (super)spacetime, only its linearly realized $SU(8)$ subgroup can be seen.

We hope that our observations on hidden $SU(8)$ symmetry of type IIB and type IIA on-shell superfields also will be useful for further development of superamplitude calculus in type II $10D$ theories. The first steps towards such application are performed in the final Section \ref{Sec=sAmpl} of this paper.

There we firstly elaborate
spinor moving frame based form of the  spinor helicity formalism and used it for the $SU(8)$ covariant description of simplest type IIB superamplitudes studied previously in \cite{Boels:2012ie}. Then we argue that these
can be used to describe scattering process also in type IIA supergravity and indicate some restriction for the description of multiparticle processes in this formalism.

Finally we discuss a way towards more complex superamplitudes of type IIA theory involving, besides supergravity multiplets, also D$0$-brane (Dirichlet superparticle) contributions, and point out some problems which are to be solved to proceed on this way. To this end we use the spinor helicity formalism and on-shell superfield description of quantum D$0$--brane which was obtained from quantization of its spinor moving frame formulation in our recent
\cite{Bandos:2025pxv}.

The paper is organized according to the table of contents. Some technical material is moved to Appendices.

\setcounter{equation}0

\section{Spinor moving frame formulation of type IIB superparticle with SU(8) symmetry}
\label{IIB}

\subsection{Lagrangian mechanics of type IIB superparticle in its spinor moving frame formulation}

\subsubsection{Standard formulation of type IIB superparticle}

Let us denote the coordinates of type IIB superspace $\Sigma^{(10|16+16)}$ by $Z^{M}_{IIB} = (x^a, \theta^{\alpha 1},  \theta^{\alpha 2})$, and, following the custom of
 e.g. \cite{Zima:1984jaf,Galperin:2001seg},  express this by

\be\label{IIB=ZM}
\Sigma^{(10|16+16)}\; = \; \{ Z^{M}_{IIB} \}= \{(x^a, \theta^{\alpha 1},  \theta^{\alpha 2}) \}\; . \ee
Here and below we use Latin letters
to denote $10$--vector indices,  $a,b,c=0,1,...,9$,  and Greek letters  to denote  $10D$ Majorana--Weyl spinor indices, $\alpha,\beta,\gamma=1,...,16$. The supervielvein of flat type IIB superspace $\Sigma^{(10|16+16)}_{IIB}$,
\be
\label{EA=IIB} E^{{\cal A}}= (E^a,  E^{\alpha 1},  E^{\alpha 2})=dZ^{M} E^{{\cal A}}_{M}(Z)
\ee
can be chosen in the form of Volkov-Akulov 1-form  and differentials of the fermionic coordinates,

\be\label{Ea=IIB}
E^a= \Pi^a = dx^a -i d\theta^1\sigma^a \theta^1 -i d\theta^2\sigma^a \theta^2\; , \qquad E^{\alpha 1}=d \theta^{\alpha 1}\; , \qquad E^{\alpha 2}=d \theta^{\alpha 2}\; , \qquad
\ee
where $\sigma^a_{\alpha\beta}=\sigma^a_{\beta\alpha}$ are $10D$ generalization of the relativistic Pauli matrices. These  matrices
together with their upper index counterparts $\tilde{\sigma}^{a \; \alpha\beta}=\tilde{\sigma}^{a \;\beta\alpha}$ obey
the algebra

\be\label{sts=}
\sigma^a\tilde{\sigma}^b +\sigma^b\tilde{\sigma}^a=
2\eta^{ab}{\bb I}_{16\times 16}\; , \qquad \ee
where $ \eta^{ab}={\rm diag}(1,-1,...,-1)$ is the Minkowski metric in mostly minus signature.

The supervielbein is invariant under the following type IIB supersymmetry transformations

\begin{eqnarray}\label{susy=IIB}
\delta_\varepsilon x^a= i \theta^1\sigma^a\varepsilon^1+  i \theta^2\sigma^a\varepsilon^2   \; , \qquad
\delta_\varepsilon \theta^{\alpha 1} =\varepsilon^{\alpha 1} \; , \qquad\delta_\varepsilon \theta^{\alpha 2} =\varepsilon^{\alpha 2} \; . \qquad
\;
\end{eqnarray}

The superparticle worldline in type IIB superspace can be defined parametrically

\be\label{IIB=cW1in}
{\cal W}^1 \;\in \; \Sigma^{(10|16+16)} \; : \qquad Z^{M}_{IIB} = Z^{M}_{IIB}(\tau)  = (x^a(\tau), \theta^{\alpha 1}(\tau),  \theta^{\alpha 2}(\tau))
\ee
with the use of bosonic and fermionic coordinate functions $Z^{M}_{IIB}(\tau)  = (x^a(\tau), \theta^{\alpha 1}(\tau),  \theta^{\alpha 2}(\tau))$ depending
on the proper time $\tau$. To simplify notation, we will denote the coordinate functions by the same symbols as coordinate; we believe that this cannot produce any confusion.

The pull--backs of the supervielbein 1--forms to the worldline are obtained by substituting the coordinate functions for the coordinates
\bea\label{EApb=IIB}
&& E^A(Z(\tau))= d\tau E^A_\tau \; , \qquad  \nonumber
%\\ \nonumber
\\ \label{Eat=IIB}  && E^a_\tau  = \dot{x}^a -i \dot{\theta}{}^1\sigma^a \theta^1 -i \dot{\theta}{}^2\sigma^a \theta^2\; , \qquad E^{\alpha 1}_\tau= \dot{\theta}{}^{\alpha 1}\; , \qquad E^{\alpha 2}_\tau =\dot{\theta}{}^{\alpha 2}\; . \qquad  %\\ \nonumber
\eea
Here and below dot denotes the
derivative with respect to the
proper time, $\dot{Z}^M = \frac {d Z^M} {d\tau}$ etc.

The type IIB (counterpart of the first order form of the) Brink--Schwarz massless superparticle action \cite{Brink:1981nb} is constructed from pull-back of bosonic supervielbein, auxiliary momentum variable $p_a(\tau)$ and Lagrange multiplier $e(\tau)$:
\be\label{S=BS}
S_{BS}=  \int d\tau \left(p_a E_\tau^a  - \frac 1 {2}\, e\, p_ap^a\right) \; .
\ee
Besides being manifestly invariant under target (super)space supersymmetry \eqref{susy=IIB}, this action is also invariant under local fermionic $\kappa$--symmetry \cite{Siegel:1983hh}\footnote{Curiously, the counterpart of this   $\kappa$--symmetry was first found for the case of massive ${\cal N}=2$ superparticle in \cite{deAzcarraga:1982dhu} published a bit earlier than  \cite{Siegel:1983hh}.}
\begin{eqnarray}\label{kappa=red}
\delta_\kappa x^a= -{i} \theta^1\sigma^a\delta_\kappa \theta^1 -{i} \theta^2\sigma^a\delta_\kappa \theta^2 \;  \; , \qquad
\delta_\kappa \theta^{\alpha 1} = p_a\tilde{\sigma}^{a\alpha\beta}\kappa^1_\beta \; , \qquad \delta_\kappa \theta^{\alpha 2} = p_a\tilde{\sigma}^{a\alpha\beta}\kappa^2_\beta \; , \qquad \\ \nonumber
%\\ \nonumber
\delta_\kappa  p_a = 0\; , \qquad  \delta_\kappa e= -4i \dot{\theta}^1\kappa^1 -4i \dot{\theta}^2\kappa^2   \; . \qquad
%\\ \nonumber
\end{eqnarray}
This symmetry is important because it guarantees that the ground state of the dynamical system preserves $1/2$ of the supersymmetry and hance is stable\footnote{ It can be  called $1/2$ BPS where $1/2$ refers to amount of preserved supersymmetry and BPS indicates that this state saturates the so-called  Bogomol'nyi--Prasad--Sommerfield (or BPS) bound.}.

This fermionic gauge symmetry is infinitely reducible and does not allow for a Lorentz covariant gauge fixing {\it in the frame of its standard formulation}. This created problem for covariant quantization of superparticles and superstrings (where the same problem appeared, again, just in the standard formulation). The search for a solution of this problem was quite stimulating for development of the new approaches to superparticle and superstring models, including, among others, the most successful in this respect pure spinor description of quantum superparticles and superstrings by Berkovits \cite{Berkovits:2000fe,Berkovits:2001rb,Berkovits:2006vi}, as well as a number of twistor-like formulations.
The set of these latter includes the spinor moving frame formulations of superparticles and superstrings \cite{Bandos:1990ji,Bandos:1991bh,Bandos:1992np,Bandos:1992ze,Bandos:1996ju,Bandos:2007mi,Bandos:2007wm}
which we will use in this paper \footnote{Actually in the case of superparticle the spinor moving frame approach was  elaborated only for a simple, ${\cal N}=1$ supersymmetry in different dimensions $D$  \cite{Bandos:1990ji,Bandos:1996ju,Bandos:2007mi,Bandos:2007wm}. However, the generalization of the action and classical mechanics of the spinor moving frame formulation  for $10D$ type II superparticles which we will describe in a moment, is quite straightforward.}.

In the spinor moving frame formulation, the $\kappa$--symmetry of the superparticle action acquires its irreducible form. But the main reason to use this formulation in our work is that the quantization in its frame results in the higher dimensional generalization of the so-called on-shell superfields of ${\cal N}=4$ SYM and ${\cal N}=4$ supergrvaity (SUGRA) which are one-particle counterpart of the superamplitudes of these theories  \cite{Bianchi:2008pu,Brandhuber:2008pf,Arkani-Hamed:2008gz,Elvang:2009wd,Elvang:2015rqa}.
For the case of $10D$ SYM and $11D$ SUGRA the generalization of these on-shell superfields and superparticle quantization were discussed in \cite{Bandos:2017eof,Bandos:2017zap}. In this paper we extend this approach  for type IIB and type IIA superparticle cases and use it to to reveal the hidden $SU(8)$ symmetry of linearized both $D=10$ type II supergravities.

%\subsection{Spinor moving frame formulation of
% massless  superparticle  }

\subsubsection{Spinor moving frame. Group-theoretical structure and Lorentz harmonic superspace }
\label{V=group-th}

In the spinor moving frame formulation target superspace of the model becomes enlarged by
(a coset of) $Spin(1,9)$ group parametrized by a pair of the constrained  $16\times 8$ matrix variables
$ v_{\alpha q}^{\; -}, v_{\alpha \dot{q}}^{\; +}$ which form Spin(1,9) valued matrix

\begin{eqnarray}\label{harmV=10}
V_{\alpha}^{(\beta)}= \left(\begin{matrix} v_{\alpha \dot{q}}^{\; +} , & v_{\alpha q}^{\; -}
  \end{matrix}\right)& \in & Spin(1,9)\; ,   \qquad
   \alpha=1,...,16\; , \qquad \dot{q}=1,...,8 \; , \qquad q=1,...,8 \; .
 \;   \qquad
\end{eqnarray}
A good name for these objects
when they are
considered as coordinates of
a target superspace  is {\it Lorenz harmonic variables} or {\it Lorenz harmonics}, and the superspace including these as coordinate is called {\it Lorentz harmonic superspace},
  \be\label{IIB=cZcM}
\Sigma^{(10+16|16+16)}\; = \; \{ {\cal Z}^{{\frak M}}_{IIB} \}= \{(x^a, \theta^{\alpha 1},  \theta^{\alpha 2};
  (v_{\alpha q}^{\; -}\, , \,  v_{\alpha \dot{q}}^{\; +})\}\; .
  \ee
(The first number $16$ for additional  bosonic variables in the notation for this superspace will be  clarified below\footnote{Besides the constraint  \eqref{harmV=10}, the explanation of this number  refers to certain gauge symmetries used as identification relations on the set of these constrained variables}).

When a particle (or a string)  is moving through
this Lorentz harmonic superspace, these Lorentz harmonic variables give rise to worldline (or worldsheet)  fields $v_{\alpha \dot{q}}^{\; +}(\tau)$ and $v_{\alpha q}^{\; -}(\tau)$  which can be called {\it  spinor moving frame variables} due to the reasons we are going to describe below.

To streamline the presentation, in this paper we will mainly use the names  "Lorentz harmonics" and  "spinor moving frame variables"
as synonyms, applying it to both coordinates of Lorentz harmonic superspace and to their pull-back to the worldline. We make the terminological difference between Lorentz harmonics
$v_{\alpha \dot{q}}^{\; +}, v_{\alpha q}^{\; -}$ and spinor moving frame variables
 $v_{\alpha \dot{q}}^{\; +}(\tau), v_{\alpha q}^{\; -}(\tau)$ only in the places where this is essential (for instance in
 footnote \ref{footnote7} below).

Notice that the condition \eqref{harmV=10} imposes strong constraints on the rectangular  blocks of  $v_{\alpha \dot{q}}^{\; +}$ and $v_{\alpha q}^{\; -}$ of the matrix
$V_{\alpha}{}^{\!\!\!(\beta)}$ (which is called {\it spinor moving frame matrix},  which is rigorous
when we speak about
$v_{\alpha \dot{q}}^{\; +}(\tau)$ and $v_{\alpha q}^{\; -}(\tau)$) reducing the number of their components from 128+128=256 to 45=dim$(SO(1,9))$=dim$(Spin(1,9))$. We shall discuss a more explicit form of these constraints below, but before make some comments about group-theoretical structure of spinor moving frame variables (or Lorentz harmonics).

The splitting of the spinor moving frame  matrix in \eqref{harmV=10} is invariant under local transformations of $SO(1,1)\times SO(8)$ subgroup of $SO(1,9)\approx Spin(1,9)$; the group
$SO(1,1)$ acts by rescaling
\be\label{SO1-1}
v_{\alpha \dot{q}}^{\; +} \mapsto e^{\beta } v_{\alpha \dot{q}}^{\; +} \, , \qquad  v_{\alpha q}^{\; -}  \mapsto e^{- \beta }  v_{\alpha q}^{\; -}
\ee
and $SO(8)$ gauge symmetry transforms $ v_{\alpha q}^{\; -}$ and $v_{\alpha \dot{q}}^{\; +}$ as an s--spinor and a c--spinor, respectively\footnote{\label{footnote7} When we speak about spinor moving frame variables, i.e. about fields on the worldline (or worldsheet),  $\tau$--dependent
$ v_{\alpha q}^{\; -}(\tau) $ and $v_{\alpha \dot{q}}^{\; +}(\tau)$ the parameter $\beta$ in \eqref{SO1-1} should depend on $\tau $, $\beta=\beta (\tau)$. However, if we are speaking about coordinates of Lorentz harmonic superspace,
$\beta$ should be a function on $Spin(1,9)$ group manifold. The same applies to the parameters of  $SO(8)$ transformations acting on the worldline fields and on the Lorentz harmonic superspace coordinates, respectively.
}.

Using these gauge symmetries as  an  identification relation on the space of Lorentz harmonics, we can identify these with homogeneous coordinates of the coset
\bea
\label{coset=string} \frac{Spin(1,9)}{SO(1,1)\times SO(8)}\; .
\eea
The spinor moving frame fields of a (super)string \cite{Bandos:1991bh,Bandos:1992np,Bandos:1992ze} are coordinate functions corresponding to this sector of Lorentz harmonic superspace, which then can be denoted by $\Sigma^{(10+16|16+16)}$ in type IIB case. Schematically we can express this by
\be\label{IIB=cZcM:}
\Sigma^{(10+16|16+16)}\; = \; \{ {\cal Z}^{{\frak M}}_{IIB} \}= \{(x^a, \theta^{\alpha 1},  \theta^{\alpha 2};
  (v_{\alpha q}^{\; -}\, , \,  v_{\alpha \dot{q}}^{\; +})_{\text{mod}[SO(1,1)\times SO(8)]})\}\;  \ee
  or simply by our original \eqref{IIB=cZcM}, but
with keeping in mind that we have to consider only superfields with definite homogeneous transformation properties
under $SO(1,1)\times SO(8)$ acting on  $v_{\alpha q}^{\; -}\, , \,  v_{\alpha \dot{q}}^{\; +}$ (this is to say, invariant functions and sections of fiber bundles)\footnote{Such type treatment originates in harmonic superspace approach  to  ${\cal N}=2$ and ${\cal N}=3$ SYM \cite{Galperin:1984av,Galperin:1984bu} (probably in the form not so explicit in the original paper, but a bit reformulated and elaborated in \cite{Bandos:1988bn}). In \eqref{IIB=cZcM} this specific restrictions on function on harmonic superspace parametrized by
Lorentz group coordinates  $(v_{\alpha q}^{\; -}\, , \,  v_{\alpha \dot{q}}^{\; +})$ is indicated by  $_{\text{mod}[SO(1,1)\times SO(8)]}$ subscript.}. This restriction is reflected  by the number of additional bosonic coordinates $16$ in the notation of superspace in l.h.s. of  \eqref{IIB=cZcM}
\footnote{Of course we could also introduce a harmonic superspace $\Sigma^{(10+45|16+16)}$ with bosonic sector
${\bb R}^{1+9} \times Spin(1,9)$ and make restriction to superfields covariant under  $SO(1,1)\times SO(8)$ at the next stage, but we find more convenient the superspace including coset \eqref{coset=string} and simplified notation
\eqref{IIB=cZcM} for this.}.

Of course, Lorentz harmonic superspace makes sense as target superspace for supersymmetric point--like or extended objects, $p$--branes, and it is usually specific for each $p$ \cite{Bandos:1994eu}. The exception are the cases of superstring \cite{Bandos:1992np,Bandos:1992ze} and massless superparticle \cite{Delduc:1991ir,Galperin:1991gk,Galperin:1992pz} for which, in the frame of above approach, one can use the same Lorentz harmonic superspace. However, for massive superparticle the Lorentz harmonic superspace is clearly different (see   \cite{Bandos:2000tg} as well as \cite{Bandos:2025pxv} and refs. therein).

As we shall see below, in the case of a massless superparticle, the Lagrangian formulation of its spinor moving frame
formulation involves only one of two spinor frame variables, $ v_{\alpha q}^{\; -}(\tau) $. This implies that the  model possesses
an additional $8$--parametric gauge symmetry ${\bb K}_8$ which acts by such shifts of $v_{\alpha \dot{q}}^{\; +}$ variables that do not break the condition  (leave invariant constraints) \eqref{harmV=10}, namely

\be\label{K8=10D}
\delta_{_{{\mathbb K}_8}} v_{\alpha q}^{\; -} =0  \; , \qquad \delta_{_{{\mathbb K}_8}} v_{\alpha \dot{q}}^{\; +} = K^i v_{\alpha p}^{\; -}  \gamma^i_{p\dot{q}}  \; , \qquad
\ee
where $ \gamma^i_{p\dot{q}}=\tilde{\gamma}^i_{\dot{q}p}$ are $SO(8)$ Clebsch-Gordan coefficients (roughly speaking, SO(8) Pauli matrices) which obey
\be\label{gitgj+=I}
\gamma^{(i}\tilde{\gamma}^{j)} = \frac 1 2 \left(\gamma^{i}\tilde{\gamma}^{j}+\gamma^{j}\tilde{\gamma}^{i}\right)  =\delta^{ij}\delta_{qp}\; , \qquad \tilde{\gamma}^{(i}{\gamma}^{j)} = \delta^{ij}\delta_{\dot{q}\dot{p}}\; .  \qquad
\ee
On one hand, if we use also  ${\mathbb K}_8$ symmetry as an identification relation on the space of spinor moving frame variables, we can consider them as generalized (constrained) homogeneous coordinates of the coset \be SO(1,9)/[(SO(1,1)\times SO(8))\subset\!\!\!\!\!\!\times {\bb K}_8] ={\bb S}^8  \ee  isomorphic to eight-sphere ${\bb S}^8$ which further can be identified with celestial sphere of $10D$ observer \cite{Delduc:1991ir,Galperin:1991gk}.

On the other hand, it is completely legitimate to consider the spinor moving frame formulation of superparticle as superparticle moving in Lorentz harmonic superspace
with internal sector \eqref{coset=string} and treat  ${\mathbb K}_8$ as an additional gauge symmetry of the action of a system moving in this superspace \footnote{In $D=4$ such an approach was taken in \cite{Bandos:1990ji}.}. Actually, this point of view is more convenient for our study  since we use essentially the possibility to change the coordinate basis in Lorentz harmonic superspace and that looks much more natural if the  latter contains the coset \eqref{coset=string}.

\subsubsection{Constraints on spinor moving frame variables,   moving frame vectors, Cartan forms and covariant harmonic derivative}

The name spinor moving frame variables is actually motivated by  the fact that the above discussed $v_{\alpha \dot{q}}^{\; +}$ and  $v_{\alpha q}^{\; -}$ can be considered as a kind of square roots of the vectors which form moving frame attached to the worldline of massless (super)particle (or to the worldsheet of  a (super)string).

Such a moving frame can be constructed from $10$ orthogonal and normalized vectors, which form the Lorentz group valued matrix
\begin{eqnarray}\label{Uab=in10}
u_a^{(b)}= \left( {1\over 2}\left( u_a^{=}+u_a^{\#}
 \right), \; u_a^j \, , {1\over 2}\left( u_a^{\#}-u_a^{=}
 \right)\right)\; \in \; SO^\uparrow (1,9)\,   , \qquad \,  j=1,\ldots, 8 . \qquad
\end{eqnarray}
Actually, we have combined one time--like and one of the spacelike vectors,
into  a
light-like $ u_a^{=}= u_a^{(0)}-u_a^{(9)}$ and $ u_a^{\#}= u_a^{(0)}+u_a^{(9)}$. This is convenient for  'attaching' the moving frame to the worldline of a massless particle by  identification of the momentum  with one of the light-like vectors,
say with $u_a^{=}$,
\be\label{p=ru--}
p_a (\tau) =\rho^{\#}(\tau)u_a^{=}(\tau)  \; .
\ee
Here the multiplier  $\rho^{\#}(\tau)$ is introduced to provide the invariance of the momentum under $SO(1,1)$ symmetry acting on light--like vectors by rescaling ({\it cf.} \eqref{SO1-1}).

\be\label{SO1-1v}
u_a^{=} \mapsto e^{-2\beta } u_a^{=} \, , \qquad  u_a^{\#} \mapsto e^{2\beta } u_a^{\#} \, , \qquad   u_a^{i} \mapsto u_a^{i} \, . \qquad
\ee
$\rho^{\#}(\tau)$ in \eqref{p=ru--} serves as a St\"uckelberg field for this gauge symmetry,

\be\label{SO1-1r}
\rho^{\#} \mapsto e^{2\beta }\rho^{\#} \, .  \qquad
\ee

The splitting in \eqref{Uab=in10} is manifestly invariant under the
direct product of the above $SO(1,1)$ and of the
$SO(8)$ symmetry acting on index $j=1,..,8$ of $u_a^{j}$ moving frame  vector by its vector (v-) representation.

The spinor moving frame variables can be considered as a kind of square roots
of the moving frame vectors in the sense of the following constraints:
\begin{eqnarray}\label{u--s=v-v-}
 && u_a^= \sigma^a_{\alpha\beta}= 2v_{\alpha q}{}^- v_{\beta q}{}^-  \; , \qquad\qquad
 v^-_{{q}} \tilde{\sigma}_{a}v^-_{{p}}= u_a^= \delta_{{q}{p}}, \qquad
 %\\ \nonumber
 \\ \label{v+sv+=u++}
&& u_{ {a}}^{\# }{\sigma}^{ {a}}_{ {\alpha} {\beta}} =2 v_{{\alpha}\dot{q}}{}^{+}v_{{\beta}\dot{q}}{}^{+}  \; , \qquad \qquad v_{\dot{q}}^+ \tilde{\sigma}_{ {a}} v_{\dot{p}}^+ =  u_{ {a}}^{\# } \delta_{\dot{q}\dot{p}}\; . \qquad
%\\ \nonumber
\\
  \label{uIs=v+v-}
&&  u_{ {a}}^{i} {\sigma}^{a}_{\alpha\beta}  = 2 v_{( {\alpha}|{q} }{}^- \gamma^i_{q\dot{q}}v_{|{\beta})\dot{q}}{}^{+} \; , \qquad
v_{{q}}^- \tilde {\sigma}_{ {a}} v_{\dot{p}}^+=u_{ {a}}^{i} \gamma^i_{q\dot{p}}\;  \qquad
%\\ \nonumber
 \end{eqnarray}
which imply that
\bea\label{u--2=0}
&& u_a^= u^{a=}=0 \; , \qquad  u_a^\# u^{a\#}=0 \; , \qquad   u_a^= u^{a\#}=2 \; , \qquad
%\\ \nonumber
\\ \label{u--ui=0} && u_a^= u^{ai}=0 \; , \qquad  u_a^\# u^{ai}=0 \; , \qquad   u_a^i u^{aj}=-\delta^{ij} \; , \qquad
\\
\label{eu...u=e}
&& \epsilon^{abc_1\ldots c_8} u_a^=u_b^{\#} u_{c_1}^{i_1}\ldots u_{c_8}^{i_8}=\epsilon^{i_1\ldots i_8}\; .
\eea
The set of the relations  \eqref{u--2=0}--\eqref{eu...u=e} is equivalent to the condition \eqref{Uab=in10}.

The constraint \eqref{u--s=v-v-} allows to treat the light-like moving frame vector $u_a^{=}$ as a composite of $v_{{\alpha}{q} }^{\; -} $, $\;u_a^{=} =\frac 1 8 v^-_q \tilde{\sigma}_av^-_q$. Then  \eqref{p=ru--} implies $p_a= \frac 1 8 \rho^\# \,  v^-_{q} \tilde{\sigma}_{a}v^-_{q}$ and
\bea\label{p=rv-sv-}
 p_a \sigma^a_{\alpha\beta}= 2\rho^\# \, v_{\alpha q}^{\; -} v_{\beta q}^{\; -}  \; , \qquad \rho^\# \,  v^-_{q} \tilde{\sigma}_{a}v^-_{p}= p_a\delta_{qp}  \; . \qquad
%\\ \nonumber
\eea
These relations guarantee that the momentum $p_a$ is light-like, $p_ap^a=0$, and provide us with  straightforward $10$ dimensional generalization of the famous Cartan-Penrose representation for the four--dimensional light--like vector in terms of bilinear of the bosonic Weyl spinors, one of the basic relations of twistor approach \cite{Penrose:1972ia,Penrose:1986ca,Ferber:1977qx,Shirafuji:1983zd} and of the $D=4$ spinor helicity formalism \cite{Bianchi:2008pu,Brandhuber:2008pf,Arkani-Hamed:2008gz,Elvang:2009wd} (see \cite{Elvang:2015rqa} for a review and more references).

Further  details on spinor moving frame variables can be found in Appendix \ref{App=harm}

\subsubsection{Type IIB superparticle action in spinor moving frame formulation and its irreducible $\kappa$--symmetry }

The action for massless type IIB  superparticle in spinor moving frame formulation can be written in the following forms
\bea\label{S=IIB}
S = \frac 1 8  \int d\tau \rho^\# \,   E^a_\tau \, v^-_{q} \tilde{\sigma}_{a}v^-_{q} \quad =\quad  \int\limits_{{\cal W}^1}  \rho^\# \,   E^a \, u_a^=  \quad = \int\limits_{{\cal W}^1}\rho^\# \,   E^= \; . \qquad
%\\ \nonumber
\eea
Here $E^a=d\tau E_\tau^a$ is the pull--back of type IIB Volkov-Akulov 1--form \eqref{Ea=IIB} and,  in the second equality, we have used the constraint \eqref{u--s=v-v-} to  re-write the action in terms of composite moving frame vector $u_a^=$. Finally, in the third equality, we have used one of the covariant singlet components $E^= = E^au_a^=$  of the (pull--back of) supervielbein \eqref{Ea=IIB}. This can be slit
in a Lorentz invariant manner
using the moving frame and spinor moving frame variables:
 \bea\label{E--=}
 && E^= = E^au_a^=\; , \qquad   E^\# = E^au_a^\#\; , \qquad  E^i = E^au_a^i\; , \qquad
 \\
 %\nonumber \\
 \label{E-q1=} && E^{-q1,2} = d\theta^{\alpha 1,2} v_{\alpha q}^{\; -} \; , \qquad  E^{+\dot{q}1,2} = d\theta^{\alpha 1,2} v_{\alpha \dot{q}}^{\; +} \; . \qquad
 \eea

The first form of the action  \eqref{S=IIB} of spinor moving frame formulation of massless IIB superparticle makes manifest that it provides $10D$ generalization of the so-called twistor--like of Ferber--Shirafuji formulation of 4D massless superparticle \cite{Ferber:1977qx,Shirafuji:1983zd}. However, the bosonic spinor variables $v_{\alpha q}^{\; -}$ involved in the action \eqref{S=IIB} (as well as in its ${\cal N}=1$ counterpart  \cite{Bandos:1996ju}) are highly constrained. At the first glance, this makes  the problem of its variation highly nontrivial. However, the clear group theoretical meaning of the spinor moving frame variables (Lorentz harmonics), which we have discussed in sec. \ref{V=group-th}, helps to solve the problem.

As it is explained in Appendix \ref{App=harm}, the derivatives and variations of spinor moving frame variables can be expressed in terms of $SO(1,9)$ Cartan forms the simplest expressions for which are
\bea\label{Om--i:==}
 \Omega^{= i}:=u_a^{=}du^{ai}\; , \qquad   \Omega^{\# i}:=u_a^{\#}du^{ai}\; , \qquad \Omega^{(0)}:={1\over 4}u_a^{=}du^{a\#} \; , \qquad \Omega^{ij}:=u_a^{i}du^{aj}\; . \qquad
\eea
These
provide a basis in the space
co-tangent to the $Spin(1,9)$ group manifold.
 Decomposing the differential on this manifold on this basis,
\bea\label{dSpin==}
d^{Spin(1,9)}:=dv_{\alpha q}^{\; -} \frac {\partial } {\partial v_{\alpha q}^{\; -} }+d{v}_{\alpha \dot{q}}^{\; +}  \frac {\partial } {\partial v_{\alpha \dot{q}}^{\; +} }= \Omega^{(0)}D^{(0)} -\frac 1 2 \Omega^{ij}D^{ij} +\Omega ^{=i}D^{\# i }+\Omega ^{\# i }D^{=i}\; ,  \qquad
\eea
we find expressions for the so-called covariant harmonic derivatives
\bea\label{D0:==}
 D^{(0)}&=& {v}_{\alpha \dot{q}}^{\; +}  \frac {\partial } {\partial v_{\alpha \dot{q}}^{\; +} } - v_{\alpha q}^{\; -} \frac {\partial } {\partial v_{\alpha q}^{\; -} } \; , \qquad  D^{ij} = \frac 1 2 {v}_{\alpha \dot{q}}^{\; +}\tilde{\gamma}{}^{ij}_{\dot{q}\dot{p}}  \frac {\partial } {\partial v_{\alpha \dot{p}}^{\; +} } + \frac 1 2 v_{\alpha q}^{\; -} \gamma^{ij}_{qp}\frac {\partial } {\partial v_{\alpha p}^{\; -} } \; , \qquad \\ \label{D++i:==}  D^{\# i } &=&  \frac 1 2 {v}_{\alpha \dot{q}}^{\; +}\tilde{\gamma}{}^{i}_{\dot{q}p}  \frac {\partial } {\partial v_{\alpha p}^{\; -} }  \; , \qquad  D^{=i} = \frac 1 2 v_{\alpha q}^{\; -}{\gamma}{}^{i}_{q\dot{p}}  \frac {\partial } {\partial v_{\alpha \dot{p}}^{\; +} } .  \qquad
\eea
They are given by such combinations of the usual partial derivatives of Lorentz harmonics which 'preserve' the constraints
 \eqref{u--s=v-v-}--\eqref{uIs=v+v-}. This is tantamount to saying that these constraints are in the kernel of the covariant harmonic derivatives \eqref{D0:==}--\eqref{D++i:==}.

The action \eqref{S=IIB} is invariant under the {\it irreducible} version of the local fermionic $\kappa$--symmetry
\eqref{kappa=red}\footnote{The transformation rules
%\eqref{kappa=irr=x} and
\eqref{kappa=irr=th12}
can be obtained from
\eqref{kappa=red} by using \eqref{p=ru--}
and \eqref{u--s=v-v-}, and denoting
%\begin{eqnarray}\label{kappa12=}
$ \kappa^{+1\dot{q}} = 2\rho^{\#} \kappa^1_\beta  v^{-\alpha}_{\dot{q}}$ and  $\kappa^{+2\dot{q}} = 2\rho^{\#} \kappa^2_\beta  v^{-\alpha}_{\dot{q}}$.
%\qquad\\ \nonumber\end{eqnarray}
}:
\begin{eqnarray}\label{kappa=irr=th12}
%\label{kappa=irr=x}
\delta_\kappa x^a= -{i} \theta^1\sigma^a\delta_\kappa \theta^1 -{i} \theta^2\sigma^a\delta_\kappa \theta^2 \;  \; , \qquad
\delta_\kappa \theta^{\alpha 1} = v^{-\alpha}_{\dot{q}} \kappa^{+1\dot{q}} \; , \qquad
%\\ \nonumber
%\\ \label{kappa=irr=th12}
\delta_\kappa \theta^{\alpha 2} = v^{-\alpha}_{\dot{q}} \kappa^{+1\dot{q}}  \; , \qquad
%\\ \nonumber
\\ \label{kappa=irr=v}
\delta_\kappa v^{\; -}_{\alpha q} = 0=\delta_\kappa  v^{-\alpha}_{\dot{q}} \; , \qquad
\delta_\kappa v^{\; +}_{\alpha \dot{q}}  = 0 = \delta_\kappa v^{-\alpha}_{\dot{q}} \; , \qquad  \delta_\kappa v^{-\alpha}_{\dot{q}} = 0\; , \qquad
\delta_\kappa \rho^{\#} = 0 \; . \qquad
%\\ \nonumber
\end{eqnarray}
Furthermore, spinor moving frame formulation allows for Lorentz invariant gauge fixing of both the above irreducible  $\kappa$--symmetry and also for the worldsheet reparametrization:
\bea\label{kappa-gauge}
\theta^{+1}_{\dot{q}} := \theta^{\alpha 1}v_{\alpha \dot{q}}^{\; +}=0\; , \qquad \theta^{+2}_{\dot{q}} := \theta^{\alpha 2}v_{\alpha \dot{q}}^{\; +}=0\; , \qquad
 %\\ \nonumber
 \\ \label{rep-gauge} X^\# := x^au_a^\# =\rho^\# \, \tau \; . \qquad
 %\\ \nonumber
\eea
Indeed, it is not difficult to check that, under the $\kappa$--symmetry
\eqref{kappa=irr=th12}-- \eqref{kappa=irr=v},
\bea\label{kappa=th+}
\delta_\kappa \theta^{+1}_{\dot{q}} =  \kappa^{+1\dot{q}} \; , \qquad \delta_\kappa \theta^{+2}_{\dot{q}} =  \kappa^{+2\dot{q}} \; , \qquad
%\\ \nonumber
\eea
which
makes it
manifest that the Lorentz invariant conditions \eqref{kappa-gauge} fix the gauge with respect to the local fermionic $\kappa$--symmetry completely. The same applies to the Lorenz invariant condition \eqref{rep-gauge} and reparametrization symmetry of the spinor moving frame formulation if the nondegeneracy condition $\dot{X^a}u_a^{\#}\not =0$ are assumed \footnote{This can be checked to be consistent with the massless superparticle dynamics and are actually necessary, as this implies that
$\dot{X^a}u_a^{=}$ and $\dot{X^a}u_a^{i}$ are expressed in terms of fermionic bilinear and thus are nilpotent. }.

\subsection{Spinor moving frame action in analytical basis of type IIB Lorentz harmonic superspace}
\label{sec=analyt}

The advantage of treating the spinor moving frame formulation of superparticle as superparticle moving in Lorentz harmonic superspace \eqref{IIB=cZcM} is the possibility to change the coordinate basis by mixing the standard superspace coordinates with Lorentz harmonics. Especially convenient is the following {\it analytical} coordinate basis

\bea\label{IIB=cZcM-}
\Sigma^{(10+16|16+16)}\; = \; \{ {\cal Z}^{{\frak M}}_{An\, IIB} \}&=& \{(X^=, X^\# , X^i_{An}\; ; \theta^{-1}_q ,  \theta^{-2}_q\; , \theta^{+1}_{\dot{q}},
   \theta^{+2}_{\dot{q}}\; ,
  (v_{\alpha q}^{\; -}\, , \,  v_{\alpha \dot{q}}^{\; +}))\}\;  ,
  \\ \nonumber
  \\ \label{X--:=}
  X^== x^au_a^=\; , \qquad X^\# = x^au_a^\# \; ,&& \qquad
 % \\ \nonumber
  \\  \label{XiAn:=} && X^i_{An} = x^au_a^i+ i\theta^{-1}_q \gamma^i_{q\dot{p}}\theta^{+1}_{\dot{p}} + i\theta^{-2}_q \gamma^i_{q\dot{p}}\theta^{+2}_{\dot{p}} \; , \qquad
  %\\ \nonumber
  \\   \label{th-12=IIB}
  \theta^{-1}_q = \theta^{\alpha 1} v_{\alpha q}^{\; -} \; , \qquad  \theta^{-2}_q = \theta^{\alpha 2} v_{\alpha q}^{\; -}  \; , &&\qquad
  %\\  \nonumber
  \\ \label{th+12=IIB}
 && \theta^{+1}_{\dot{q}} =\theta^{\alpha 1}  v_{\alpha \dot{q}}^{\; +} \; , \qquad \theta^{+2}_{\dot{q}} =\theta^{\alpha 2}  v_{\alpha \dot{q}}^{\; +}
   \; . \qquad
%\\ \nonumber
  \eea
As usually when defining an analytical basis \cite{Ogievetsky:1975nu,Galperin:2001seg,Sokatchev:1985tc,Sokatchev:1987nk}, the choice of its  coordinates is aimed to simplify the search for  invariant sub--superspaces. In our case the choice of
$X^i_{An}$ in \eqref{XiAn:=} is made in such a way that the invariant sub-superspace $\Sigma^{(9+16|8+8)}\subset \Sigma^{(10+16|16+16)}$ is parametrized by
\be\label{fZcM=IIB}
{\frak Z}^{{\cal M}}=  (X^=, X^i_{An}\; ; \theta^{-1}_q ,  \theta^{-2}_{q}\; ,
  (v_{\alpha q}^{\; -}\, , \,  v_{\alpha \dot{q}}^{\; +}))\; .
\ee
Indeed, one can easily check that $\Sigma^{(9+16|8+8)}=\{ {\frak Z}^{{\cal M}}\}$ is invariant  under spacetime supersymmetry
\bea\label{susy=IIB-9+16+}
\delta_{\varepsilon} X^= = 2i \theta^{-1}_q\varepsilon^{-1}_q+ 2i \theta^{-2}_{\dot{q}}\varepsilon^{-2}_{\dot{q}}\;,  \qquad \delta_{\varepsilon} X^i_{An} = 2i \theta^{-1}_q\gamma^{i}_{q\dot{p}}\varepsilon^{+1}_{\dot{q}}+ 2i \theta^{-2}_{{q}}{\gamma}^i_{q\dot{p}} \varepsilon^{+2}_{\dot{p}}\; ,  \qquad
%\\ \nonumber
\\  \delta_{\varepsilon} \theta^{-1}_q= \varepsilon^{-1}_q\; ,  \qquad  \delta_{\varepsilon} \theta^{-1}_q= \varepsilon^{-1}_q\; ,  \qquad
%\\ \nonumber
\\
 \delta_{\varepsilon} v_{\alpha q}^{\; -}=0 \, , \qquad   \delta_{\varepsilon} v_{\alpha \dot{q}}^{\; +}=0 \, , \qquad
 %\\ \nonumber
 %\eea where \bea
\\ \text{where} \qquad \varepsilon^{1-}_q= \varepsilon^{\alpha 1} v_{\alpha q}^{\; -} \, , \qquad   \varepsilon^{1+}_{\dot{q}}= \varepsilon^{\alpha 1}  v_{\alpha \dot{q}}^{\; +} \; , \qquad
%\\ \nonumber \\
\varepsilon^{2-}_q= \varepsilon^{\alpha 2} v_{\alpha q}^{\; -} \, , \qquad   \varepsilon^{2+}_{\dot{q}}= \varepsilon^{\alpha 2}  v_{\alpha \dot{q}}^{\; +} \; . \qquad
%\\ \nonumber
\eea

It is important that making the change of variables \eqref{X--:=}--\eqref{th+12=IIB}  in the action \eqref{S=IIB} we find that it depends on the pull--backs of the   coordinates (coordinate functions) from  invariant superspace \eqref{fZcM=IIB} only:
\bea \label{cL1=IIB-An}
S^{IIB} &=&  \int_{{\cal W}^1} {\cal L}_1^{IIB} \; , \qquad {\cal L}_1^{IIB}=\rho^\# E^= = \rho^\# \left( DX^= -2i D\theta^{-1}_q\, \theta^{-1}_q -2i D\theta^{-2}_q\, \theta^{-2}_q - \Omega^{= i} X_{An}^i\right)=:d\tau L^{IIB}
\; . \qquad
%\\ \nonumber
\eea
Here the covariant derivatives are defined with the use of the composite $SO(1,1)$ and $SO(8)$ connection given by Cartan forms $\Omega^{0}$ and $\Omega^{ij}$ \eqref{Om--i:==} ({\it cf.}  Eqs. \eqref{Dv-q=}--\eqref{Dv-1+q}) in  Appendix \ref{App=harm}),
\bea\label{DX--:=}
 DX^=&=& dX^= +2 X^= \Omega^{(0)}  \; , \qquad
 %\\ \nonumber
 \\ \label{Dth1-:=}
 D\theta^{-1}_q &=&  d\theta^{-1}_q+ \Omega^{(0)} \theta^{-1}_q + {1\over 4} \Omega^{ij}
\theta^{-1}_p\gamma_{{p}{q}}^{ij} \; , \qquad
%\\ \nonumber
         \\ \label{Dth2-:=}
 D\theta^{-2}_q&=&  d\theta^{-2}_q+ \Omega^{(0)} \theta^{-2}_q + {1\over 4} \Omega^{ij}
\theta^{-2}_p\gamma_{{p}{q}}^{ij}
\; . \qquad
%\\ \nonumber
\eea

The 1 bosonic and $8+8=16$ fermionic coordinates (coordinate functions) absent in the actions are $X^\#$, $\theta^{+1}_{\dot{q}}$ and $\theta^{+2}_{\dot{q}}$ which are transformed additively by (are St\"uckelberg fields for)  reparametrizations and  local fermionic kappa--symmetry \eqref{kappa=irr=th12}, respectively.
Their disappearance after changing coordinate basis can be treated as automatic gauge fixing of these symmetries.

\subsection{Hamiltonian mechanics}

\subsubsection{Canonical momenta and primary constraints}

In the central coordinate basis \eqref{IIB=cZcM} the canonical momenta conjugate to usual bosonic and fermionic coordinate functions \eqref{IIB=cW1in} are defined by  \be\label{PM:=} P_M:= (p_a, \pi_{\alpha}^{1}, \pi_{\alpha}^{2}) = \frac {\partial L} {\partial \dot{Z}^M}\equiv \frac {\partial {\cal L}_1} {\partial d{Z}^M} \; .
\ee
For the Lagrangian of the spinor moving frame action \eqref{S=IIB}, \be\label{cL1=IIB} {\cal L}_1 = d\tau L= \rho^\# E^au_a^==\frac 1 8 \rho^\# v^-_q\tilde{\sigma}_av^-_q E^a \; , \qquad  \ee  calculation of these momenta result in the following set of primary constraints

\bea\label{p-ru--=0}
\frak{f}_a= p_a -\rho^\# u_a^= \approx 0\; ,\qquad \\ \label{fd1=0}
\frak{d}_\alpha^1= \Pi_{\alpha}^{1}+ip_a \sigma^a_{\alpha\beta}\theta^{\beta 1}  \approx 0\; , \qquad \\ \label{fd2=0}  \frak{d}_\alpha^2= \Pi_{\alpha}^{2}+ip_a \sigma^a_{\alpha\beta}\theta^{\beta 2}  \approx 0\; . \qquad
\eea
Notice that the constraint \eqref{p-ru--=0} reproduces \eqref{p=ru--} and hance, due to the constraint \eqref{u--s=v-v-}, also
the $10D$ generalization \eqref{p=rv-sv-} of the Cartan--Penrose representation for $4D$ light-like momentum.

The Poisson brackets for coordinate functions  are defined as graded--antisymmetric linear operation with the following basic relations
\bea\label{PB=PMZN}
{} [ P_M\, , Z^N\}_{_{P.B.}} = - (-)^{\epsilon (M)\, \epsilon (N)}  [  Z^N\, ,  P_M\}_{_{P.B.}} =-\delta_{M}^N\; , \qquad  {} \qquad [ Z^M\, , Z^N\}_{_{P.B.}} =0\; , \qquad   [ P_N\, ,  P_M\}_{_{P.B.}} =0\; , \qquad
%\\ \nonumber
\eea
where $\epsilon (M):= \epsilon (Z^M)$ is Grassmann parity of the coordinates (also called fermionic parity):
$\epsilon (a):= \epsilon (x^a)=0$, $\epsilon (\alpha)=: \epsilon (\theta^{\alpha 1,2})=1$.
Using \eqref{PB=PMZN} we easily find the following algebra of fermionic constraints

\bea\label{PB=fd1fd1}
{} [\frak{d}_\alpha^1\, , \frak{d}_\beta^1\}_{_{P.B.}} =-2ip_a\sigma^a_{\alpha\beta}\approx -4i \rho^\# v_{\alpha q}^{\; -} v_{\beta q}^{\; -} \; , \qquad \\ \label{PB=fd1fd2}
{} [\frak{d}_\alpha^1\, , \frak{d}_\beta^2\}_{_{P.B.}} =0\;  , \qquad \\ \label{PB=fd2fd2}
{} [\frak{d}_\alpha^2\, , \frak{d}_\beta^2\}_{_{P.B.}} =-2ip_a\sigma^a_{\alpha\beta}\approx - 4i \rho^\# v_{\alpha q}^{\; -} v_{\beta q}^{\; -} . \qquad
%\\ \nonumber
\eea
Appearance of the $16\times 16$ matrix of rank $8$ in the r.h.s.-s of  \eqref{PB=fd1fd1} and \eqref{PB=fd2fd2} clearly indicates that each of $\frak{d}_\alpha^1$ and $\frak{d}_\alpha^2$  is the mixture of the second class and  first class constraints. (These latter are the generators of the $\kappa$-symmetry \eqref{kappa=irr=th12}
%, \eqref{kappa=irr=x}
in the frame of Hamiltonian formalism by Dirac \cite{Dirac:1963}).

Neither  \eqref{p-ru--=0} is the first class constraint, despite
its Poisson brackets with fermionic constraints vanish. To see this, we should take into account one more primary constraint,

\be
P^{(\rho)}_\# :=  \frac {\partial L} {\partial \dot{\rho}^{\#}}\equiv \frac {\partial {\cal L}_1} {\partial d\rho^\#}  \approx 0\; .  \qquad
\ee
Its Poisson brackets with  \eqref{p-ru--=0}, ${}[\frak{f}_a , P^{(\rho)}_\#]_{P.B.}=-u_a^=$ shows that it forms the pair of conjugate second class constraint with $u^{a\#}\frak{f}_a$.

To proceed we need to calculate also the  momenta for the spinor moving frame variables. Their canonical momenta and nonvanishing basic  Poisson brackets can be defined similarly to \eqref{PM:=},
\bea\label{Pv-:=}
P_{-q}^{\; \alpha}  = \frac {\partial L} {\partial \dot{v}_{\alpha q}^{\; -} } =  \frac {\partial {\cal L}_1} {\partial dv_{\alpha q}^{\; -} }\; ,&& \; \qquad P_{+\dot{q}}^{\; \alpha}  = \frac {\partial L} {\partial \dot{v}_{\alpha \dot{q}}^{\; +} } =  \frac {\partial {\cal L}_1} {\partial dv_{\alpha \dot{q}}^{\; +} }\; , \qquad
%\\ \nonumber
\\ \label{PB=Pvv-}
{} [P_{-q}^{\; \alpha}  \, , \, v_{\beta q}^{\; -}]_{_{P.B.}} =- [ v_{\beta p}^{\; -}\, , \, P_{-q}^{\; \alpha}   ]_{_{P.B.}} &=& - \delta^\alpha{}_\beta \delta_{qp} \;  , \qquad
%\\ \nonumber
\\ \label{PB=Pvv+} && {} [P_{+\dot{q}}^{\; \alpha}  \, , \, v_{\beta \dot{p}}^{\; +}]_{_{P.B.}} =- [  v_{\beta \dot{p}}^{\; +}\; , \; P_{+\dot{q}}^{\; \alpha}]_{_{P.B.}} = - \delta^\alpha{}_\beta  \delta_{\dot{q}\dot{p}} \;  . \qquad
%\\ \nonumber
\eea

Calculation of the canonical momenta \eqref{Pv-:=} with the Lagrangian form \eqref{cL1=IIB} clearly gives the constraints

\be\label{Pv-=0=Pv+} P_{-q}^{\; \alpha} \approx 0  \qquad \text{and}\qquad  P_{+\dot{q}}^{\; \alpha} \approx 0 \; .  \qquad \ee

If working
 with these canonical momenta and
 these constraints, one should also consider the defining relations of the spinor moving frame variables \eqref{u--s=v-v-}--\eqref{uIs=v+v-} (Lagrangian constraints) as constraints in the Hamiltonian formalism. The number of independent relations among these Lagrangian constraints is clearly
$256-45=211$. Putting aside a nontrivial problem how
to describe these independent constraints hidden in  \eqref{u--s=v-v-}--\eqref{uIs=v+v-} \footnote{See \cite{Bandos:1993yc} for discussing this problem for the case of spinor moving frame variables for M2--brane, that time called by its original name  $11D$ supermembrane.}, this should be canonically conjugate to some $211$ linear combinations of the constraints \eqref{Pv-=0=Pv+}. Then, in principle, one can find these $211$ constraints, pass to the suitable Dirac brackets and consider these $211$ constraints coming from \eqref{Pv-=0=Pv+} as well as \eqref{u--s=v-v-}--\eqref{uIs=v+v-} as strong relation, this is to say as the relations which can be used before calculating all the Dirac brackets \cite{Dirac:1963}.

Fortunately there exists a shortcut which allows to avoid the above described very technically involving way. It consists in  using, instead of canonical momenta \eqref{Pv-:=},  only the so-called covariant momenta of the spinor moving frame variables.

\subsubsection{Covariant momenta of spinor moving frame variables}
\label{CovMom=sususe}
Covariant momenta for spinor moving frame variables can be expressed through canonical momenta
by

\bea\label{fd0:=}
\frak{d}^{(0)}&=& {v}_{\alpha \dot{q}}^{\; +}  P_{+\dot{q}}^{\alpha } - v_{\alpha q}^{\; -}  P_{-q}^{\alpha }\; , \qquad  \frak{d}^{ij} = \frac 1 2 {v}_{\alpha \dot{q}}^{\; +}\tilde{\gamma}{}^{ij}_{\dot{q}\dot{p}}  P_{+\dot{p}}^{\alpha } + \frac 1 2 v_{\alpha q}^{\; -} \gamma^{ij}_{qp} P_{-p}^{\alpha } \; , \qquad  \\ \label{fd++i:=} \frak{d}^{\# i }&=&  \frac 1 2 {v}_{\alpha \dot{q}}^{\; +}\tilde{\gamma}{}^{i}_{\dot{q}p}   P_{-p}^{\alpha }\; , \qquad\label{fd--i:=} \frak{d}^{=i} =  \frac 1 2 v_{\alpha q}^{\; -}{\gamma}{}^{i}_{q\dot{p}} P_{+\dot{}}^{\alpha } \; . \qquad
%\\ \nonumber
\eea
Their characteristic property is that they have vanishing Poisson brackets with all the constraints \eqref{u--s=v-v-}--\eqref{uIs=v+v-}.
Thus the above described Dirac brackets (if constructed) would coincide with their Poisson brackets and, hence, the constraints \eqref{u--s=v-v-}--\eqref{uIs=v+v-} can be considered as strong equalities if only the covariant momenta \eqref{fd0:=}--\eqref{fd++i:=} are used.

These covariant momenta are "classical counterparts" of the covariant harmonic derivatives \eqref{D0:==}--\eqref{D++i:==}. Particularly, their quantum version in coordinate representation is given by operators proportional to these covariant derivatives.
They can be obtained by decomposing the canonical 1-form of the spinor moving frame sector of the phase (super)space on the Cartan forms ({\it cf.} \eqref{dSpin==}) \footnote{The r.h.s. of \eqref{dvPv=Omfd} can be obtained from its l.h.s. by using the admissible derivatives of $v_{\alpha q}^{\; -}$ and ${v}_{\alpha \dot{q}}^{\; +}$, Eqs. \eqref{Dv-q=} and \eqref{Dv+q=}.}

\bea\label{dvPv=Omfd}
dv_{\alpha q}^{\; -} P_{-q}^{\alpha } +d{v}_{\alpha \dot{q}}^{\; +}   P_{+\dot{p}}^{\alpha }= \Omega^{(0)}\frak{d}{}^{(0)} -\frac 1 2 \Omega^{ij}\frak{d}^{ij} +\Omega ^{=i}\frak{d}^{\# i }+\Omega ^{\# i }\frak{d}^{=i}\; . \qquad
\eea
which, in the light of \eqref{Pv-:=}  is tantamount to saying that they can be obtained by taking derivative of the Lagrangian one-form with respect to Cartan 1-forms,

\bea\label{fd=dL-Om}
 \frak{d}{}^{(0)} = \frac {\partial {\cal L}_1} {\partial\Omega^{(0)} }\; , \qquad  \frak{d}^{ij} =- \frac {\partial {\cal L}_1} {\partial\Omega^{ij} }\; , \qquad \frak{d}^{\# i }=  \frac {\partial {\cal L}_1} {\partial\Omega^{=i} }\; , \qquad \frak{d}^{=i}= \frac {\partial {\cal L}_1} {\partial\Omega ^{\# i } }\; . \qquad
\eea

The calculation of these covariant momenta using the Lagrangian form of the spinor moving frame formulation of superparticle in central coordinate basis, Eq. \eqref{cL1=IIB} results in the constraints

\bea\label{fd0=0}
 \frak{d}{}^{(0)} \approx 0 \; , \qquad  \frak{d}^{ij} \approx 0\; , \qquad \frak{d}^{\# i } \approx 0\; , \qquad \frak{d}^{=i}\approx 0\;  . \qquad
\eea
These simple set of constraints however does not imply that all the spinor moving frame variables are pure gauge (St\"uckelberg) fields, because not all of these constraints are of the first class.

We will not study further the Hamiltonian mechanics of the type IIB massless superparticle in the central basis of Lorentz harmonic superspace because the experience (of e.g. \cite{Bandos:2024hnj} and refs. therein) suggests that the problem can be essentially simplified by passing to the analytical basis  described in sec. \ref{sec=analyt}.

\subsubsection{Hamiltonian mechanics in the analytical basis }

In the analytical basis of Lorentz harmonic superspace the configuration  space of our dynamical system is automatically reduced and include
the subset of coordinate functions corresponding to the coordinates of invariant sub--superspace
$\Sigma^{(9+16|8+8)}$ of Eq. \eqref{fZcM=IIB},
\be\label{fZtcM=IIB}
{\frak Z}^{{\cal M}}(\tau)=  \{(X^=(\tau), X^i_{An}(\tau)\; ; \theta^{-1}_q(\tau) ,  \theta^{-2}_{q}(\tau)\; ,
  (v_{\alpha q}^{\; -}(\tau)\, , \,  v_{\alpha \dot{q}}^{\; +}(\tau)))\}\; , \qquad
\ee
Lagrange multiplier $\rho^\# (\tau)$,
and
their canonical and covariant momenta,
\bea
\label{p--:=}
& P_= := \frac {\partial {\cal L}_1} {\partial dX^=}\; , \qquad  P_i := \frac {\partial {\cal L}_1} {\partial d X^i_{An}}\; , \qquad
\Pi^{1-}_q := \frac {\partial {\cal L}_1} {\partial d\theta^{-1}_q} \; , \qquad
\Pi^{2-}_q := \frac {\partial {\cal L}_1} {\partial d\theta^{-2}_q} \; , \qquad P^{(\rho)}_\#  := \frac {\partial {\cal L}_1} {\partial d\rho^{\#}} \; , \qquad
\eea and
\bea
\label{fd=dL-Om=An}
 & \frak{d}{}^{(0)} = \frac {\partial {\cal L}_1} {\partial\Omega^{(0)} }\; , \qquad  \frak{d}^{ij} =- \frac {\partial {\cal L}_1} {\partial\Omega^{ij} }\; , \qquad \frak{d}^{\# i }=  \frac {\partial {\cal L}_1} {\partial\Omega^{=i} }\; , \qquad \frak{d}^{=i}= \frac {\partial {\cal L}_1} {\partial\Omega ^{\# i } }\; . \qquad
\eea
The derivatives in \eqref{fd=dL-Om=An} are taken with keeping fixed
$X^=, X^i_{An}, \theta^{-1}_q, \theta^{-2}_q$.

\if{}%%%%%
We nevertheless keep the same notation for covariant momenta as  used in \eqref{fd=dL-Om}, calculated with the use of Lagrangian form of the central coordinate basis \eqref{cL1=IIB} (with keeping fixed $x^a, \theta^{\alpha 1}, \theta^{\alpha 2}$), as we do not expect this might lead to a confusion.
\fi %%%%%%%%%

The above definitions also imply the following form of the Legendre transform from the Lagrangian to (almost) canonical Hamiltonian

\bea\label{H=canonical=An}
d\tau H=  dX^= P_= + d X^i_{An} P_i+d\theta^{-1}_q
\Pi^{1}_{-q}+d\theta^{-2}_q
\Pi^{2}_{-q} +   d\rho^{\#}P^{(\rho)}_\#  +
\Omega^{(0)}\frak{d}{}^{(0)}-  \frac 1 2 \Omega^{ij}\frak{d}^{ij } +\Omega^{=i} \frak{d}^{\# i }+\Omega ^{\# i }\frak{d}^{=i}- {\cal L}_1\; . \qquad
\eea

With the Lagrangian form \eqref{cL1=IIB-An}, the calculation of all the canonical and covariant momenta result in the constraints. It is convenient to arrange them in the following sequence

\bea\label{p--=r++} \rho^{\#} - p_{=}\approx 0 \; , \qquad P^{(\rho)}_\# \approx 0 \; , &\qquad
%\\ \nonumber
\\ \label{pi=0} &&  X^i_{An} + \frac 1  {\rho^{\#}}\frak{d}^{\# i} \approx 0 \; , \qquad p_{i}\approx 0 \; , \qquad
%\\ \nonumber
\\ \label{fd-1=0}
 \frak{d}^{1}_{-q} = \Pi^{1}_{-q} +2i p_= \theta^{-1}_q\approx 0 \; , && \qquad
 %\\ \nonumber
 \\ \label{fd-2=0}
&& \frak{d}^{2}_{-q} = \Pi^{2}_{-q} +2i p_= \theta^{-2}_q\approx 0 \; , \qquad
%\\ \nonumber
\\  \label{fd0-==0}
 \frak{d}^{(0)}- 2X^=p_= - \theta^{-1}_q\Pi^{1}_{-q}  - \theta^{-2}_q\Pi^{2}_{-q} \approx 0 \; , &&\qquad
%\\ \nonumber
\\ \label{fdij==0}
&& \frak{d}^{ij}- \frac 1 4  \theta^{-1}\gamma^{ij}\Pi^{1}_{-}- \frac 1 4  \theta^{-2}\gamma^{ij}\Pi^{2}_{-}\approx 0 \; , \qquad
%\\ \nonumber
\\ \label{fd--i==0}
 \frak{d}^{=i}\approx 0 \; . &&  \qquad
 %\\ \nonumber
\eea

Eqs. \eqref{p--=r++} and \eqref{pi=0} are pairs of explicitly solved second class constraints which hance can be used to reduce the phase space of our dynamical system by removing
$(\rho^\# , P_\#^{(\rho)})$ and $(X_{An}^i, p_i)$ pairs of canonically conjugate variables.
The constraints \eqref{fd0-==0}-- \eqref{fd--i==0} are of the first class\footnote{The first class constraints \eqref{fd0-==0} and \eqref{fdij==0} are the linear combination of the results of straightforward calculations of the covariant momenta and the fermionic second class constraints \eqref{fd-1=0} and \eqref{fd-2=0}.}. They generate $SO(1,1)$, $SO(8)$ and
${\bb K}_8$ gauge symmetries, respectively, on the reduced phase space.

Finally,  \eqref{fd-1=0} and  \eqref{fd-2=0} are the second class fermionic constraints which are imaginary and have nondegenerate Poisson brackets with themselves,

\bea\label{d1d1=p--}
\{ \frak{d}^{1}_{-q}, \frak{d}^{1}_{-p}  \}_{PB} = -4i p_=\delta_{qp}\; , \qquad
\{ \frak{d}^{1}_{-q}, \frak{d}^{2}_{-p}  \}_{PB} = 0\; , \qquad
\{ \frak{d}^{2}_{-q}, \frak{d}^{2}_{-p}  \}_{PB} = -4i p_=\delta_{qp}\; .  \qquad
%\\  \nonumber
\eea
(The logic beyond using for these constraints the same symbol $ \frak{d}$ as for covariant momenta of spinor moving frame variables  is that they are actually classical counterpart of fermionic {\it covariant} derivatives, see Eqs. \eqref{fd-A=-iD} --\eqref{bD-A:=} below).
To proceed with the massless type IIB superparticle quantization in
its spinor moving frame formulation, we should decide how to deal with these constraints.

Let us note that the canonical Hamiltonian \eqref{H=canonical=An} vanishes on the surface of constraints, as it should be for any reparametrization-invariant theory.

\subsubsection{Complex structure of type IIB superspace, conversion vs. Gupta --Bleuler,
and prequel of the hidden SU(8) symmetry}

One of the ways to deal with the above real octuplets of fermionic constriants is to combine them in two  complex octuplets related by complex conjugation,
\bea\label{fd-=1-i2}
\bar{{\frak d}}^{-q} =\frac 1 {\sqrt{2}}\left({\frak d}^1_{-q}+i{\frak d}^2_{-q}\right) \, \; , \qquad
{\frak d}_{-q} =\frac 1 {\sqrt{2}}\left({\frak d}^1_{-q}-i{\frak d}^2_{-q}\right) =- (\bar{{\frak d}}^{-q}) \, \; , \qquad q,p=1,\ldots , 8\; .
%\\ \nonumber
\eea
These obey (see \eqref{d1d1=p--})

\bea\label{d1bd=p--}
\{ \frak{d}_{-q}, \frak{d}_{-p}  \}_{PB} =0 \; , \qquad
\{ \frak{d}_{-q}, \bar{\frak{d}}^p_{-}  \}_{PB} = -4i p_=\delta_{q}{}^{p}\; , \qquad
\{ \bar{\frak{d}}^q_{-}, \bar{\frak{d}}^p_{-}  \}_{PB} = 0 \;   \qquad
\eea
which allows to make the {\it conversion} of the second class constraints into the first class one and to use {\it Gupta-Bleuler procedure} in quantization.

The {\it conversion method} consists in  skipping one of two conjugate second class constraints, e.g. $\frak{d}_{-q}$ in our case, thus converting the other, $ \bar{\frak{d}}^p_{-}$ in our case, into the first class constraint \cite{Faddeev:1986pc, Egorian:1993sc,Batalin:1989dm}. The {\it Gupta--Bleuler method} (see e.g. \cite{deAzcarraga:1982dhu} and refs. therein) implies making the same at the stage of quantization, namely
imposing as a condition on the state vector the quantum version of only one of two canonically conjugate second class constraints. When this is  related to the other one by hermitian conjugation, the matrix elements of both quantum constraints between physical states vanish.

Notice that the conversion method is more general as, in principle,  it can be used also for canonically conjugate second class constraints which are not related by complex conjugation: rather one of two conjugate second class constraints  is considered as a gauge fixed condition for the gauge symmetry generated by the new first class constraints  created  by the conversion procedure.

The above complex constraints can be written in terms of complex fermionic coordinates of the reduced superspace
\be\label{Th=1+i2}
\Theta^{- q}= \frac 1 {\sqrt{2}}\left(\theta^{1-}_q+i\theta^{2-}_q\right) \; , \qquad \bar{\Theta}{}^-_q= \frac 1 {\sqrt{2}}\left(\theta^1_q-i\theta^2_q\right) \; , \qquad q,p=1,\ldots , 8\; , \ee
and their conjugate momenta
\be\label{Pi=1-i2}
\Pi_{- q}= \frac 1 {\sqrt{2}}\left(\Pi^1_{-q}-i\Pi^2_{-q}\right) \; , \qquad \bar{\Pi}{}^{-q}= \frac 1 {\sqrt{2}}\left(\Pi^1_{-q}+i\Pi^2_{-q}\right) \; , \qquad q,p=1,\ldots , 8\; , \ee
namely
\bea\label{fd-==1-i2}
\bar{{\frak d}}^{-q} = \bar{\Pi}{}^{-q} +2i p_= \Theta^{- q} \approx 0\; , \qquad {\frak d}_{-q} = {\Pi}_{-q} +2i p_=\bar{\Theta}{}^-_q \approx 0\; . \qquad
%\\ \nonumber
\eea

Notice that we have begun to write
upper and lower indices $q,p$ on
the complex fermionic coordinates, momenta and constraints. This is the prequel of the discussion on the $SU(8)$ symmetry of type IIB supergravity which is hidden in the Lagrangian \eqref{cL1=IIB-An} when written in terms of real  fermionic variables. The indications of this hidden invariance will be even more manifest after the quantization, if we perform this in terms of the complex fermionic coordinates \eqref{Th=1+i2}. However, the classical Lagrangian written in   terms of these variables does not possess such symmetry. In the next (sub)section we shall discuss the gauge equivalent version of the model which does possess the manifest $SU(8)$ symmetry.

\subsection{Massless type IIB superparticle mechanics in the analytical basis with manifest complex structure and SU(8) symmetry  }

\subsubsection{SU(8)/SO(8) coset and complex structure}

To make $SU(8)$ symmetry manifest at the level of the  classical Lagrangian, we can introduce new  variables  which
parametrize the space of possible complex structures.
These can be introduced in the reduced type IIB Lorentz harmonic superspace \eqref{fZtcM=IIB}. These are described by $SU(8)$ valued 8$\times 8$ complex matrix ${ w}_q{}^A=(\bar{{w}}_{qA})^*$.
\bea\label{wbw=inSU8}
{w}_q{}^A\bar{{w}}_{pA}=\delta_{qp} \; , \qquad  \bar{{w}}_{qA}{w}_q^B=\delta_{A}{}^B \;,  \qquad {\rm det }({w}_q{}^A)=1\; , \qquad
\eea
which are transformed by $SO(8)$ on its $p,q$ indices. The complex fermionic coordinates can be introduced as

\be\label{Th=1w+i2w}
\Theta^{- A}= \frac 1 {\sqrt{2}}\left(\theta^{1-}_q+i\theta^{2-}_q\right) {w}_q{}^A\; , \qquad \bar{\Theta}{}^-_A= \frac 1 {\sqrt{2}}\left(\theta^1_q-i\theta^2_q\right) \bar{{ w}}_{qA}\; , \qquad A,B=1,\ldots , 8\; .  \ee
The original guess for complex coordinates, \eqref{Th=1+i2}, is reproduced by setting
$\bar{{w}}_{qA}=\delta_{qA}= {w}_q{}^A$.

In the generic case, as $\theta^1_q$ and $\theta^2_q$ are transformed by local $SO(8)$ symmetry, which is one of the characteristic gauge symmetries acting on the spinor moving frame variables $v_{\alpha q}^{\; -}$, the newly introduced $\bar{w}_{qA}$ and $ {w}_q{}^A$ should be worldline fields rather than constant matrices. Then the $SU(8)$ symmetry acting on their $A,B$ indices by its fundamental representation can also be made local. Actually this local $SU(8)$ will be a gauge symmetry of the Lagrangian form which we obtain below.  Of course, in this case $\bar{w}_{qA}$ and $ {w}_q{}^A$ providing the bridge between $SU(8)$ and $SO(8)$ representations become pure gauge or St\"uckelberg fields. This fact reflects the freedom in the choice of complex structure in the (reduced) type IIB Lorentz harmonic superspace.

Hence we can conclude that the $SU(8)$ symmetry manifests itself when the model is described in (is reduced to) the form in which the explicit choice of complex structure is not essential. We will see that exactly this happens with our model after a suitable quantization.

The $SO(8)$ invariant Cartan form which plays the role of $SU(8)$ connection is defined with the use of $SO(8)$ covariant derivatives of our bridge variables,
\bea\label{mho=wDw}  {\mho}_{B}{}^{A}= \bar{{w}}_{qB}D {w}_q{}^A= \bar{{w}}_{qB}d {w}_q{}^A - \frac 1 4 \Omega^{ij}\bar{{w}}_{qB}\gamma^{ij}_{qp} {w}_p{}^A \; .
\qquad
%\\ \nonumber
\eea
This implies
\bea\label{Dw=IIB}
 %\begin{cases}
 D {w}_q{}^A= d {\rm w}_q{}^A - \frac 1 4 \Omega^{ij}\gamma^{ij}_{qp} {w}_p{}^A =
 {w}_q{}^A{\mho}_{B}{}^{A}\; , \qquad
 %\cr {} \cr
D \bar{{\rm w}}_{qB} = d\bar{{\rm w}}_{qB} - \frac 1 4 \Omega^{ij}\gamma^{ij}_{qp}\bar{{\rm w}}_{pB}=-{\mho}_{B}{}^{A}  \bar{{\rm w}}_{qA}\; . \qquad
%\end{cases} \\ \nonumber
\eea
These  equations can be written in terms of a complete $SU(8)\times SO(8)$ covariant derivatives
as covariant constancy of the bridge variables  \footnote{These conditions provide  an equivalent form of the definition of the composite $SU(8)$ connection \eqref{mho=wDw} similarly to how  the covariant constancy of metric identifies affine connection with Christoffel symbol in General Relativity.},

\bea\label{cDw=0}  %\begin{cases}
{\cal D} {w}_q{}^A= d {w}_q{}^A - \frac 1 4 \Omega^{ij}\gamma^{ij}_{qp} {w}_p{}^A -
 {w}_q{}^A{\mho}_{B}{}^{A} =0 \; , \qquad
 %\cr \cr
{\cal D} \bar{{w}}_{qB} = d\bar{{w}}_{qB} - \frac 1 4 \Omega^{ij}\gamma^{ij}_{qp}\bar{{w}}_{pB}+{\mho}_{B}{}^{A}  \bar{{w}}_{qA}=0\; .
%\end{cases}
\qquad
%\\ \nonumber
\eea

Notice that the matrix 1--form \eqref{mho=wDw} is traceless, $\mho_A{}^A=0$, due to the third condition in \eqref{wbw=inSU8}. More details on the composite connection
$\mho_A{}^B=0$ can be found in Appendix \ref{App=dmho}.

Covariant derivatives in Lorentz harmonic superspace complemented by $SU(8)$ group manifold parametrized by  coordinates $\bar{w}_{qA}, w_q^A$  are defined by decomposition
\bea
d^{v, w,\bar{w}}\equiv d^{Spin(1,9)\otimes SU(8)}&=& dw_q^A \partial_{qA}+d\bar{w}_{qA}\bar{\partial}_{q}^A + dv_{\alpha q}^{\; -} \frac {\partial } {\partial v_{\alpha q}^{\; -} }+d{v}_{\alpha \dot{q}}^{\; +}  \frac {\partial } {\partial v_{\alpha \dot{q}}^{\; +} }= \nonumber
%\\ \nonumber
\\ &=& {\mho}_{B}{}^{A} D_{A}{}^{B}+ \Omega^{(0)}D^{(0)} -\frac 1 2 \Omega^{ij}\tilde{D}{}^{ij} +\Omega ^{=i}D^{\# i }+\Omega ^{\# i }D^{=i}
%\\  \nonumber
\eea
where the differential operators
\bea\label{DAB:=}
D_{A}{}^{B} = w_q^B \partial_{qA}-\bar{w}_{qA}\bar{\partial}_{q}^B \; , &&\qquad
%\\  \nonumber
\\  \label{tDij:=D+}
\text{and}\qquad \tilde{D}{}^{ij}&=& {D}{}^{ij} +\frac 1 2  \left( (w^A\gamma^{ij})_q\partial_{qA}+(\bar{w}_{A}\gamma^{ij})_q\bar{\partial}_{q}^A\right)
%\\  \nonumber
\eea
provide representations of $SU(8)$ and $SO(8)$ groups and   $D^{(0)}$, $D^{ij}$, $D^{\# i }$ and $D^{=i}$ are covariant derivatives of Lorentz harmonic variables defined in \eqref{D0:==}--\eqref{D++i:==}.
Now these  are also
covariant derivatives in the Lorentz harmonic superspace enlarged (with respect to \eqref{IIB=cZcM:}) by $SU(8)$ group manifold.

\subsubsection{Lagrangian of type IIB superparticle with manifest $SU(8)$ symmetry}

Substituting the expressions for real fermionic coordinates
\bea
& \theta^{1-}_q=  \frac 1 {\sqrt{2}}\left(\bar{\Theta}{}^-_A {w}_q{}^A+ \Theta^{- A} \bar{{w}}_{qA}\right) \; , \qquad \theta^{2-}_q=  \frac i {\sqrt{2}}\left(\bar{\Theta}{}^-_A {w}_q{}^A- \Theta^{- A} \bar{{w}}_{qA}\right)\;  \qquad
\eea
into
\eqref{cL1=IIB-An}, we  find the Lagrangian form in terms of complex coordinates:
\bea
\label{cL1=IIB-An=c} && {\cal L}_1^{IIB}=\rho^\# E^= = \rho^\# \left( DX^= -2i {\cal D}\Theta^{-A}\, \bar{\Theta}^{-}_A +2i \Theta^{-A} {\cal D}\bar{\Theta}^{-}_A\, - \Omega^{= i} X_{An}^i\right)
\; ,  \qquad
\eea
where
\bea\label{cDThA:=}
{\cal D}\Theta^{-A}:= d\Theta^{-A}+\Omega^{(0)}\Theta^{-A} - \Theta^{-B }{\mho}_{B}{}^{A} \; , \qquad {\cal D}\bar{\Theta}{}^-_B:=d\bar{\Theta}{}^-_B +\Omega^{(0)}\bar{\Theta}{}^-_A +{\mho}_{B}{}^{A}\bar{\Theta}{}^-_A
\eea
are $SU(8)\times SO(1,1) $ covariant derivatives ({\it cf.} \eqref{cDw=0} and \eqref{Dth1-:=}). Actually, the terms with $\Omega^{(0)}$  coming from two covariant derivatives in \eqref{cL1=IIB-An=c} cancel, so that effectively $\Omega^{(0)}$ defined in \eqref{Om--i:=} does not enter thie Lagrangian form. We however included it in \eqref{cDThA:=} to make derivative covariant with respect to all symmetries of the model.

Besides the manifest $SU(8)\times SO(8)\times SO(1,1)$ gauge symmetry,  the Lagrangian 1--form \eqref{cL1=IIB-An=c}
is also invariant under  rigid $U(1)$ symmetry  which originates in $SO(2)$ invariance of the  massless IIB superparticle in its standard formulation  (but broken by type IIB superstring).

\subsubsection{$SU(8)$ invariant form of the Hamiltonian mechanics}

The Legendre transform of the Lagrangian form \eqref{cL1=IIB-An=c} defining its corresponding  canonical Hamiltonian

\bea\label{H=canonical=An-c}
d\tau H=  dX^= P_= + d X^i_{An} P_i+d\theta^{-1}_q
\Pi^{1}_{-q}+d\theta^{-2}_q
\Pi^{2}_{-q} +   d\rho^{\#}P^{(\rho)}_\#  +
\Omega^{(0)}\frak{d}{}^{(0)}-  \frac 1 2 \Omega^{ij}\tilde{\frak{d}}{}^{ij } +\Omega^{=i} \frak{d}^{\# i }+\Omega ^{\# i }\frak{d}^{=i}+ \nonumber \\
%\nonumber \\
+  {\mho}_B{}^A \frak{d}_B{}^A- {\cal L}_1\; , \qquad
%\nonumber \\
\eea
differs from  \eqref{H=canonical=An} by additional term ${\mho}_B{}^A \frak{d}_B{}^A$,
which contains a new covariant momentum
\bea
\frak{d}{}_{A}{}^{B} &=& w_q^B P^{w}_{qA}-\bar{w}_{qA}\bar{P}{}^{\bar{w}}{}_{q}^B \; , \qquad
%\\  \nonumber
\eea
and by modification of the covariant momentum of spinor moving frame variables generating  $SO(8)$ symmetry
\bea
\tilde{\frak{d}}{}^{ij}&=& \frak{d}{}^{ij} +\frac 1 2  \left( (w^A\gamma^{ij})_q\frak{d}_{qA}+(\bar{w}_{A}\gamma^{ij})_q\bar{\frak{d}}_{q}^A\right)\; .
%\\  \nonumber
\eea
This has non vanishing Poisson brackets also with the new variables  $ w_q^A$ and $ \bar{w}_{qA}$.

The set of constraints, which follows from the calculation of
the canonical and covariant momenta, includes, as before,  explicitly resolved pairs of second class constraints \eqref{p--=r++} and \eqref{pi=0} which we use to reduce the phase space by omitting $\rho^\#$, $X^i_{An}$ directions in the configuration space and their canonical momenta.

After some algebra, the remaining set of constraints
can be separated in the first class constraints
\bea
\label{fd0-=0}
&& \frak{d}^{(0)}- 2X^=p_= - \Theta^{-A}\Pi_{-A}  - \bar{\Theta}^{-}_A\bar{\Pi}{}^{A}_{-} \approx 0 \; , \qquad
%\\ \nonumber
\\ \label{fdij=0}
&& \tilde{\frak{d}}{}^{ij}\approx 0 \; , \qquad
%\\ \nonumber
\\ \label{fd--i=0}
&& \frak{d}^{=i}\approx 0 \; . \qquad
%\\ \nonumber
\eea
generating $SO(1,1)$, $SO(8)$ and ${\bb K}_8$ gauge symmetry, respectively,
the first class constraint
\bea\label{fdAB-=0}
&& \frak{d}{}_B{}^{A}+ \bar{\Theta}^{-}_B\bar{\Pi}{}^{A}_{-}-  \Theta^{-A}\Pi_{-B}   - \frac 1 8 \left(\bar{\Theta}^{-}_C\bar{\Pi}{}^{C}_{-}- \Theta^{-C}\Pi_{-C}   \right) \approx 0 \; , \qquad
%\\ \nonumber
\eea
generating $SU(8)$ gauge symmetry, as well as fermionic second class constraints
\bea \label{fd-A=0}
&& \frak{d}_{-A} = \Pi_{-A} +2i p_= \bar{\Theta}^{-}_A\approx 0 \; , \qquad
 \bar{\frak{d}}{}^{A}_{-} = \bar{\Pi}{}^{A}_{-} +2i p_= \Theta^{-A}\approx 0 \; . \qquad
 %\\ \nonumber
\eea
The canonical Hamiltonoian vanishes as it should be for the theory with reparametrization invariance.

Constraints \eqref{fd-A=0} obey the $d=1$, $n=8$ supersymmetry algebra with $SU(8)$ automorphism group.
\bea\label{fdbfdd1=p--}
\{ \frak{d}_{-A}, \frak{d}_{-A}  \}_{PB} = 0 \; , \qquad \{ \frak{d}_{-A}, \bar{\frak{d}}{}^{B}_{-} \}_{PB} =  -4i p_=\delta_{A}{}^B\; , \qquad
\{\bar{\frak{d}}{}^{A}_{-}, \bar{\frak{d}}{}^{B}_{-}  \}_{PB} = 0\; .   \qquad
%\\  \nonumber
\eea
For these constraints we will use Gupta-Bleuler quantization prescription which can be also formulated as conversion at the classical level. This is to say, we can omit constraint $\frak{d}_{-A}$ and thus arrive at the system with one octuplet of complex fermionic first class constraints
\bea \label{bfd-A=0=I}
&&
 \bar{\frak{d}}{}^{A}_{-} = \bar{\Pi}{}^{A}_{-} +2i p_= \Theta^{-A}\approx 0 \;  \qquad
 %\\ \nonumber
\eea
generating complex gauge supersymmetry. Upon quantization, this first class fermionic constraint can be imposed on the state vector.

\setcounter{equation}0

\section{Quantization of type IIB superparticle  and on-shell superfields of linearized type IIB supergravity }

\subsection{Quantization }

The quantization procedure in the generalized coordinate representation implies the replacement of phase space variables by quantum operators and Poisson brackets by graded commutators (i.e. commutators or anti-commutators),

\begin{equation}\label{PB=quant}
[...,...\}_{PB}\quad \mapsto \quad \frac 1 i\, [...,...\}\; . \qquad
\end{equation}
In the coordinate representation this prescription is realized by replacing the canonical momenta by derivatives,
\bea
p_{=}&\mapsto & -i \partial_{=} \equiv -i \frac  \partial {\partial X^=} \; , \qquad
%\\ \nonumber \\
\Pi_{-A}\mapsto -i \partial_{-A} \equiv -i \frac  \partial {\partial \Theta^{-A}} \; , \qquad
%\\ \nonumber \\
 \bar{\Pi}{}^{A}_{-} \mapsto  -i \bar{\partial}{}^{A}_{-} \equiv -i \frac  \partial {\partial \bar{\Theta}^{-}_A} \; , \qquad
 % \\ \nonumber
 \eea
 and covariant momenta by covariant derivatives
\eqref{tDij:=D+},  \eqref{D0:=}--\eqref{D--i:=}, \eqref{DAB:=}
\bea
\frak{d}^{(0)}&\mapsto&  -iD^{(0)}   \; , \qquad
%\\ \nonumber \\
\tilde{\frak{d}}{}^{ij}\mapsto -i\tilde{D}{}^{ij}= -i\left(D^{ij}-\frac 1 2  (w^A\gamma^{ij})_q\partial_{qA}-\frac 1 2  (\bar{w}_{A}\gamma^{ij})_q\bar{\partial}_{q}^A\right)
 \; , \qquad
 %\\ \nonumber
 \\
 && {}\qquad \frak{d}^{\# i}\mapsto  -iD^{\# i}   \; , \qquad
 %\\ \nonumber \\
 \frak{d}^{= i}\mapsto  -iD^{= i}   \; , \qquad
%\\ \nonumber
\\
 \frak{d}_{A}{}^{B}&\mapsto&  -i D_{A}{}^{B}   \; , \qquad
 %\\ \nonumber
 \eea

Notice that with this prescription the fermionic constraints \eqref{fd-A=0} are represented by covariant fermionic derivatives in the reduced superspace
\bea \label{fd-A=-iD}
\frak{d}_{-A} &\mapsto& -i D_{-A}  \; , \qquad
%\\ \label{D-A:=}&&
D_{-A} = \partial_{-A} +2i \bar{\Theta}^{-}_A\partial_=\; , \qquad
\\
%\nonumber \\ \label{bfd-A=-ibD}
 \bar{\frak{d}}{}^{A}_{-} &\mapsto& -i\bar{D}{}^{A}_{-} \; , \qquad   \label{bD-A:=}
 \bar{D}{}^{A}_{-}  = \bar{\partial}{}^{A}_{-} +2i  \Theta^{-A}\partial_= \; . \qquad %\\ \nonumber
\eea
These covariant derivatives obey the $SU(8)$ supersymmetry algebra
\bea\label{D-AbD-B=d--}
\{ D_{-A}, D_{-A}  \} = 0 \; , \qquad \{ D_{-A}, \bar{D}{}^{B}_{-} \} =  +4i \partial_=\delta_{A}{}^B\; , \qquad
\{\bar{D}{}^{A}_{-}\, , \, \bar{D}{}^{B}_{-}  \} = 0\; .   \qquad %\\  \nonumber
\eea

In the reduced phase space and after conversion of fermionic constraints, we have the system with only  the first class constraints, bosonic  \eqref{fd0-=0}-- \eqref{fdAB-=0} and fermionic \eqref{bfd-A=0=I}. According to the Dirac prescription \cite{Dirac:1963}, the quantum version of these constraints have to be imposed on the state vector. Thus our state vector superfield
\be
\Xi= \Xi(x^=, v, w ; \Theta^{-A}, \bar{\Theta}^-_A)
\ee
should satisfy the following set of equations
\bea \label{D0-Xi=0}
&& \left(D^{(0)}- 2X^=\partial_= - \Theta^{-A}\partial_{-A}  - \bar{\Theta}^{-}_A\bar{\partial}{}^{A}_{-}\right) \Xi= 0 \; , \qquad
%\\ \nonumber
\\ \label{Dij-Xi=0} \tilde{D}{}^{ij} \Xi &=& \left(D^{ij}+\frac 1 2  \left( (w^A\gamma^{ij})_q\partial_{qA}+(\bar{w}_{A}\gamma^{ij})_q\bar{\partial}_{q}^A\right)  \right) \Xi= 0  \; , \qquad
%\\ \nonumber
\\ \label{D--iXi=0}
D^{=i} \Xi &=& 0 \; , \qquad
%\\ \nonumber
\\ \label{DAB-Xi=0}
&& \left(D_B{}^{A}+ \bar{\Theta}^{-}_B\bar{\partial}{}^{A}_{-}-  \Theta^{-A}\partial_{-B}   - \frac 1 8 \left(\bar{\Theta}^{-}_C\bar{\partial}{}^{C}_{-}- \Theta^{-C}\partial_{-C}   \right) \right) \Xi= 0 \; , \qquad
%\\ \nonumber
\eea
and
\bea \label{bD-AXi=0}
&& \bar{D}{}^{A}_{-}\Xi= \left( \bar{\partial}{}^{A}_{-} +2i  \Theta^{-A}\partial_= \right) \Xi= 0 \; . \qquad
%\\ \nonumber
\eea

The constraints \eqref{D0-Xi=0} and  \eqref{Dij-Xi=0}  imply that the state vector superfield should be invariant under $SO(1,1)$ and $SO(8)$ symmetry which act, in the first case on the Lorentz harmonics and
$X^=$, $\Theta^{-A}$ and $\bar{\Theta}^{-}_{A}$ coordinates, while in the second case on the Lorentz harmonics and
the auxiliary bridge variables
$w_q^A$ and $\bar{w}_{qA}$.  Eq. \eqref{D--iXi=0} implies that $\Xi$ should depend only on
$v_{\alpha q}^{\; -}$, but not on $v_{\alpha \dot{q}}^{\; +}$ Lorentz harmonic coordinate.

Eq. \eqref{DAB-Xi=0} implies that the state vector superfield is invariant under $SU(8)$ transformations acting on
bosonic $w_q^A$ and $\bar{w}_{qA}$ and fermionic $ \Theta^{-A}$ and $\bar{\Theta}^{-}_B$ coordinates. This can be used to reproduce the value of superfield at arbitrary $w_q^A$ and $\bar{w}_{qA}$ starting from its 'value' at some fixed point of $SU(8)$ group manifold, e.g. at $w_q^A=\delta_q^A=\bar{w}_{qA}$. We shall turn to the
discussion on this point later.

Finally,  \eqref{bD-AXi=0} is the chirality condition imposed on our state vector superfield. It is solved by
\bea \label{Xi=Phi}
\Xi (x^=,v,w; \Theta^-,\bar{\Theta}^{-}) = e^{2i \Theta^{-A}\bar{\Theta}^{-}_A\partial_= }\Phi (x^=,v,w; \Theta^- )=\; \Phi (x^=_{L},v,w; \Theta^- ) , \qquad
 \\ \nonumber
\eea
where $
%\bea
x^=_L:=x^=+2i \Theta^{-A}\bar{\Theta}^{-}_A = e^{2i \Theta^{-A}\bar{\Theta}^{-}_A\partial_= }x^=$
%\eea
and  superfield $\Phi (x^=,v,w; \Theta^- )$ depends on $\Theta^{-A}$ but not on its c.c. $\bar{\Theta}^{-}_A$. Hence its decomposition on fermionic variables reads

\bea\label{Phi=phi+}
& \Phi= \phi+ \Theta^{-A} \psi^+_A + \frac 1 2 \Theta^{-B}\Theta^{-A} \phi^{\#}_{AB} +
\frac 1 {3!} \Theta^{-C} \Theta^{-B} \Theta^{-A} \psi^{[+3]}_{ABC}
%+ \qquad  \nonumber \\ \nonumber \\ &&
+ \frac 1 {4!} \Theta^{-D}  \Theta^{-C} \Theta^{-B} \Theta^{-A} \phi^{[+4]}_{ABCD} +  \nonumber
%\\ \nonumber
\\
& +\frac 1 {5!\, 3!} \epsilon_{A_1\ldots A_5B_1B_2B_3} \Theta^{-A_5}  \ldots \Theta^{-A_1} \tilde{\psi}{}^{B_1B_2B_3+5} +   \frac 1 {6!\, 2!} \epsilon_{A_1\ldots A_4B_1B_2} \Theta^{-A_6}  \ldots \Theta^{-A_1} \tilde{\phi}{}^{B_1B_2+6} + \qquad   \nonumber
%\\ \nonumber
\\
& + \frac 1 {7!} \epsilon_{A_1\ldots A_7B} \Theta^{-A_7}  \ldots \Theta^{-A_1} \tilde{\psi}{}^{B+7} + \frac 1 {8!} \epsilon_{A_1\ldots A_7B} \Theta^{-A_8}  \ldots \Theta^{-A_1} \tilde{\phi}^{+8} \; .  \qquad   \\ \nonumber
\eea
where all the fields depend on $x^=$,
on the
Lorentz harmonics $v_{\alpha q}^{\; -}$ and on the auxiliary $w$--coordinates.

Up to the type of variables the field components depend on, the above superfield  is similar to the
chiral four dimensional ${\cal N}=8$ superfield which can be obtained by quantization of the massless superparticle in
$D=4$, ${\cal N}=8$. However, as in this case, and also in the case of light-cone description of the linearized
type IIB supergravity \cite{Green:1982tk,Howe:1983sra}, a superfield that obeys only the chirality condition
does not describe irreducible representation of supersymmetry (type IIB supersymmetry in our  case). Then, as it is done with the aim  to obtain the
so-called on-shell superfields for D=4 ${\cal N}=8$ supergravity theory \cite{Arkani-Hamed:2008gz,Brandhuber:2008pf,Elvang:2015rqa} and light-cone superfield description of linearized type IIB supergravity \cite{Green:1982tk,Howe:1983sra} we have to impose additional condition to make the representation irreducible\footnote{See \cite{Bandos:2025pxv} for discussion of similar problem in the quantization of type IIA D$0$-brane model. }.

We impose the following duality  relations between half-maximal derivative of our state vector superfield $\Xi$ and of its conjugate $\bar{\Xi}$:
\bea\label{D-4Xi=ebD-4bXi}
 D_{-A_1}\ldots D_{-A_4}\Xi  =\frac 1 { 4!} \epsilon_{A_1\ldots A_4B_1B_2B_3B_4} \bar{D}{}_-^{B_1}\ldots \bar{D}{}_-^{B_4}\bar{\Xi} \; , \qquad  (\Xi)^* =\bar{\Xi}\; .
 %\\ \nonumber
\eea
Taking into account the chirality of the state vector superfield \eqref{bD-AXi=0} and anti-chirality of its  conjugate
\bea
D_{-A}\bar{\Xi}=0
\; ,
%\\ \nonumber
\eea
and using the algebra \eqref{D-AbD-B=d--}, we find that higher components of our state vector superfield are now expressed through the conjugate to the lower components\footnote{One should not be worried  by appearance of inverse power of bosonic derivative $\partial_=$ as, on one hand, the same happens in the light-cone approach \cite{Green:1982tk} and, on the other hand, because the main use is for the Fourier image of this superfield with respect to $x^=$, in which $\partial_=$ is replaced by $p_=$. Just this is the so-called on-shell superfield   which can be considered as
one particle counterpart of the superamplitudes and $p_=$ in it appears in the expression for light-like momenta $p_a = p_= 1/8\,v_{ q}^{\; -} \tilde{\sigma}_av_{ p}^{\; -} $  (see \eqref{p=rv-sv-}).  }
\bea\label{tpsi=psi*}
& \tilde{\psi}{}^{B_1B_2B_3[+5]}=\propto \frac 1 {\partial_=} ({\psi}^{[+3]}_{B_1B_2B_3})^*\; , \qquad \tilde{\phi}{}^{B_1B_2[+6]}=\propto \frac 1 {(\partial_=)^2}  ({\phi}_{B_1B_2}^{[+2]})^*\; , \qquad
%\\ \nonumber
\\ & \tilde{\psi}{}^{B[+7]} =\propto \frac 1 {(\partial_=)^3}  ({\psi}_{B}^{[+]})^* \; ,  \qquad \tilde{\phi}^{[+8]}=\propto \frac 1 {(\partial_=)^4} ({\phi})^*\; , \qquad
\eea
as well as the duality relations between the intermediate component and its c.c.,

\be\label{phi4=ep-phi4*}
 \phi^{[+4]}_{A_1\ldots A_4} =\frac 1 {4!\, 4!} \epsilon_{A_1\ldots A_4B_1B_2B_3B_4} \tilde{\phi}{}^{B_1B_2B_3B_4[+4]} \; , \qquad  (\phi^{[+4]}_{B_1\ldots B_4})^* =\tilde{\phi}{}^{B_1B_2B_3B_4}\; .
\ee

Let us show that the independent components of our on-shell superfields describe  linearized type IIB supergravity, or, to be more precise, the solutions of type IIB supergravity equations.

\subsection{On-shell superfield describes type IIB supergravity }

To prove that our on-shell superfield obeying \eqref{bD-AXi=0} and \eqref{D-4Xi=ebD-4bXi}  describes linearized type IIB supergravity we shall follow the line of the analysis of the light--cone superfield approach   by Green and Schwartz \cite{Green:1982tk}.
We  begin by restricting the consideration
to the point $w_q^A=\delta_q^A=\bar{w}_{qA}$ of the $SU(8)$ sector of
the
enlarged Lorentz harmonic superspace. This allows
us to use straightforwardly the properties of $SO(8)$ '$\gamma$'--matrices \cite{Green:1983hw}  which we collect in Appendix \ref{SO8gammas}.

At this '$SU(8)$ point'  our state vector superfield and its complex conjugate obey

\bea\label{bDPhi==0}
\bar{D}_{-}^{q}  \Xi=0\; , \qquad \bar{D}_{-}^{q} =\bar{\partial}{}_{-}^{q} +2i{\Theta}^{-q}\partial_{=}\; , \qquad
%\\ \nonumber
\\ \label{DbPhi==0} D_{-q} \bar{\Xi}=0\; , \qquad D_{-q} =\partial_{-q}  +2i\bar{\Theta}^-_q\partial_{=}\; , \qquad
%\\ \nonumber
\eea
where ${\Theta}^{-q}$ and $\bar{\Theta}^-_q$ are defined in \eqref{Th=1+i2},
and also
\bea\label{D+4P=ebD4P}
D_{-q_1}\ldots D_{-q_4} \Xi =\frac 1 {4!} \epsilon_{q_1\ldots q_4p_1\ldots p_4} \bar{D}^{p_1}_-\ldots \bar{D}_-^{p_4} \bar{\Xi} \; .
%\\ \nonumber
\eea
As a result, the set of independent components of our on-shell superfield includes  a complex scalar and
unrestricted antisymmetric spin-tensors
\bea\label{comp=IIB}
\phi= \Xi\vert_0 \; , \qquad \psi^+_q= {D}_{-q} \Xi\vert_0 \; , \qquad    \phi^{\#}_{qp}={D}_{-q} \bar{D}_{-p} \Xi\vert_0 \; , \qquad  \psi^{[+3]}_{qpr}={D}_{-q}{D}_{-p}{D}_{-r} \Xi\vert_0 \; , \qquad
\eea
as well as self-dual antisymmetric spin--tensor
\bea\label{comp4=IIB}
\phi^{[+4]}_{q_1...q_4}=
D_{-q_1}\ldots D_{-q_4} \Xi \vert_0  =\frac 1 {4!} \epsilon_{q_1\ldots q_4p_1\ldots p_4} \bar{D}_-^{p_1}\ldots \bar{D}_-^{p_4} \bar{\Xi} \vert_0 =\frac 1 {4!} \epsilon_{q_1\ldots q_4p_1\ldots p_4} (\phi^{[+4]}_{p_1...p_4})^*  \; .
%\\ \nonumber
\eea
Here and below $\vert_0 $ denotes the leading component of superfield i.e. its 'value' at $\Theta^{-q}=0=\bar{\Theta}{}^-_q$.

\subsubsection{Bosonic fields of type IIB supergravity from spin-tensors}

The bosonic sector of type IIB supergravity contains i) an axion and
a
dilaton, which can be collected in one complex scalar, ii)
a
complex 2-form field containing NS-NS and RR sector 2-forms, ii)
a
self-dual antisymmetric 4-form gauge field (from RR sector) and iv) graviton. Their linearized equations can be solved, respectively,  in terms  of i)  a
complex scalar depending on the light-like momentum, which in our case can be associated with the
leading component of our complex state vector superfield $\phi (x^=,v_{\alpha q}^{\; -})$, ii) a complex antisymmetric $SO(8)$ tensor field
$c^{ij}= -c^{ji}=c^{[ij]}$, iii) a real antisymmetric self-dual 4-th rank SO(8) tensor   $c^{ijkl}=\frac 1 {4!}
\epsilon^{ijkli'j'k'l'} c^{i'j'k'l'}$ and iv) a symmetric traceless second rank tensor $h^{ij}=h^{ji}$, $h^{ii}=0$.
All these fields should depend on $(x^=,v_{\alpha q}^{\; -})$, transform covariantly under $SO(8)$ symmery,  and should be inert under $SO(1,1)$.
We however will not concentrate on this latter property as the weight of
the
fields with respect to $SO(1,1)$ can be easily compensated by corresponding powers of $\partial_=$ in $x^=$--coordinate representation or by
powers of $p_=$ when passing to the Fourier images of the fields.

Let us begin from the 2-nd rank antisymmetric spin-tensor $ \phi_{qp}^{\#}$. Using the identity
%\bea  \label{gijgij==} &
$\delta_{q[r}\delta_{s]p}=-\frac 1 {16}  \gamma^{ij}_{qp}\gamma^{ij}_{rs}$
%\eea
(Eq. \eqref{gijgij=} in Appendix  \ref{SO8gammas}), we find that
\be\label{phiqp=phiij}
 \phi_{qp}^{\#} = -\frac 1 {16} \gamma^{ij}_{qp} \, \gamma^{ij}_{rs} \,  \phi_{rs}^{\#}  =  \gamma^{ij}_{qp} c^{ij}\; ,\qquad c^{ij}=- \frac 1 {16} \gamma^{ij}_{qp} \phi_{qp}^{\#}\; .
\ee
Thus the complex antisymmetric spin-tensor component of our state vector superfield is in one-to one correspondence with the complex antisymmetric $SO(8)$ tensor $c^{ij}(x^=,v_{\alpha q}^{\; -})$ describing the solution of free equations for complex 2-form field $c_{\mu\nu}(x^\rho)$, $\partial^{\mu }\partial_{[\mu }c_{\nu\rho ]}=0$,  after the suitable gauge fixing.

To be more precise, as $c^{ij}$ should be invariant under $SO(1,1)$ gauge symmetry acting on the Lorentz harmonics, the r.h.s. of  \eqref{phiqp=phiij} should contain
operator $(\partial_=)^{-1}$ or, after Fourier transform with respect to $x^=$, multiplier
$p_= = \rho^\#$. In this section we, however, conventionally omit such multipliers to simplify and to make more transparent the structure of   \eqref{phiqp=phiij} as well as  of Eqs. \eqref{phiq1-q4==}--\eqref{ephiq1-q4*=} and \eqref{Psi3=PsiIdq} below.

Let us turn to the complex  antisymmetric 4-rank $SO(8)$ spin-tensor component of our state vector superfield, which is related to its c.c. by the duality relation \eqref{comp4=IIB}. First we use the same identity \eqref{gijgij=} to find
\bea\label{phiij-kl=}
\phi^{[+4]}_{q_1\ldots q_4} = \frac 1 {16}  \gamma^{ij}_{[q_1q_2}  \gamma^{kl}_{q_3q_4]} \phi^{ij,kl}\;  , \qquad
 \phi^{ij,kl}=  \frac 1 {16}  \gamma^{ij}_{[q_1q_2}  \gamma^{kl}_{q_3q_4]} \phi^{[+4]}_{q_1\ldots q_4}\; .
 %\\ \nonumber
\eea
Now we
decompose
$ \gamma^{ij}_{[q_1q_2}  \gamma^{kl}_{q_3q_4]}$ onto irreducible parts (see \eqref{gijgkl=t+t}, \eqref{t44:=}, \eqref{t24:=} in Appendix \ref{SO8gammas})
\bea\label{gijgkl==t+t}
 \gamma^{ij}_{[qp}\gamma^{kl}_{rs]}&=& t^{ijkl}_{qprs}-\frac 2 3 \delta^{[i|[k}  t^{l]|j]}_{qprs}\; ,  \qquad \\
  \label{t44:==}
&& t^{ijkl}_{qprs} := \gamma^{[ij}_{[qp}\gamma^{kl]}_{rs]}\; , \qquad
t^{ij}_{qprs} :=  \gamma^{ik}_{[qp}\gamma^{jk}_{rs]} = t^{ji}_{qprs}\;  \qquad \eea
which are, respectively  totally antisymmetric and anti--self--dual in both quadruplets of indices,
\bea
\label{t44==t*44} t^{ijkl}_{qprs}  &= &-\frac 1{4!} \epsilon^{qprsq'p'r's'}t^{i'j'k'l'}_{q'p'r's'} =-\frac 1{4!} \epsilon^{ijkli'j'k'l'}t^{i'j'k'l'}_{q'p'r's'} \; , \qquad
%\\ \nonumber
\eea
and symmetric traceless on vector indices while self-dual on quadruplet of spin-tensor indices,
\bea\label{t24==t2*4} t^{ij}_{qprs}=t^{ji}_{qprs}\, , \qquad t^{ii}_{qprs}=0\; , \qquad
 t^{ij}_{qprs}  = \; \frac 1{4!} \epsilon^{qprsq'p'r's'}t^{ij}_{q'p'r's'} \;  \qquad
 %\\ \nonumber
\eea
(see  Appendix \ref{SO8gammas}).

Using \eqref{gijgkl==t+t} and the above properties we find

\bea\label{phiij-kl==}
&& \phi^{ij,kl} = i c^{ijkl}- \frac 2 3 \delta^{[i|[k}h^{l]|j]} \; , \qquad
%\\ \nonumber
\\ \label{Hij:==}
h^{ij}=\phi^{ik,jk}=\frac 1 {16}  \gamma^{ik}_{[q_1q_2}  \gamma^{jk}_{q_3q_4]} \phi^{[+4]}_{q_1\ldots q_4} && \qquad \Rightarrow  \qquad h^{ij}=h^{ji}\; , \qquad h^{ii}=0\; , \;
%\\ \nonumber
\\ \label{Cijkl:==}
&& ic^{ijkl}=\frac {1} {16}  \gamma^{[ij}_{[q_1q_2}  \gamma^{kl]}_{q_3q_4]} \phi^{[+4]}_{q_1\ldots q_4}
 = -\frac 1 {4!} \epsilon_{ijkli'j'k'l'}\; ic^{i'j'k'l'}\; .
\eea
Eqs. \eqref{phiij-kl=} and \eqref{phiij-kl==} imply
\bea\label{phiq1-q4==}
\phi^{[+4]}_{q_1\ldots q_4} = \frac 1 {16}\left( ic^{ijkl} t^{ijkl}_{q_1q_2q_3q_4}  +\frac 2 3  t^{ij}_{q_1q_2q_3q_4} h^{ij}  \right)\;
%\\ \nonumber
\eea
and $
%\bea\label{bphiq1-q4==}
(\phi^{[+4]}_{q_1\ldots q_4})^*  =  \frac 1 {16}\left( -i(c^{ijkl})^* t^{ijkl}_{q_1q_2q_3q_4}  +\frac 2 3  t^{ij}_{q_1q_2q_3q_4} (h^{ij})^*  \right)\;
%\eea
$
so that, due to \eqref{t24==t2*4} and  \eqref{t44==t*44},

\bea\label{ephiq1-q4*=}
\frac 1 {4!}\epsilon^{q_1q_2q_3q_4p_1\ldots p_4}(\phi^{[+4]}_{q_1\ldots q_4})^*  =  \frac 1 {256}\left( i(c^{ijkl})^* t^{ijkl}_{q_1q_2q_3q_4}  +\frac 2 3  t^{ij}_{q_1q_2q_3q_4} (h^{ij})^*  \right)\;
\eea
and hence  \eqref{D+4P=ebD4P} implies the reality of both fields,

\be
(c^{ijkl})^*=c^{ijkl}\; , \qquad (h^{ij})^*= h^{ij}\; . \qquad
\ee

Thus we have shown that the bosonic components of the state vector superfield  describe the bosonic fields of type IIB supergravity multiplet.

\subsubsection{Gravitino and dilatino of type IIB supergravity from fermionic spin tensors }

Let us turn to the fermionic  components of our state vector superfield, complex s--spinor  $\psi^+_{q} =D_{-q} \Xi\vert_0  $ and  complex spin-tensor $\psi^{[+3]}_{q_1q_2q_3} =D_{-q_1}D_{-q_2}D_{-q_3} \Xi\vert_0  $. The first one,
$\psi^+_{q}(x^=,v^-)$ (or, more precisely
its Fourier image  $\psi^+_{q}(p_=,v^-)=\int dx^= e^{-ip_=x^=} \psi^+_{q}(x^=,v^-)$)
clearly describes the complex solution $\chi_{\alpha}=(\chi_{\alpha}^1+i\chi_{\alpha}^2)=
v_{\alpha q}^{\; -}\psi^+_{q}(p_=,v^-)$ of the $10D$ Weyl equation  in momentum representation,
$p_\mu \tilde{\sigma}^{\mu \alpha\beta} \chi_{\beta}=0$ with light-like momentum composed from spinor harmonics as in \eqref{p=rv-sv-}. Thus $\psi^+_{q}(x^=,v^-)$ describes dilatini of type IIB supergravity.

To clarify the meaning of the complex fermionic fields $\psi_{q_1q_2q_3} $, we observe
that it can be expressed in terms of  $\gamma$ - traceless SO(8) vector--spinor field $\psi^{i}_{\dot{q}}$ by

\be\label{Psi3=PsiIdq}
\psi^{[+3]}_{q_1q_2q_3}  = -\frac 1 8  t^{i}_{q_1q_2q_3\dot{q}}\psi^{i}_{\dot{q}} = \frac 1 8  \gamma^{ij}_{[q_1q_2}\gamma^j_{q_3]\dot{q}}\psi^{i}_{\dot{q}}\; .
\ee

To be convinced
that the r.h.s. gives the general structure of an antisymmetric spin-tensor of rank 3, one can notice
that the  number of independent components in the expression on the l.h.s
is $\left\{ \begin{matrix}3 \cr 8\end{matrix}\right\}=56$, and the same is the number of components of gamma-traceless vector-tensor $\psi^{i}_{\dot{q}}$, $64-8=56$. One may note that the r.h.s. of \eqref{Psi3=PsiIdq} can include an arbitrary vector spinor  $\psi^{i}_{\dot{q}}$  with 64 components. However,
substituting for  $\psi^{i}_{\dot{q}}$  the expression $\psi^{i}_{\dot{q}}+\chi_p\gamma^{i}_{p\dot{q}}$ in the r.h.s. of \eqref{Psi3=PsiIdq} does not change it. Then equivalence relation  $\psi^{i}_{\dot{q}}\approx\psi^{i}_{\dot{q}}+\chi_p\gamma^{i}_{p\dot{q}}$  can be treated as a counterpart of gauge symmetry and can be used
to impose $8$ conditions

\be\label{giPsii=0}
\gamma^{i}_{p\dot{q}}\psi^{i}_{\dot{q}}=0
\ee
thus equating the number of degrees of freedom in the l.h.s. and r.h.s. of \eqref{Psi3=PsiIdq}.

A more explicit proof of one-to one correspondence consists in using
\eqref{titj=I+g2} from Appendix \ref{SO8gammas} to solve  Eq. \eqref{Psi3=PsiIdq} by

\be\label{PsiIdq=Psi3}
\psi^{i}_{\dot{q}}= \psi^{i}_{\dot{q}}- \frac 1 8 (\tilde{\gamma}{}^i{\gamma}{}^j\psi^{j})_{\dot{q}}=
- \frac 1 6 \psi_{q_1q_2q_3} t^i_{q_1q_2q_3\dot{q}} =  \frac 1 6  \gamma^{ij}_{[q_1q_2}\gamma^j_{q_3]\dot{q}}\psi_{q_1q_2q_3}\; ,
\ee
where the first equality is written with the use of \eqref{giPsii=0}, while the second is valid also for unrestricted  $\psi^{i}_{\dot{q}}$ (see also \eqref{tigi=0}).

Thus we have shown that the fermionic components of our state vector superfield are in one-to one correspondence with the solutions of dilatini and gravitini equations of linearized  type IIB supergravity. This completes the proof that our state vector superfield describes linearized type IIB supergravity.

\subsubsection{SU(8) symmetric form of correspondence with type IIB supergravity fields }

Although we have derived the above results at a point $\bar{w}_{qA}=\delta_{qA}={w}_q^{A}$ of SU(8) sector of the extended Lorentz harmonic superspace, it can be lifted to the SU(8) covariant expressions. To this end let us
introduce the following $SO(8)$ tensors carrying $SU(8)$ indices
\bea
\gamma^{ij}_{AB}= \bar{w}_{qA}\bar{w}_{pB}\gamma^{ij}_{qp}\; , \qquad \bar{\gamma}^{ij\; AB}= {w}_q^{A}{w}_{p}^{B}\gamma^{ij}_{qp} = (\gamma^{ij}_{AB})^* \; , \qquad
%\\ \nonumber
\eea
 (see also \eqref{gij=SU8}) as well as ({\it cf. } \cite{Bandos:2025pxv})
 \bea
 t^{ijkl}_{ABCD}&=& \gamma^{[ij}_{[AB} \gamma^{kl]}_{CD]}  =\bar{w}_{qA}\bar{w}_{pB}\bar{w}_{rC}\bar{w}_{sD}\gamma^{ik}_{[qp} \gamma^{jk}_{rs]}= (\bar{t}^{ijkl\, ABCD})^* \; , \qquad
 %\\ \nonumber
 \\  t^{ij}_{ABCD}&=& \gamma^{ik}_{[AB} \gamma^{jk}_{CD]}  =\bar{w}_{qA}\bar{w}_{pB}\bar{w}_{rC}\bar{w}_{sD}\gamma^{ik}_{[qp} \gamma^{jk}_{rs]} = \; (\bar{t}^{ij\, ABCD})^* \; . \qquad
 \eea

Then it is easy to observe that Eqs. \eqref{phiqp=phiij},  \eqref{phiq1-q4==} and their c.c.   can be obtained as ''values'' at the point $\bar{w}_{qA}=\delta_{qA}=w_q^A$ of the $SU(8)$ sector of the enlarged Lorentz harmonic superspace of the following expressions for the components of the state vector superfield

\bea\label{phiAB=phiij}
 \phi_{AB}^{\#} &=&  - \frac 1 {16} \gamma^{ij}_{AB} c^{ij}\; , \qquad c^{ij}=- \frac 1 {16} \bar{\gamma}^{ijAB} \phi_{AB}^{\#}\; , \\  \nonumber
\\ \label{phiAB*=phiij}
 \bar{\phi}^{\# AB} &=&  - \frac 1 {16}  \bar{\gamma}^{ijAB}  c^{ij *}= (\phi_{AB}^{\#})^* \; , \qquad c^{ij*}=- \frac 1 {16} \gamma^{ij}_{AB}  \bar{\phi}^{\# AB}\; , \\  \nonumber
\\
 \label{phiA1-A4==}
\phi^{[+4]}_{A_1\ldots A_4} &=& \frac 1 {16}\left( ic^{ijkl} t^{ijkl}_{A_1A_2A_3A_4}  +\frac 2 3  t^{ij}_{A_1A_2A_3A_4} h^{ij}  \right)\;  \\ \nonumber
\\
\label{bphiA1-A4==}
\bar{\phi}{}^{[+4]A_1\ldots A_4} &=&(\phi^{[+4]}_{A_1\ldots A_4})^*  =  \frac 1 {16}\left( -ic^{ijkl} \bar{t}^{ijkl\, ABCD}  +\frac 2 3  \bar{t}^{ij\, ABCD} h^{ij}  \right)\; .
\eea
Notice that the  last two fields
obey \eqref{comp4=IIB} due to the following identities

\bea\label{t44=-ebt44=IIB}
& \bar{t}{}^{ijkl\, ABCD}= -\frac 1 {4!}  \epsilon^{ABCDEFGH}
t^{ijkl}_{EFGH}= -\frac 1 {4!}  \epsilon^{ijkli'j'k'l'} \bar{t}{}^{i'j'k'l'\, ABCD} = ( t^{ijkl}_{ABCD})^*\; , \qquad
%\\ \nonumber
\\ \label{t24=ebt24=IIB}
& \bar{t}{}^{ij\; ABCD}= +\frac 1 {4!}  \epsilon^{ABCDEFGH}
t^{ij}_{EFGH} = (t^{ij}_{ABCD} )^*\;  \qquad
%\\ \nonumber
\eea
which follow from  \eqref{t44==t*44}, \eqref{t24==t2*4} and
${\rm det}(w_q^A)=1$ (see \eqref{wbw=inSU8}).

The generic fermionic field components of the state vector superfield are  $\psi^+_A$ and $\psi^{[+3]}_{ABC}=\psi^+_{[ABC]}$. The first of these complex fields  and its c.c. are related to  $\psi^+_q$ and $\bar{\psi}^{+ q}$ by
\be
\psi^+_A = \bar{{\rm w}}_{qA}\psi^+_q \; , \qquad \bar{\psi}^{+ A} = {\rm w}_q^A(\psi^+_q)^* \; .  \qquad
\ee
To deal with  $\psi^{[+3]}_{[ABC]}$ and its c.c., we need to introduce a companion of our bridge variables $({\rm w}_q{}^A\in SU(8)$ \eqref{wbw=inSU8}, ${w}_{\dot{q}}{}^A =( \bar{{\rm w}}_{\dot{q}A})^*\in SU(8)$,  which connect fundamental and anti-fundamental representations of $SU(8)$ with c-spinor representation of $SO(8)$,

\bea\label{wbw=inSU8-c}
{\rm w}_{\dot{q}}{}^A\bar{{\rm w}}_{\dot{p}A}=\delta_{\dot{q}\dot{p}} \; , \qquad  \bar{{\rm w}}_{\dot{q}A}{\rm w}_{\dot{q}}{}^B=\delta_{A}{}^B \;,  \qquad {\rm det }({\rm w}_{\dot{q}}{}^A)=1\; , \qquad
\eea
This allows us to define
\bea\label{giAB=}
\gamma^i_{AB} =\bar{{\rm w}}_{{q}A} \gamma^i_{q\dot{p}} \bar{{\rm w}}_{\dot{p}B}=: \tilde{\gamma}{}^i_{BA}\; , \qquad \tilde{\bar{\gamma}}{}^{i\, BA}:=\bar{\gamma}{}^{i\, AB} ={\rm w}_{q}^{A} \gamma^i_{q\dot{p}}{\rm w}_{\dot{p}}^{B}\; \qquad
%\\ \nonumber
\eea
which obey
\bea
\gamma^i_{AC}\tilde{\bar{\gamma}}{}^{j\, CB} =\delta^{ij} \delta_{A}{}^{B}+  \gamma^{ij}{}_{A}{}^{B}\; , \qquad   \gamma^{ij}{}_{A}{}^{B}= \bar{{\rm w}}_{{q}A} \gamma^{ij}_{qp}{\rm w}_p{}^{B}=-( \gamma^{ij}{}_{B}{}^{A})^*\; .
%\\ \nonumber
\eea

Now we can check that Eq. \eqref{Psi3=PsiIdq} and its c.c. can be obtained at the point ${\rm w}_q^A=\delta_q^A$, ${\rm w}_{\dot{q}}^A=\delta_{\dot{q}}^A$ of the SU(8) sector of the enlarged Lorentz harmonic superspace from the following equations
\be\label{psi3=psiiA}
\psi^{[+3]}_{ABC}  =  \frac 1 8  \gamma^{ij}_{[AB}\gamma^j_{C]D}\psi^{i D}\; , \qquad \bar{\psi}^{[+3]ABC}  = (\psi^{[+3]}_{ABC})^*  = \frac 1 8  \bar{\gamma}^{ij[AB}\gamma^{C]Dj}\bar{\psi}^{i}_{ D}\; .
\ee
with
\be\label{psiDi=} \psi^{i D}=\psi^{i}_{\dot{q}} w_{\dot{q}}^{ D}\qquad \text{and}\qquad \bar{\psi}^{i}_{ D}=(\psi^{i D})^* =\bar{\psi}^{i}_{\dot{q}} \bar{w}_{\dot{q}D}
\ee  obeying
\bea\label{psiAigiAB=0}
\gamma^i_{AB}\psi^{i B}=0=\bar{\gamma}{}^{iAB}\bar{\psi}^{i}_{B}\; .
%\\ \nonumber
\eea

The inverse relations read
\bea\label{psiiA=psi3}
\psi^{iD} =  \frac 1 6  \bar{\gamma}{}^{ij[AB}\bar{\gamma}{}^{C]Dj}\psi^{[+3]}_{ABC}\; , \qquad \bar{\psi}{}^{i}_{D} =  \frac 1 6  \gamma^{ij}_{[AB}\gamma^j_{C]D}\bar{\psi}{}^{[+3]ABC}\; , \qquad
%\\ \nonumber
\eea
and, with  \eqref{psi3=psiiA}, reproduce \eqref{PsiIdq=Psi3} at the point
${\rm w}_q^A=\delta_q^A$, ${\rm w}_{\dot{q}}^A=\delta_{\dot{q}}^A$ of the SU(8) sector of the enlarged Lorentz harmonic superspace  \footnote{Eq. \eqref{psiAigiAB=0} is obeyed by \eqref{psiiA=psi3} due to the identity
$\gamma^{ij}_{[AB}\gamma^{ij}_{C]}{}^{D}=0$ following from $\gamma^{ij}_{[qp}\gamma^{ij}_{r]s}=0$ (which in its turn follows from \eqref{gijgij=}).
}. Of course, at that point the $SU(8)$ symmetry is not seen as $\gamma^i_{q\dot{p}}$ are invariants of $SO(8)$ group only.

The SU(8) transformations that act
on ${\rm w}_q^A$, ${\rm w}_{\dot{q}}^A$
can be used to connect any 'point' of the $SU(8)$ sector of
the
enlarged Lorentz harmonic superspace to ${\rm w}_q^A=\delta_q^A$, ${\rm w}_{\dot{q}}^A=\delta_{\dot{q}}^A$ 'point', so that
in the superparticle model ${\rm w}_q^A(\tau)$, ${\rm w}_{\dot{q}}^A(\tau)$ are {\it St\"uckelberg fields}.
After quantization, these St\"uckelberg fields describe the complex structure used in our quantization prescription and thus the possible complex structures are in one-to-one correspondence with the points of $SU(8)/SO(8)$ coset. The hidden $SU(8)$ symmetry of the linearized type IIB supergravity is thus seen in the constructions which are independent on the choice of this complex structure. These include the on-shell superfield \eqref{Phi=phi+} and its antisymmetric spin tensor components, but not their decompositions in terms of more ''standard'' fields
\eqref{phiAB=phiij}, \eqref{phiA1-A4==}, \eqref{psi3=psiiA}.
The indication of the relevance of this $SU(8)$ symmetry is that the superamplitudes of type IIB supergravity should be given by multiparticle generalization of these on-shell superfields,
as it is observed in the case of $4D$ supergravity
\cite{Brandhuber:2008pf,Arkani-Hamed:2008gz,Elvang:2015rqa} and $11D$ supergravity \cite{Bandos:2017eof,Bandos:2017zap}.

As type IIB and type IIA supergravity are related by T-duality, we should expect this $SU(8)$ symmetry to be seen also in the linearized type IIA supergravity as described by quantization of the type IIA superparticle in its spinor moving frame formulation. We perform such a quantization in the next section.

%\bigskip

\setcounter{equation}0

%\section{Quantum state spectrum of Type IIA superparticle
%and hidden SU(8) symmetry of linearized IIA supergravity}

\section{Type IIA superparticle quantization and hidden SU(8) symmetry of linearized IIA supergravity }

The action of type IIA massless superparticle in its spinor moving frame formulation can be written in the form of Eq. \eqref{S=IIB},
\bea\label{S=IIA}
S^{IIA} &=& \frac 1 8  \int d\tau \rho^\# \,   E^a_\tau \, v^-_{q} \tilde{\sigma}_{a}v^-_{q} \quad =\quad  \int\limits_{{\cal W}^1}  \rho^\# \,   E^a \, u_a^=  \quad = \int\limits_{{\cal W}^1}\rho^\# \,   E^= \; , \qquad
%\\ \nonumber
\eea
but with pull--back  $E^a=d\tau E_\tau^a$ of the bosonic  type IIA Volkov-Akulov 1--form, the bosonic forms of the supervielbein of flat type IIA superspace
\bea\label{Ea=IIA}
E^a= \Pi^a = dx^a -i d\theta^1\sigma^a \theta^1 -i d\theta^2\tilde{\sigma}{}^a \theta^2\; , \qquad E^{\alpha 1}=d \theta^{\alpha 1}\; , \qquad E_{\alpha}^{ 2}=d \theta_{\alpha}^{ 2}\; .  \qquad
\eea
This differential forms are, and hence the action \eqref{S=IIA} is, invariant under type IIA supersymmetry which acts on the coordinate functions by
\begin{eqnarray}\label{susy=IIA}
\delta_\varepsilon x^a= i \theta^1\sigma^a\varepsilon^1+  i \theta^2\tilde{\sigma}{}^a\varepsilon^2   \; , \qquad
\delta_\varepsilon \theta^{\alpha 1} =\varepsilon^{\alpha 1} \; , \qquad\delta_\varepsilon \theta_{\alpha}^2 =\varepsilon_{\alpha}^2 \;  \qquad
\;
\end{eqnarray}
and leaves
the spinor moving frame variables and
the field $\rho^\#$  intact.

This dynamical system can be considered as particle moving in the type IIA Lorentz harmonic superspace ({\it cf.} \eqref{IIB=cZcM}
  \be\label{IIA=cZcM}
\Sigma^{(10+16|16+16)}_{IIA}\; = \; \{ {\cal Z}^{{\frak M}}_{IIA} \}= \{(x^a, \theta^{\alpha 1},  \theta_{\alpha}^{ 2};
  (v_{\alpha q}^{\; -}\, , \,  v_{\alpha \dot{q}}^{\; +})\}\;
  \ee
which contains the same Lorentz harmonic sector $[SO(1,9)/SO(1,1)\times SO(8)]$ as IIB Lorentz harmonic superspace \eqref{IIB=cZcM}
(see discussion and comments on the notation around Eq.  \eqref{IIB=cZcM}).
In the central coordinate basis  $\{ {\cal Z}^{{\frak M}}_{IIA} \}$, the only difference with  type IIB case is  the chirality
of the second fermionic coordinate reflected by the position of its Majorana-Weyl (MW) spinor index ($ \theta_{\alpha}^{ 2}$ {\it vs} $ \theta^{\alpha 2}$ in type IIB case) which is now different from the chirality of the first fermionic coordinate $ \theta^{\alpha 1}$.

One can not rise or lower
the $10D$ MW  spinor indices
in an $SO(1,9)$ covariant manner which indicates that to describe
T-duality relating the flat type IIA and type IIB superspaces we have
to break the Lorentz covariance.
If we single-out some direction in the bosonic sector of type II superspace by choosing  a constant vector $k_a$, we can use it to write the relation between type IIB and type IIA coordinates
\be\label{th2=Tdual}
\theta_{\alpha}^{ 2}= k^a\sigma_{a\alpha\beta }
\theta^{\beta 2}_{IIB} \; .
\ee
This is invertible,
$\theta^{\alpha 2}_{IIB} = \frac 1 {k_ak^a}k_b\tilde{\sigma}{}^{b\alpha\beta } \theta_{\beta}^{ 2}$ if the vector is not light-like, $k_ak^a\not =0$. Actually the standard approach to T-duality is to chose the $k_a$ to be spacelike $k_ak^a<0$, however timelike T-duality has been discussed since the late 90s
\cite{Hull:1998vg}.

As the chirality of two fermionic supervielbein  forms of type IIA superspace \eqref{Ea=IIA}, in the split form of the pull--backs of type IIA flat supervielbein
 \bea\label{E--=IIA}
 && E^= = E^au_a^=\; , \qquad   E^\# = E^au_a^\#\; , \qquad  E^i = E^au_a^i\; , \qquad
 %\\ \nonumber
 \\ \label{E-q1=IIA} && E^{-q1} = d\theta^{\alpha 1} v_{\alpha q}^{\; -} \; , \qquad  E^{+\dot{q}1} = d\theta^{\alpha 1} v_{\alpha \dot{q}}^{\; +} \; , \qquad
 %\\ \nonumber
 \\  \label{E-dq2=IIA}
 && E^{-\dot{q}2} = d\theta_{\alpha}^{2} v^{-\alpha}_{\dot{q}} \; , \qquad  E^{+q2} = d\theta_{\alpha}^{2}
 v^{+\alpha}_{q} \;  \qquad
 %\\ \nonumber
 \eea
the components of two fermionic $1$--forms with the same $SO(1,1)$ weight carry indices of different (s-spinor and c-spinor) representation of $SO(8)$ group ({\it cf.} \eqref{E--=}, \eqref{E-q1=})\footnote{Of course, this can be related by triality using also  $SO(8)$ vector (v-) representation, in particular
$k^i=k_\mu u^{\mu i}$ (see below).}. This is what makes the  quantization of type IIA superparticle a bit different from type IIB case.

\subsection{Type IIA superparticle action in analytical coordinate basis}

The detailed discussion of type IIB case in sec. \ref{IIB} allows us to proceed much faster and, in particular,  to pass  directly
to the study of type IIA superparticle in the analytical basis of Lorentz harmonic superspace
\bea\label{IIA=cZcM-}
\Sigma^{(10+16|16+16)}_{IIA}\; &= &\; \{ {\cal Z}^{{\frak M}}_{An\, IIA} \}= \{(X^=, X^\# , X^i_{An}\; ; \theta^{-1}_q ,  \theta^{-2}_{\dot{q}}\; , \theta^{+1}_{\dot{q}},
   \theta^{+2}_{q}\; ,
  (v_{\alpha q}^{\; -}\, , \,  v_{\alpha \dot{q}}^{\; +}))\}\;  , \\ \nonumber \\ \label{X--:=IIA}
 && X^== x^au_a^=\; , \qquad X^\# = x^au_a^\# \; , \qquad
 %\\ \nonumber
 \\  \label{XiAn:=IIA} && X^i_{An} = x^au_a^i+ i\theta^{-1}_q \gamma^i_{q\dot{p}}\theta^{+1}_{\dot{p}} + i\theta^{+2}_q \gamma^i_{q\dot{p}}\theta^{-2}_{\dot{p}} \; , \qquad
  %\\ \nonumber
  \\   \label{th-12=IIA}
&&  \theta^{-1}_q = \theta^{\alpha 1} v_{\alpha q}^{\; -} \; , \qquad  \theta^{+1}_{\dot{q}} =\theta^{\alpha 1}  v_{\alpha \dot{q}}^{\; +} \; , \qquad
%\\  \nonumber
\\ \label{th+12=IIA}
 &&  \theta^{-2}_{\dot{q}} =\theta_{\alpha}^{ 2}  v^{-\alpha}_{\dot{q}}
   \; ,  \qquad \theta^{+2}_q = \theta_{\alpha}^{ 2}  v^{+\alpha}_q\; . \qquad
%\\ \nonumber
  \eea

The action \eqref{S=IIA} in this coordinate basis reads
\bea
%\label{S-An=IIA}
S^{IIA} &=&  \int_{{\cal W}^1} {\cal L}_1^{IIA} \; , \qquad
%\\ &&
\label{cL1=IIA-An} {\cal L}_1^{IIA}=\rho^\# E^= = \rho^\# \left( DX^= -2i D\theta^{-1}_q\, \theta^{-1}_q -2i D\theta^{-2}_{\dot{q}}\, \theta^{-2}_{\dot{q}} - \Omega^{= i} X_{An}^i\right)
\; , \qquad
%\\ \nonumber
\eea
where  the covariant derivatives are defined with the use of the composite $SO(1,1)$ and $SO(8)$ connection given by Cartan forms $\Omega^{0}$ and $\Omega^{ij}$, Eq. \eqref{Om--i:=},
({\it cf.}  \eqref{DX--:=}--\eqref{Dth2-:=})
\bea\label{DX--:=IIA}
 DX^=&=& dX^= +2 X^= \Omega^{(0)}  \; , \qquad
 %\\ \nonumber
 \\ \label{Dth1-:=IIA}
 D\theta^{-1}_q &=&  d\theta^{-1}_q+ \Omega^{(0)} \theta^{-1}_q + {1\over 4} \Omega^{ij}
\theta^{-1}_p\gamma_{{p}{q}}^{ij} \; , \qquad
%\\ \nonumber
\\ \label{Dth2-:=IIA}
 D\theta^{-2}_{\dot{q}}&=&  d\theta^{-2}_{\dot{q}}+ \Omega^{(0)} \theta^{-2}_q + {1\over 4} \Omega^{ij}
\theta^{-2}_{\dot{p}}\tilde{\gamma}^{ij}_{\dot{p}\dot{q}}
\; . \qquad
%\\ \nonumber
\eea

The action \eqref{cL1=IIA-An} involves only coordinate functions from the sub-superspace  $\Sigma^{(9+16|8+8)}\subset \Sigma^{(10+16|16+16)}$,

\be\label{fZcM=IIA}
\Sigma^{(9+16|8+8)} =\{{\frak Z}_{IIA}^{{\cal M}}\}=  \{(X^=, X^i_{An}\; ; \theta^{-1}_q ,  \theta^{-2}_{\dot{p}}\; , \;
  (v_{\alpha q}^{\; -}\, , \,  v_{\alpha \dot{q}}^{\; +}))\}\;
\ee
which is invariant under the  target superspace supersymmetry \eqref{susy=IIA}  realized
on its coordinates as follows
\bea\label{susy=IIA-9+16+}
\delta_{\varepsilon} X^= = 2i \theta^{-1}_q\varepsilon^{-1}_q+ 2i \theta^{-2}_{\dot{q}}\varepsilon^{-2}_{\dot{q}}\;,  \qquad \delta_{\varepsilon} X^i_{An} = 2i \theta^{-1}_q\gamma^{i}_{q\dot{p}}\varepsilon^{+1}_{\dot{q}}- 2i \theta^{-2}_{\dot{q}}\tilde{\gamma}^i_{\dot{q}p} \varepsilon^{+2}_p\; ,  \qquad
%\\ \nonumber
\\  \delta_{\varepsilon} \theta^{-1}_q= \varepsilon^{-1}_q\; ,  \qquad  \delta_{\varepsilon} \theta^{-2}_{\dot{q}}=  \varepsilon^{-2}_{\dot{q}}\; ,   \qquad
%\\ \nonumber
\\
 \delta_{\varepsilon} v_{\alpha q}^{\; -}=0 \, , \qquad   \delta_{\varepsilon} v_{\alpha \dot{q}}^{\; +}=0 \, , \qquad  \\ \nonumber
\text{where} \qquad
\varepsilon^{1-}_q= \varepsilon^\alpha v_{\alpha q}^{\; -} \, , \qquad   \varepsilon^{1+}_{\dot{q}}= \varepsilon^\alpha  v_{\alpha \dot{q}}^{\; +} \; , \qquad
\varepsilon^{-2}_{\dot{p}} =\varepsilon^{2}_\alpha  v^{-\alpha}_{\dot{p}} \; ,  \qquad
   \varepsilon^{+2}_q =\varepsilon^{2}_\alpha  v^{+\alpha}_{q} \; . \qquad
   %\\ \nonumber
\eea

The Hamiltonian formalism constructed in this analytical basis differs with the one we have described for IIB case only in Grassmann sector:
\bea
\theta^{-2}_q{}_{(IIB)}\mapsto \theta^{-2}_{\dot{q}}\; , \qquad \Pi^{2}_{-q}{}_{(IIB)}\mapsto \Pi^{2}_{-\dot{q}}\; . \qquad
%\\ \nonumber
\eea
This implies the change of the SO(8) representation of the second fermionic constraint \eqref{fd-2=0} which now reads
\bea
\label{fd-2IIA=0}
&& \frak{d}^{2}_{-\dot{q}} = \Pi^{2}_{-\dot{q}} +2i p_= \theta^{-2}_{\dot{q}}\approx 0 \; . \qquad
%\\ \nonumber
\eea
so that there is no an $SO(8)$ covariant way to pass to their complex combination as in \eqref{fd-==1-i2}, neither $SO(8)$ covariant way of passing to complex coordinates, as in \eqref{Th=1+i2} exists.

The algebra of the fermionic constraints of type IIA superparticle is
\bea\label{d1d1=p--=IIA}
\{ \frak{d}^{1}_{-q}, \frak{d}^{1}_{-p}  \}_{PB} = -4i p_=\delta_{qp}\; , \qquad
\{ \frak{d}^{1}_{-q}, \frak{d}^{2}_{-\dot{q}} \}_{PB} = 0\; , \qquad
\{ \frak{d}^{2}_{-\dot{q}}, \frak{d}^{2}_{-\dot{p}}  \}_{PB} = -4i p_=\delta_{\dot{q}\dot{p}} \; .  \qquad
%\\  \nonumber
\eea

\subsection{Introducing the complex structure in the type IIA superparticle mechanics }

To follow the way towards quantization of type IIA superparticle similar to that we have used for its type IIB cousin, we need  to introduce a complex structure and split the set of real fermionic coordinates onto a pair of complex conjugate coordinates.

A possibility is to use to this end the $SO(8)/[SO(6)\times U(1)]$ harmonics which were introduced for the description of $10D$ SYM  amplitudes in \cite{Bandos:2017eof}. Their introduction provides a covariant counterpart of the simplest construction which explicitly breaks $SO(8)$ (gauge) symmetry down to its
$SO(6)\times U(1)\approx SU(4) \times U(1)\approx U(4)$ subgroup.
However, in type IIA case a more natural way to introduce complex structure, both in sector of bosonic spinor variables and for Grassmann coordinates, corresponds to a covariant counterpart of breaking of $SO(8)$ symmetry down to its $SO(7)$ subgroup. Then, as we will see, the $SO(7)$ symmetry can be enhanced till $SU(8)$ in the manner similar (but not identical) to how it happens in the case of quantization of D$0$--brane performed in \cite{Bandos:2025pxv}.

The $SO(7)$ subgroup of $SO(8)$ can be defined by the conditions of preservation of some constant $SO(8)$ $8$--vector $k^i$. Then, choosing this vector to be of unit length, we can construct the $SO(8)$ valued $8\time 8$ matrix
\bea\label{ki}
k\!\!\!/{}_{q\dot{p}}:=k^i\gamma^i_{q\dot{p}}=: \tilde{k}\!\!\!/{}_{\dot{p}q}\; , \qquad  k^ik^i=1 \quad \Leftrightarrow \quad k\!\!\!/{} \tilde{k}\!\!\!/{}={\bb I}_{8\times 8}\,  \qquad
%\\ \nonumber
\eea
which can be used to relate  s- and c- spinor representations for the transformations of $SO(7)\in SO(8)$. Then the conjugate complex octuplets of fermionic coordinates can be constructed as
\bea\label{bTh-q=IIA}
\Theta^{- q}= \frac 1 {\sqrt{2}} (\theta^{1-}_q+ik\!\!\!/{}_{q\dot{p}} \theta^{2-}_{\dot{q}})\; , \qquad  \bar{\Theta}{}^{-}_q=   \frac 1 {\sqrt{2}} (\theta^{1-}_q-ik\!\!\!/{}_{q\dot{p}} \theta^{2-}_{\dot{q}})\; . \qquad
\eea
Althought these complex coordinates carry $SO(8)$ s-spinor indices, their construction remains invariant under such $SO(7)$ subgroup of $SO(8)$ which leaves invariant the $SO(8)$ vector $k^i$. The $SO(8)/SO(7)$ transformation can be used to rotate $k^i$ and thus describe changes of complex structure defined by \eqref{bTh-q=IIA}.

Notice that in Eqs. \eqref{bTh-q=IIA} we have used upper and lower s-spinor indices ($\Theta^{- q}$ vs. $\bar{\Theta}{}^{-}_q= (\Theta^{- q})^*$), thus suggesting $SO(7)$ symmetry enhancement till $SU(8)$.

Such an  enhancement of $SO(7)$ symmetry to  $SU(8)$ can be reached  by introducing the $SU(8)$ valued matrix variables
${w}_q^A = (\bar{{w}}_{qA})^*$ which serve as a bridge between $SU(8)$ and $SO(7)$ group transformations. These obey the constraints \eqref{wbw=inSU8}
\be\label{w=inSU8}
 \bar{{w}}_{qA}{w}_q^B=\delta_A{}^B\qquad \Leftrightarrow \qquad
 {w}_p^A\bar{{w}}_{qA}=\delta_{pq}\; , \qquad \det ({w}_q^A)= 1 \; . \qquad
\ee
and are St\"uckelberg fields/variables.
Until now, we consider their $q,p$ indices to be transformed by $SO(7)$ subgroup of $SO(8)$ which preserves the unit $SO(8)$ vector
$k^i$ defining the complex structure through \eqref{bTh-q=IIA}. This restriction will be removed below, but presently is kept to simplify the discussion for now.

Using the bridge variables
\eqref{wbw=inSU8} we can write a formally  $SU(8)$ invariant counterpart of the complex fermionic  coordinates \eqref{bTh-q=IIA}:
\bea\label{Th-A==IIA}
\Theta^{- A}= \frac 1 {\sqrt{2}} {w}_q^A (\theta^{1-}_q+ik\!\!\!/{}_{q\dot{p}} \theta^{2-}_{\dot{q}})\; , \qquad  \bar{\Theta}{}^{-}_A=  \frac 1 {\sqrt{2}} \bar{{w}}_A^q  (\theta^{1-}_q-ik\!\!\!/{}_{q\dot{p}} \theta^{2-}_{\dot{q}})\; . \qquad
\eea
Next, let us use the constant vector $k^i$ to construct
\bea\label{wAc=wAsk}
{w}^A_{\dot{q}}= {w}^A_pk\!\!\!/{}_{p\dot{q}}\; , \qquad \bar{{w}}_{\dot{q}A}= \bar{{w}}_{pA} k\!\!\!/{}_{p\dot{q}}\; .
\eea
This $8\times 8$ matrix is also unitary and unimodular, i.e. $SU(8)$ valued: one can easily check that it obeys the counterpart of \eqref{wbw=inSU8-c}.

Furthermore, using \eqref{wAc=wAsk} together with \eqref{w=inSU8}, we can factorize the matrix \eqref{ki},
\bea\label{kg=wbw}
k\!\!\!/{}_{q\dot{p}}=  { w}_q^A\bar{{w}}_{\dot{p}A}=  { w}^A_{\dot{p}} \bar{{ w}}_{qA} \; .
\eea
Substituting this factorization into \eqref{Th-A==IIA}, we obtain
\bea\label{Th-A=IIA}
\Theta^{- A}= \frac 1 {\sqrt{2}} ({w}_q^A \theta^{1-}_q+i{ w}_{\dot{q}}^A \theta^{2-}_{\dot{q}})\; , \qquad  \bar{\Theta}{}^{-}_A=  \frac 1 {\sqrt{2}}
(\bar{{w}}_{pA} \theta^{1-}_q-i\bar{{ w}}_{\dot{q}A} \theta^{2-}_{\dot{q}})\; . \qquad
%\\ \nonumber
\eea
In contrast to  \eqref{Th-A==IIA}, these relations can be considered  to be  invariant under the complete $SO(8)$ acting on the $q,p$ and $\dot{q},\dot{p}$ indices. The $SO(8)/SO(7)$ transformation would change the constant unit vector $k^i$ \eqref{ki} in
\eqref{wAc=wAsk}, but this does not matter as $k^i$  is not included in \eqref{Th-A=IIA}.

Actually, the drawback of the above construction has been the requirement for $SO(8)$ vector $k^i$ to be constant as this  breaks explicitly $SO(8)$ symmetry which is essential for the definition of spinor moving frame variables.
At this stage it becomes evident that this can be avoided  by requiring instead that this vector is  {\it covariantly constant} \footnote{Notice also that, as far as one dimensional gauge field is always purely gauge, Eq.  \eqref{Dki=0} implies that the vector field $k^i(\tau)$ can be obtained by $SO(8)$ gauge
transformations starting from a constant unit vector $k^i_0$ ($dk^i_0=0$).},
\be\label{Dki=0}
Dk^i = dk^i+k^j\Omega^{ji}=0\; . \ee
% One more advantage of this is that the 10-vector
%$k_\mu =u_\mu^i k^i $ with covarianly constant $k^i$ is
%constant, $dk_\mu = 0$ as it should be for the vector
% defining the direction of T-duality

The bridge variables $w_q^A=(\bar{w}_{qA})^*$ and $w_{\dot{q}}^A=(\bar{w}_{\dot{q}A})^*$, which define the complex structure  \eqref{Th-A=IIA}, are then transformed by  {\it the same} fundamental (and anti-fundamental) representation of $SU(8)$ with  s- and c-spinor representation of $SO(8)$ group acting as well on the Lorentz harmonics. Thus the space of possible complex structures which can be used to quantize type IIA massless superparticle can be identified with coset $SU(8)/SO(8)$, the same  one that  we have found in type IIB case.

\subsubsection{Lagrangian of type IIA superparticle with manifest $SU(8)$ symmetry}

Eqs. \eqref{wAc=wAsk} and \eqref{kg=wbw} with real covariantly constant $k^i$  can be used to solve Eqs. \eqref{Th-A=IIA} with respect to the 'original' real coordinates of the analytical basis \eqref{IIA=cZcM-},
\bea\label{th1-q=Thbw+=IIA}
 \theta^{1-}_q =  \frac 1 {\sqrt{2}}\left(\Theta^{- A}
\bar{w}_{qA} + \bar{\Theta}{}^{-}_A w_q^A\right) \; , \qquad
\theta^{2-}_{\dot{q}}= \frac 1 { \sqrt{2} i}\left(\Theta^{- A}
\bar{w}_{\dot{q}A} - \bar{\Theta}{}^{-}_A w_{\dot{q}}^A\right) \;. \qquad
%\\ \nonumber
\eea
To obtain the Lagrangian of massless type IIA superparticle with manifest $SU(8)$ symmetry, we have to substitute these  expressions in the Lagrangian \eqref{cL1=IIA-An}.

To present the result of such substitution in a compact and suggestive form, it is important to observe  that the covariant constancy  of the $SO(8)$ vector $k^i$,
 Eq. \eqref{Dki=0},  implies that  the derivatives of $\bar{w}_{\dot{q}A}$ and $w_{\dot{q}}^A$, defined in Eq. \eqref{wAc=wAsk}, are expressed in terms of the same $SU(8)$  Cartan forms  \eqref{mho=wDw} as derivatives of $\bar{w}_{{q}A}$ and $w_{{q}}^A$, so that
\bea\label{Dwdq=IIB}
 %\begin{cases}
 {\cal D} {w}_{\dot{q}}{}^A= d {w}_{\dot{q}}{}^A - \frac 1 4 \Omega^{ij}\tilde{\gamma}^{ij}_{\dot{q}\dot{p}} {\rm w}_{\dot{p}}{}^A-
 {\rm w}_{\dot{q}}{}^A{\mho}_{B}{}^{A}=0\; , \qquad
 {\cal D} \bar{{\rm w}}_{\dot{q}B} = d\bar{{\rm w}}_{\dot{q}B} - \frac 1 4 \Omega^{ij}\gamma^{ij}_{\dot{q}\dot{p}}\bar{{\rm w}}_{\dot{p}B}+{\mho}_{B}{}^{A}  \bar{{\rm w}}_{\dot{q}A}=0\; .
%\end{cases} \\ \nonumber
\eea

With this in mind, we find that in terms of complex fermionic variables \eqref{Th-A=IIA}, the Lagrangian form of type IIA superparticle reads
\bea
\label{cL1=IIA-An=c} && {\cal L}_1^{IIA}=\rho^\# E^= = \rho^\# \left( DX^= -2i {\cal D}\Theta^{-A}\, \bar{\Theta}^{-}_A +2i \Theta^{-A} {\cal D}\bar{\Theta}^{-}_A\, - \Omega^{= i} X_{An}^i\right)
\; ,  \qquad
\eea
where ${\cal D}\Theta^{-A}$ and its c.c. are $SU(8)$ covariant derivatives defined in \eqref{cDThA:=},
\bea\label{cDThA:==}
{\cal D}\Theta^{-A}:= d\Theta^{-A}- \Theta^{-B }{\mho}_{B}{}^{A} \; , \qquad {\cal D}\bar{\Theta}{}^-_B:=d\bar{\Theta}{}^-_B +{\mho}_{B}{}^{A}\bar{\Theta}{}^-_A\; .
\eea

Notice that the Eq. \eqref{cL1=IIA-An=c} coincides with \eqref{cL1=IIB-An=c} so that the difference between type IIA and type IIB particle is in how the complex octuplet of fermionic coordinates and its complex conjugate are related to the original Majorana--Weyl fermionic coordinate functions.

Notice that this relation in the case of type IIA superparticle inevitably involves a covariantly constant  $SO(8)$ vector $k^i$.  This can be related with spacelike vector defining T-duality transformations \eqref{th2=Tdual} by using spacelike moving frame vectors,
\be\label{ka=kiuai}
k^\mu= k^i u^{\mu i}\; .
\ee
With such an identification, the covariant constancy of $k^i$, Eq. \eqref{Dki=0} follows from constancy of the space--like vector of the T-duality transformations, $dk^a=0$. Using  the expression for derivatives of moving frame vectors in terms of Cartan forms (see Eq.
\eqref{Dui} in Appendix \ref{App=harm}) one easily finds that this latter condition also implies
\be k^i\Omega^{=i}=0\qquad {\rm and}\qquad k^i\Omega^{\# i}=0\; .
\ee
The first of these equations is satisfied on-shell as equations of motion for $X^i_{{\rm An}}$ field are $\rho^{\#}\Omega^{=i}=0$. The second equation can be obtained by gauge fixing (complete of partial) of the ${\bb K}_8$ gauge symmetry.

Keeping this in mind, we can study the dynamical system described by the action with Lagrangian form \eqref{cL1=IIA-An=c} exactly in the same line as we did this with type IIB superparticle model described by Lagrangian form \eqref{cL1=IIB-An=c}. In particular, the Hamiltonian mechanics in the complex analytical basis is identical to that described in sec. \ref{IIB} and  quantization results in the chiral (analytical) superfield \eqref{Phi=phi+}.
The difference between type IIA and type IIB superparticle manifests itself when one turns to their standard spacetime description.
Namely, this happens when we associate the components of this superfields carrying antisymmetric sets of $SU(8)$ indices, bosonic
\bea\label{IIA-phi}
 \phi\; , \qquad  \phi^{\#}_{AB}=  \phi^{\#}_{[AB]} \; , \qquad   \phi^{[+4]}_{A_1\ldots A_4} =\frac 1 {4!\, 4!} \epsilon_{A_1\ldots A_4B_1B_2B_3B_4} (\phi^{[+4]}_{B_1\ldots B_4})^* \; \eea
and fermionic
\bea\label{IIA-psi}
 \psi^+_A \; , \qquad  \psi^{[+3]}_{ABC} =\psi^{[+3]}_{[ABC]}\; , \qquad \\ \nonumber \eea
with the solutions of the linearized equations for the field of supergravity multiplet.

In the spinor moving frame (Lorentz harmonic) formalism the field equations of type IIA supergravity multiplet can be solved in terms of {\it real} bosonic and fermionic functions
\bea \phi^{IIA}(p_=,v_{\alpha q}^{\; -})\; , \qquad c^i(p_=,v_{\alpha q}^{\; -})  \; , \qquad   c^{ijk}=c^{[ijk]} (x^=,v_{\alpha q}^{\; -})\; , \qquad  h^{ij} =h^{(ij)}(x^=,v_{\alpha q}^{\; -})\; , \qquad h^{ii}=0\; , \qquad
%\\ \nonumber
\\ \psi^+_{\dot{q}}(p_=,v_{\alpha q}^{\; -})   \; , \qquad  \psi^+_{{{q}}}(p_=,v_{\alpha q}^{\; -} )    \; , \qquad \psi^{i+}_{\dot{q}}(p_=,v_{\alpha q}^{\; -} )  \; , \qquad  \psi^{i+}_{{q}}(p_=,v_{\alpha q}^{\; -})     \; , \qquad
 \eea
by
\bea
C_\mu = -c^i u_\mu^i\; , \qquad C_{\mu\nu\rho} = -c^{ijk} u_\mu^iu_\nu^ju_\rho^k\; , \qquad h_{\mu\nu}=
u_\mu^iu_\nu^j h^{ij}\; , \qquad
%\\ \nonumber
\\
\psi^1_\alpha = \psi^+_q v_{\alpha q}^{\; -}\; , \qquad \psi_\mu^{\alpha 1} = -u_\mu^i\psi^{i+}_{\dot{q}} v_{\dot{q}}^{- \alpha }\; , \qquad   \gamma^{i}_{q\dot{q}}\psi^{i-}_{\dot{q}}=0\; , \qquad
%\\ \nonumber
\\ \psi^{2\alpha} = \psi^+_{\dot{q}} v^{-\alpha}_{\dot{q}}\; , \qquad\psi^2_{\mu\alpha} = -u_\mu^i\psi^{i+}_q v_{\alpha q}^{\; -}\; , \qquad  \psi^{i+}_q\gamma^{i}_{q\dot{q}}=0\; . \qquad
%\\ \nonumber
\eea
The signs are introduced in such a way that the "inverse" expressions for  the on-shell fields with $SO(8)$ indices in terms of spacetime fields have the natural form
\bea
 c^i= C_\mu u^{\mu i}\; , \qquad c^{ijk}= C^{\mu\nu\rho}  u_\mu^iu_\nu^ju_\rho^k\; , \qquad  h^{ij}=h^{\mu\nu}
u_\mu^iu_\nu^j\; , \qquad \\
%\nonumber \\
 \psi^+_q =\psi^1_\alpha v^{+\alpha}_{q}\; , \qquad \psi^{i+}_{\dot{q}}= \psi^{\mu\alpha 1} u_\mu^i v_{\alpha \dot{q}}^{\; +}\; , \qquad
 %\\ \nonumber
 \\ \psi^+_{\dot{q}}= \psi^{2\alpha} v_{\alpha\dot{q}}^{\; +}\; , \qquad \psi^{i+}_q=\psi^2_{\mu\alpha} u_\mu^i v^{+\alpha}_{ q}\; . \qquad
 %\\ \nonumber
\eea

Keeping in mind that the description of quantum spectrum of type IIA superparticle by complex chiral superfield
with field content given by \eqref{IIA-phi} and  \eqref{IIA-psi} is reached by introducing complex structure with the use of a unit covariantly constant SO(8) vector $k^i$, one should not be surprised by that the correspondence with the solutions of the linearized type IIA equations in spacetime  should also use this vector. Furthermore, in the light of observation that this vector can be used to define T-duality map between type IIB and type IIA superspaces, the problem of finding
the correspondence with components of state vector superfield can be equivalently formulated as a search for  T-duality map between the above described solutions of linearized type IIA field equations and their type IIB counterparts discussed before.

Such a correspondence can be written with the use of unit $SO(8)$  vector $k^i$ as follows

\bea\label{phiIIB=phi+ck}
\phi^{IIB} =\Re{\rm e} \phi^{IIB}+ i \Im{\rm m} \phi^{IIB}  & \qquad \longleftrightarrow \qquad & \phi^{IIA} +ic^ik^i\; , \qquad
%\\ \nonumber
\\
c^{ij} =\Re{\rm e} c^{ij}+ i \Im{\rm m} c^{ij}  & \qquad \longleftrightarrow \qquad & c^{ijl}k^{l}  + ic^{[i}k^{j]}\; , \qquad
%\\ \nonumber
\\ c^{ijkl} = -\frac 1 {4!}\epsilon^{ijkli'j'k'l'}c^{i'j'k'l'}  & \qquad \longleftrightarrow \qquad  & c^{[ijk}k^{l]}  - \frac 1 {4!}\epsilon^{ijkli'j'k'l'}c^{i'j'k'}  k^{l'}\; , \qquad
%\\ \nonumber
\\  h^{ij\, IIB} \qquad  & \qquad \longleftrightarrow \qquad & {}\qquad h^{ij}\; , \qquad
%\\ \nonumber
\\
\psi_q^{+IIB} =\Re{\rm e} \psi_q^{+IIB} + i \Im{\rm m} \psi_q^{+IIB}   & \qquad \longleftrightarrow \qquad & \psi_q^{+} + i k{}\!\!\!/{}_{q\dot{q}} \psi^+_{\dot{q}}  \; , \qquad
%\\ \nonumber
\\
\psi_{\dot{q}} ^{+i\; IIB} =\Re{\rm e} \psi_{\dot{q}} ^{+i\; IIB}  + i \Im{\rm m} \psi_{\dot{q}} ^{+i\; IIB}    & \qquad \longleftrightarrow \qquad  &
\psi_{\dot{q}} ^{+i} + i
\psi_q^{+i} k{}\!\!\!/{}_{q\dot{q}}   \; .  \qquad
%\\ \nonumber
\eea

Then the correspondence of the components of state vector superfield
\eqref{IIA-phi}, \eqref{IIA-psi} and on shell solutions of the linearized type IIA equations is given by (here we, for simplicity, again ignore the degrees of $\rho^{\#}$ or $\partial_=$ necessary to equalize the $SO(1,1)$ weights of l.h.s. and r.h.s. of equations; {\it cf.} \eqref{phiAB=phiij}-- \eqref{bphiA1-A4==} and \eqref{psi3=psiiA})

\bea\label{phi==IIA}
\phi &=& \phi^{IIA} +ic^ik^i \; , \qquad
 %\\  \nonumber
\\
\label{phiAB=IIA}
 \phi_{AB}^{\#} &=&  +\frac 1 {16} \gamma^{ij}_{AB} c^{ijl}k^{l}+ \frac i {16}c^{i} \gamma^{ij}_{AB}k^{j}\; , \qquad
 %\\  \nonumber
\\
 \label{phiA1-A4=IIA}
\phi^{[+4]}_{ABCD} &=& \frac 1 {8}\left( i c^{ijk}k^{l} t^{ijkl}_{ABCD}\;     +\frac 1 3  t^{ij}_{ABCD} h^{ij}  \right)\; ,
%\\ \nonumber
\\ \label{psi3=IIA}
\psi^{[+3]}_{ABC} & =&  \frac 1 8  \gamma^{ij}_{[AB}\gamma^j_{C]D} (\psi_{\dot{q}} ^{+i} w_{\dot{q}}{}^{D} + i
\psi_q^{+}  w_{{q}}{}^{D})\; , \qquad
%\\ \nonumber
\\ \label{psiA=IIA}
\psi^{[+]}_{A} & =&  \psi_q^{+}  \bar{w}_{{q}A}+ i \psi_{\dot{q}} ^{+} \bar{w}_{\dot{q}A}\; , \qquad
%\\ \nonumber
\eea
When writing the above equations we have used \eqref{wAc=wAsk} and (anti-)self-duality properties of $t$-symbols \eqref{t44=-ebt44=IIB}.

These relations can be resolved by expressions for  solutions of linearized type IIA equations in terms of components of the state vector superfield,
\bea
h^{ij} &=& \frac 1 {16} \bar{t}^{ij\; ABCD} \phi^{[+4]}_{ABCD} \; , \qquad
%\\ \nonumber
\\
c^i&=&  \Im{\rm m} \left(2\bar{\gamma}{}^{ij\,AB}k^j\phi^{[+2]}_{AB}+\phi k^i \right)\; , \qquad
%\\ \nonumber
\\
c^{ijk}&=& \frac i 4 k^l \bar{t}^{lijk\; ABCD} \phi^{[+4]}_{ABCD}  + 3k^{[i} \, \Re{\rm e} ( \bar{\gamma}{}^{jk]\,AB}\phi^{[+2]}_{AB})\; , \qquad
%\\ \nonumber
\\
\psi^{+i}_{\dot{q}} &=& \frac 1 6 \Re{\rm e} ( \bar{\gamma}{}^{ij\,AB} \bar{\gamma}{}^{j\,CD}\psi^{[+3]}_{ABC} \bar{w}_{\dot{q}D}) \; , \qquad \psi^{+i}_{{q}} = \frac 1 6 \Im{\rm m} ( \bar{\gamma}{}^{ij\,AB} \bar{\gamma}{}^{j\,CD}\psi^{[+3]}_{ABC} \bar{w}_{{q}D}) \; , \qquad
%\\ \nonumber
\\
\psi^+_q &=& \Re{\rm e} ( \psi^{[+]}_Aw_{{q}}{}^{A}) \; , \qquad
\psi^+_{\dot{q}} = \Re{\rm e} ( \psi^{[+]}_Aw_{\dot{q}}{}^{A}) \; , \qquad
%\\ \nonumber
\eea

Thus, we have established the one-to-one correspondence between
the field content of
the
state vector superfield and
the
solutions of the linearized equations of type IIA supergravity. This correspondence inevitably involves a unit covariantly constant  SO(8) vector $k^i$ which was  also used  to introduce complex structure in the phase space of  type IIA superparticle  and, through \eqref{ka=kiuai},   to define T-duality transformations between type IIA and type IIB supergravities. The presence of this unit $8$--vector makes the relation between spacetime fields and components of state vector superfield a bit
convoluted. In particular the degrees of freedom of vector and third rank antisymmetric tensor are redistributed between different spin-tensor fields.

Notice that in the on-shell superfield description of quantum state spectrum  of type IIA superparticle, when quantized in the frame of its spinor moving frame formulation by the method we followed in this paper, possesses $SU(8)$ symmetry which is hidden in its standard spacetime (and standard superspace) description.

\setcounter{equation}{0}
\section{On amplitudes and superamplitudes in type II theories}

\label{Sec=sAmpl}

In this section we will firstly discuss and reproduce in our notation  some simple amplitudes and superamplitudes of type IIB supergravity studied in  \cite{Boels:2012ie,Boels:2012zr} and  discuss their type IIA counterparts. Then we discuss some simple amplitudes of type IIA theory involving D$0$--branes and point out the issue which hampers the way to  superamplitudes  in this case.

\subsection{Spinor moving frame approach to spinor helicity formalism}

The spinor moving frame approach to the spinor helicity formalism in $D=10$ (and in  $D=11$) was developed in \cite{Bandos:2017zap,Bandos:2017eof}. The spinor helicity variables and polarization spinors of ${\rm i}$-th of $n$ massless  scattered particles (${\rm i}=1,...,n$), $\lambda_{\alpha q{\rm i}}$ and $\lambda^\alpha_{\dot{q}{\rm i}}$  are related to the spinor moving frame $( v^-_{\alpha {q}{\rm i}}, v^+_{\alpha \dot{q}{\rm i}})$ attached to the light--like momentum of this particle in the sense of
\be\label{pi=ru--i} p_{\mu {\rm i}}= \rho^{\#}_{{\rm i}} u_{\mu {\rm i}}^= \; , \ee
and to its inverse defined in  \eqref{harmV-1=10}, by
\bea
\lambda_{\alpha{q}{\rm i}}=\sqrt{|\rho^{\#}_{{\rm i}}|} v^-_{\alpha {q}{\rm i}}\; , \qquad \lambda^\alpha_{\dot{q}{\rm i}}=  \sqrt{|\rho^{\#}_{{\rm i}}|}v^{-\alpha}_{\dot{q}{\rm i}}\; . \qquad
\eea

The complementary set of spinors used in \cite{Boels:2012ie},  $\xi_{\alpha \dot{q}{\rm i}}$ and $\xi^\alpha_{{q}{\rm i}}$ in our index notation, can be associated with second rectangular block of spinor moving frame matrix \eqref{harmV=10} and its inverse \eqref{harmV-1=10},
\bea
\xi_{\alpha{q}{\rm i}}\; \propto \;  v^+_{\alpha \dot{q}{\rm i}}\; , \qquad \xi^\alpha_{\dot{q}{\rm i}}\propto v^{+\alpha}_{q{\rm i}}\; .
\eea

Each spinor moving frame is defined up to its associated $\{[SO(1,1)\otimes SO(8)]\subset\!\!\!\!\!\!\times {\bb K}_8\}$ gauge symmetry, which we denote by $\{[SO(1,1)\otimes SO(8)]\subset\!\!\!\!\!\!\times {\bb K}_8\}_{{\rm i}}$ for ${\rm i}$-th particle. These can be used (see \cite{Bandos:2017eof}) to gauge fix its decomposition

\begin{eqnarray}\label{v-i=v-+Kigv+}
v_{\alpha q\, {\rm i}}^{\; -} &=&
v_{\alpha  q }^{\; -}+ {1\over 2} K^{=i}_{{\rm i}}  \gamma^i_{q\dot{p}} v_{\alpha  \dot{p}}^{\; +}
\; , \qquad
v_{\alpha \dot{q}\, {\rm i}}^{\;+}= v_{\alpha  \dot{q}}^{\; +}
 \; , \qquad \\
\label{v-1+i=v-Kigv} v^{+\alpha}_{q\, {\rm i}}&=&
v^{+\alpha}_{q}
 \; ,  \qquad
v^{-\alpha}_{\dot{q}\,{\rm i}}=
v^{-\alpha}_{\dot{q}}-  {1\over 2} K^{= i}_{{\rm i}}  v^{+\alpha}_{q}\gamma^i_{{q}\dot{q}}
\;  \qquad
\end{eqnarray}
with respect to some reference spinor moving frame  $( v^-_{\alpha {q}}, v^+_{\alpha \dot{q}})$ which, in its turn, is defined up to associated $[SO(1,1)\otimes SO(8)]\subset\!\!\!\!\!\!\times {\bb K}_8$ gauge transformations. Only this symmetry preserves the gauge \eqref{v-i=v-+Kigv+}.

Notice that, in particular, the spinor moving frame of one of the scattered particles can be chosen as reference frame.

The corresponding gauge fixed version of ${\rm i}$-th vector frame reads
\begin{eqnarray}\label{u--=Kiui}
u^=_{\mu {\rm i}}= u^=_{\mu}   + K^{=j}_{ {\rm i}}u^j_{\mu} + \frac 1 4 (\vec{K}^{=}_{ {\rm i}} )^2 u^\#_{\mu}
 \; , \qquad \\
 \label{ui=Kiu++}
u^j_{\mu {\rm i}} = u^j_{\mu} + \frac 1 2  K^{= j }_{ {\rm i}}\, u^\#_{\mu}\; , \qquad \\
\label{u++=u0}
u^\#_{\mu {\rm i}} =u^\#_{\mu} \; . \qquad
\end{eqnarray}

In the gauge \eqref{v-i=v-+Kigv+} the characteristic contractions of the
helicity spinors and polarization spinors is expressed by
\bea\label{v-iv-j=}
<{\rm i}\; ^-_q \; | \; {\rm j}\; ^-_{\dot{p}} \; >\; := \;
v_{\alpha q\, {\rm i}}^{\; -}
v^{-\alpha}_{\dot{p}\,{\rm j}}\; = \; \frac 1 2 \; K^{= k}_{{\rm i}{\rm j}}\gamma^k_{q\dot{p}}\; ,
\eea
  where
\be\label{Kij=Ki-Kj}
\vec{K}{}^{=}_{{\rm ij}}:= \vec{K}{}^{=}_{{\rm i}}-\vec{K}{}^{=}_{{\rm j}}= \frac 1 4 \,
<{\rm i}\; ^-_q \; | \; {\rm j}\; ^-_{\dot{p}} \;  > \; \vec{\gamma}_{q\dot{p}}\;  \; .
\ee
Notice also that $u_{\mu{\rm i}}^=u^{\mu =}_{{\rm j}}= \frac 1 2\, (\vec{K}_{{\rm ij}}^=)^2 $.

The above construction was elaborated and generalized to $11D$  in \cite{Bandos:2017eof}. However,  for our present purposes it is more convenient to introduce complex helicity spinors and polarization spinors, along the lines
initiated in \cite{Bandos:2017zap} and continued in \cite{Bandos:2019zqp}. In the light of our above  observation on hidden $SU(8)$ symmetry of the linearized supergravity which appears naturally in superparticle quantization, here we further elaborate it in its $SU(8)$ covariant form.

The complex helicity spinors and polarization spinors are defined by
\bea\label{lA:=}
{\lambda}_{\alpha A{\rm i}}=\sqrt{|\rho^{\#}_{{\rm i}}|} \bar{v}{}^-_{\alpha A{\rm i}}:=\sqrt{|\rho^{\#}_{{\rm i}}|} v^-_{\alpha {q}{\rm i}}\bar{w}_{qA {\rm i}}\; , \qquad \lambda^{\alpha A}_{{\rm i}}=  \sqrt{|\rho^{\#}_{{\rm i}}|}v^{-\alpha A}_{{\rm i}}=  \sqrt{|\rho^{\#}_{{\rm i}}|}v^{-\alpha}_{\dot{q}{\rm i}}w_{\dot{q} {\rm i}}^{\; A}\; .
\eea
These variables are sufficient to form the complete basis of spinors of opposite chiralities with ${v}{}^{+ A}_{\alpha {\rm i}}:= v^+_{\alpha \dot{q}{\rm i}}w_{\dot{q} {\rm i}}^{\; A}$ and  $
\bar{v}{}^{+\alpha}_{ A{\rm i}}:= v{}^{+\alpha}_{{q}{\rm i}}\bar{w}_{qA {\rm i}}$. However, the expression for momenta in terms of spinors \eqref{lA:=} includes also their complex  conjugates, so that we shall use the element of all the overfull basis
\bea\label{v-Ai:=}
&& \bar{v}{}^-_{\alpha A{\rm i}}= v^-_{\alpha {q}{\rm i}}\bar{w}_{qA {\rm i}}\; , \qquad \bar{v}{}^+_{\alpha A{\rm i}}= v^{+}_{\alpha \dot{q}{\rm i}}\bar{w}_{\dot{q} A {\rm i}}\; , \qquad {}\qquad
{v}{}^{-A}_{\alpha {\rm i}}= v^-_{\alpha {q}{\rm i}}{w}^{\, A}_{q{\rm i}}\; , \qquad {v}{}^{+A}_{\alpha {\rm i}}= v^{+}_{\alpha \dot{q}{\rm i}}{w}^{\;A}_{\dot{q} {\rm i}} \; , \qquad \\ \label{v-1-Ai:=}
&& v^{-\alpha A}_{{\rm i}}=  v^{-\alpha}_{\dot{q}{\rm i}}w_{\dot{q} {\rm i}}^{\; A}\; , \qquad
v^{+\alpha A}_{{\rm i}}=  v^{+\alpha}_{{q}{\rm i}}w_{{q} {\rm i}}^{\; A}\; , \qquad {}
\qquad
\bar{v}{}^{-\alpha}_{A{\rm i}} =
v^{-\alpha}_{\dot{q}{\rm i}}
\bar{w}_{q  A {\rm i}}\; , \qquad
\bar{v}{}^{+\alpha}_{A{\rm i}}=  v^{+\alpha}_{{q}{\rm i}}\bar{w}_{{q} A{\rm i}}^{\; A} \; . \qquad
\eea

The gauge fixing of symmetries the set of which in this case includes also $SU(8)_{{\rm i}}$ groups, allows us to identify all the bridge variables for each of scattered particles with  the reference bridge variables,
\bea
\bar{w}_{qA {\rm i}}=\bar{w}_{qA }\; , \qquad w_{{q} {\rm i}}^{\; A}=w_{{q}}^{\; A} \; , \qquad {} \qquad \bar{w}_{\dot{q}A {\rm i}}=\bar{w}_{\dot{q}A }\; , \qquad w_{\dot{q} {\rm i}}^{\; A}=w_{\dot{q}}^{\; A}\; ,\qquad
\eea
and to arrive at the following expressions for complex spinor frame variables in terms of reference complex spinor frame
\begin{eqnarray}\label{bv-=bv+Kv=G=10}
\bar{v}{}^-_{\alpha A{\rm i}}
 &=&
\bar{v}{}^-_{\alpha A}+ {1\over 2} K^{=j}_{{\rm i}}  \gamma^j_{AB} v_{\alpha}^{+B}
\; , \qquad v_{\alpha {\rm i}}^{+B}=  v_{\alpha}^{+B}
 \; , \qquad\\
\label{bv+-1=G=10} \bar{v}{}^{+\alpha}_{A {\rm i}}&=&
\bar{v}{}^{+\alpha}_{A}
 \; ,  \qquad
v^{-A \alpha}_{{\rm i}}=
v^{-A\alpha}-  {1\over 2} K^{= j}_{{\rm i}}  \tilde{\bar{\gamma}}{}^{j\, AB} \bar{v}^{+ \alpha}_{B}
\; . \qquad
\end{eqnarray}
Here $\gamma^i_{AB}$ and $\bar{\gamma}{}^{iAB}$ carrying $SO(8)$ vector index and two indices of fundamental anti-fundamental representations of $SU(8)$ are defined in \eqref{giAB=}. Eqs. \eqref{bv-=bv+Kv=G=10} and \eqref{bv+-1=G=10} are covariant with respect to only one copy of $SU(8)\otimes SO(8)\otimes SO(1,1)$ symmetry acting on the reference spinor moving frame and  related reference bridge variables only.

The convenience of such a complex choice of helicity spinors and associated spinor moving frame is that it can be used to write a simple representation of complex generators of type IIB supersymmetry acting on the on-shell superfields and also on superamplitudes:
\bea\label{bbbQ=}
\bar{{\bb Q}}_{\alpha {\rm i}}&=&\frac 1 {\sqrt{2}} (Q_{\alpha {\rm i}}^1+iQ_{\alpha {\rm i}}^2) = v_{\alpha q\, {\rm i}}^{\; -} w_{q {\rm i}}^A \frac {\partial} {\partial \Theta_{{\rm i}}^{-A}}= v_{\alpha {\rm i}}^{- A} \frac {\partial} {\partial \Theta_{{\rm i}}^{-A}} \; , \qquad    \\ \label{bbQ=}
{\bb Q}_{\alpha {\rm i}}&=&\frac 1 {\sqrt{2}}
(Q_{\alpha {\rm i}}^1-iQ_{\alpha {\rm i}}^2) = 4\rho^{\#}_{ {\rm i}}v_{\alpha q\, {\rm i}}^{\; -} \bar{w}_{qA {\rm i}}  \Theta_{{\rm i}}^{-A}= 4\rho^{\#}_{ {\rm i}}\bar{v}_{\alpha A {\rm i}}^{\; - }  \Theta_{{\rm i}}^{-A}
%\\ \nonumber \\ \label{bbQi=-4iQi}
%&=:& =4 {Q}_{\alpha {\rm i}}
\; .   \qquad
\eea

The relation of ${\bb Q}_{\alpha {\rm i}}$ and $\bar{{\bb Q}}_{\alpha {\rm i}}$ through Hermitian conjugation is not straightforward in this representation (like it is in its 4D counterpart, see e.g. \cite{Brandhuber:2008pf,Arkani-Hamed:2008gz} as well as \cite{Boels:2012ie} and refs therein). However, the superalgebra of type IIB supersymmetry \eqref{susy=IIB} is reproduced
\bea\label{QiQj=}
\{ {\bb Q}_{\alpha {\rm i}}\; ,\; {{\bb Q}}_{\beta {\rm j}}\}= 0
 \; , \qquad \{\bar{{\bb Q}}_{\alpha {\rm i}}\; ,\;
\bar{{\bb Q}}_{\beta {\rm j}} \} =0\; , \qquad
%\\ \nonumber  \\
\{ {\bb Q}_{\alpha {\rm i}}\; ,\; \bar{{\bb Q}}_{\beta {\rm j}}\}=2 \delta_{{\rm i}{\rm j}} \sigma^a_{\alpha\beta} p_{a {\rm i}} \; , \qquad   p_{a {\rm i}} =  \rho^\#_{{\rm i}}  u^=_{a {\rm i}} \; .  \qquad
%\\ \nonumber
\eea
Notice that \eqref{QiQj=} reflects also the fact that supercharges of different particles anticommute. This is important, for instance, when calculating the algebra of complete supercharges of the system of $n$ scattered massless (super)particles (below referred to as ''supergravitons'')
\bea
{\bb Q}_{\alpha } =
\sum\limits_{{\rm i}=1}^{n}
{\bb Q}_{\alpha {\rm i}} \; , \qquad  \bar{{\bb Q}}_{\beta} =
\sum\limits_{{\rm j}=1}^{n}
\bar{{\bb Q}}_{\beta {\rm j}}\;  \qquad
\eea
which is
 \bea\label{QbQ=}
\{ {\bb Q}_{\alpha }\; ,\; {{\bb Q}}_{\beta }\}= 0
 \; , \qquad \{ {\bb Q}_{\alpha} \; ,\; \bar{{\bb Q}}_{\beta }\}= 2 \sigma^\mu_{\alpha\beta} {\bb P}_\mu \; , \qquad \{\bar{{\bb Q}}_{\alpha }\; ,\;
\bar{{\bb Q}}_{\beta} \} =0\; , \qquad
%\\ \nonumber  \nonumber
\eea
with the total momentum given by
\bea\label{Pmu=}
{\bb P}_\mu:= \sum\limits_{{\rm i}=1}^{n} p_{\mu {\rm i}}= \sum\limits_{{\rm i}=1}^{n} \rho^\#_{{\rm i}}u^=_{\mu {\rm i}}\; . \qquad
\eea

Notice that in discussion of $10D$ SYM and $11D$ SUGRA amplitudes in \cite{Caron-Huot:2010nes,Bandos:2017zap,Bandos:2017eof}
the counterpart of ${\bb Q}_{\alpha } = 4\sum\limits_{{\rm i}=1}^{} \rho^{\#}_{{\rm i}} v^-_{\alpha q{\rm i}} \bar{w}_{qA}\Theta^{-A}_{{\rm i}}$ was called supermomentum.
Under supersymmetry it transforms as
\bea
\delta_{susy} {\bb Q}_{\alpha } =  \sqrt{2}\, \left( \sum\limits_{{\rm i}=1}^{n} p_{\mu {\rm i}}\right)\, \sigma^\mu_{\alpha\beta} (\epsilon^{1\beta } +i\epsilon^{2\beta })
\eea
and, hence, is invariant under supersymmetry if the momentum is conserved,   $\sum\limits_{{\rm i}=1}^{n} p_{\mu {\rm i}}=0$. This statement is tantamount to that for conserved momentum \eqref{QbQ=} represents complex Grassmann algebra.

The counterpart of \eqref{v-iv-j=} in complex spinor frame variables reads
\bea\label{bv-iAv-jB=}
&& <{\rm i}\; ^-_A \; | \; {\rm j}\; ^{-}_B\;  \; >\; := \;
\bar{v}{}_{\alpha A {\rm i}}^{\; -}
\bar{v}{}^{-\alpha}_{B\;{\rm j}}\; = \; \frac 1 2 \; K^{= k}_{{\rm i}{\rm j}}\gamma^k_{AB}\; , \qquad
<{\rm i}\; ^-_A \; | \; {\rm j}\; ^{-B}\;  \; >\; := \;
\bar{v}{}_{\alpha A {\rm i}}^{\; -}
{v}{}^{-B\alpha}_{\;{\rm j}}\; = 0\; , \qquad
%\\ \nonumber
\\ \label{v-iAv-jB=}
&& <{\rm i}\; ^{-A} \; | \; {\rm j}\; ^{-}_B\;  \; >\; := \;
v_{\alpha  {\rm i}}^{-A}
\bar{v}{}^{-\alpha}_{B\;{\rm j}}\; = 0\; , \qquad
<{\rm i}\; ^{-A} \; | \; {\rm j}\;^{-B}\;  \; >\; := \;
{v}{}_{\alpha {\rm i}}^{-A}
{v}{}^{-B\alpha}_{\;{\rm j}}\;= \; \frac 1 2 \; K^{= k}_{{\rm i}{\rm j}}\tilde{\gamma}^{k\, BA}\; . \qquad
%\\ \nonumber
\eea

\subsection{Supersymmetric Ward identities and their solutions}

\label{sec=IIBsuperA}
As it was discussed in \cite{Boels:2012ie}, \cite{Wang:2015jna} and  \cite{Kallosh:2024lsl}, following the $4D$ approach of  \cite{Arkani-Hamed:2008gz,Elvang:2009wd}, the Ward identities for n--particle superamplitude of type IIB supergravity, ${\cal A}_n$,
\be\label{QAn=0}
{\bb Q}_{\alpha }\; {\cal A}_n =0\; , \qquad
\bar{{\bb Q}}_{\beta} \; {\cal A}_n =0\; , \qquad
\ee
for the case of $n\geq 4$ are solved by
\be\label{cAn=delta16Bn}
 {\cal A}_n = \delta^{16} ({\bb Q}_{\alpha })\, \delta^{10} ({\bb P}_{\mu})\,   \tilde{{\cal B}}_n \equiv  \delta^{16} ({Q}_{\alpha })\, \delta^{10} ({\bb P}_{\mu})\,    {\cal B}_n ,
\ee
where $ {\cal B}_n=   {\cal B}_n (\rho^\#_{{\rm i}}, v^-_{{\rm i}}, \Theta^-_{{\rm i}})$ obeys
\be\label{bQBn=0}
\bar{{\bb Q}}_{\alpha }\,    {\cal B}_n =0\; .
\ee
In \eqref{cAn=delta16Bn}  $ \delta^{16} ({\bb Q}_{\alpha })$ is Grassmann delta function,
\be
\delta^{16} ({\bb Q}) = ({\bb Q})^{16}= {\bb Q}_{1 }\ldots{\bb Q}_{16} = \frac 1 {16!} \epsilon^{\alpha_1\ldots \alpha_{16}} {\bb Q}_{\alpha_1 }\ldots{\bb Q}_{\alpha_{16} }\; .
\ee

Notice that the Ward identities \eqref{QAn=0} imply momentum preservation in the scattering process under consideration,
\bea\label{Pmu=0}
{\bb P}_\mu:= \sum\limits_{{\rm i}=1}^{n} p_{\mu {\rm i}}= \sum\limits_{{\rm i}=1}^{n} \rho^\#_{{\rm i}}u^=_{\mu {\rm i}}=0\; , \qquad
\eea
and this is reflected by including $\delta^{10} ({\bb P}_{\mu})$ into their
solution \eqref{cAn=delta16Bn}.

In \cite{Wang:2015jna} and  \cite{Kallosh:2024lsl}
three solutions of  \eqref{bQBn=0}, or better to say of \eqref{QAn=0}, were considered:
\begin{itemize}
\item The first, called 1/2 BPS in  \cite{Kallosh:2024lsl},  contains $\Theta^{-A}$ -- independent ${\cal B}_n=   {\cal B}_n (\rho^\#_{{\rm i}}, v^-_{{\rm i}})$.
\item The second is with supersymmetric conjugate amplitude
\be\label{cbAn=delta16}
\overline{ {\cal A}}_n = \, \delta^{10} ({\bb P}_{\mu})\, \bar{{\bb Q}}^{16}   \overline{\tilde{{\cal B}}}_n (\rho^\#_{{\rm i}}, v^-_{{\rm i}}, \partial /\partial \Theta^-_{{\rm i}}) \prod\limits_{{\rm i}=1}^n (\rho^\#_i)^4 (\Theta^-_{{\rm i}})^{\wedge 8}\; ,
\ee
where
\be (\Theta^-_{{\rm i}})^{\wedge 8}= \frac 1 8 \epsilon_{A_1\ldots A_8}\Theta^{-A_1}_{{\rm i}} \ldots \Theta^{-A_8}_{{\rm i}} \; . \ee
\item The third is with  ${\cal B}_n$ given by the image of maximal power of
${\bar{{\bb Q}}}_\alpha $ operator,
\be\label{cBn=QcCn}
{\cal B}_n = (\bar{{\bb Q}})^{16 }\, {\cal C}{}_n
\; .
\ee
\end{itemize}

The case of $n=3$   is special in several respects. In particular, the dimensional analysis excludes appearance  of a complete Grassmann delta functions in the three point superamplitude. The study of this case in the frame of spinor moving frame approach, besides being illuminating,  provides
us with a basis for further discussion of type IIA amplitudes.

\subsection{Three particle kinematics and three particle on-shell amplitudes of type IIB SUGRA}
\label{sec=IIBsuperA3}

Let us apply spinor moving frame formalism to the analysis of three particle kinematics.
Following \cite{Bandos:2017eof}, we find that the momentum conservation in scattering of three massless particles, Eq. \eqref{Pmu=0} with $n=3$,
 gives three equations two of which are solved by
\bea\label{r1=-r2-r3}
 \rho^\#_{{\rm 1}}=-(\rho^\#_{{\rm 2}}+\rho^\#_{{\rm 3}})\; ,& \qquad & \\ \label{K21=rrK31}
&K^{=i}_{{\rm 21}}:= K^{=i}_{{\rm 2}}-K^{=i}_{{\rm 1}}=
- \frac {\rho^\#_{{\rm 3}}}{\rho^\#_{{\rm 2}}}K^{=i}_{{\rm 31}}\; , & \qquad \eea and the third one
requires
\bea \label{vK--2=0} && (\vec{K}{}^{=}_{{\rm 31}})^2:= K^{=i}_{{\rm 31}}K^{=i}_{{\rm 31}}=0\; . \qquad
\eea

Clearly, \eqref{vK--2=0} has a nontrivial solution if
the $SO(8)$ vector $K^{=i}_{{\rm 31}}$ is complex. Then, if we exclude collinear scattering, this complex null-vector is non vanishing,
$K^{=i}_{{\rm 31}}\not=0$, and hence $|\vec{K}{}^{=}_{{\rm 31}}|^2:=K^{=i}_{{\rm 31}}(K^{=i}_{{\rm 31}})^*\not=0$.
With this restriction, the complex matrix $\hat{K}_{q\dot{p}}:= \frac 1 {|\vec{K}{}^{=}_{{\rm 31}}|}\, K^{=j}_{{\rm 31}}\gamma^j_{q\dot{p}}$ is  nilpotent $\hat{K}(\hat{K})^T =0$, but in
its product with its Hermitian conjugate factorizes a real projector,
\bea (\hat{K}\hat{K}^\dagger)_{qp} = 2{\cal P}_{qp}\, , \qquad  ((\hat{K})^\dagger\hat{K})_{\dot{q}\dot{p}} = \tilde{{\cal P}}_{\dot{q}\dot{p}} \; , \qquad
%\\ \nonumber
\\
{\cal P}{\cal P}={\cal P}\; , \qquad \tilde{{\cal P}}\tilde{{\cal P}}=\tilde{{\cal P}}\; .
\eea
This implies that matrix $\hat{K}_{q\dot{p}}$ has rank 4 {\it i.e.}
it can be factorized as

\be\label{hK=sbs}
\hat{K}_{q\dot{p}}:= \frac 1 {|\vec{K}{}^{=}_{{\rm 31}}|}\, K^{=j}_{{\rm 31}}\gamma^j_{q\dot{p}}= 2\beta_{{\rm 31}}{\rm s}_q{}^{\tilde{A}}\bar{\tilde{{\rm s}}}_{\dot{p}\tilde{A}}\; \qquad \beta_{{\rm 31}}=\pm= {\rm{sign}} (\rho^{\#}_{{\rm 3}}\rho^{\#}_{{\rm 1}})
\; , \qquad \tilde{A}=1,...,4\; , \qquad
\ee
where the role of sign multiplier
$ \beta_{{\rm 31}}:= {\rm{sign}} (\rho^{\#}_{{\rm 3}}\rho^{\#}_{{\rm 1}})$ will become clear latter and
$ {\rm s}_q{}^{\tilde{A}}$ and $\bar{{\rm s}}_{\dot{p}\tilde{A}}$ are complex $8\times 4$ matrices which obey
\bea\label{sbs=1}
{\rm s}_q{}^{\tilde{A}}{\rm s}_q{}^{\tilde{B}}=0\; , \qquad
{\rm s}_q{}^{\tilde{A}}
\bar{{\rm s}}_{q\tilde{B}} = \delta_{\tilde{B}}{}^{\tilde{A}}  \; , \qquad \bar{{\rm s}}_{q\tilde{A}}\bar{{\rm s}}_{q\tilde{B}} = 0  \; , \qquad \\
 \tilde{{\rm s}}_{\dot{p}}{}^{\tilde{A}} \tilde{{\rm s}}_{\dot{p}}{}^{\tilde{B}}=0\; , \qquad  \tilde{{\rm s}}_{\dot{p}}{}^{\tilde{B}} \bar{\tilde{{\rm s}}}_{\dot{p}\tilde{A}}=
\delta_{\tilde{A}}{}^{\tilde{B}}\; , \qquad
\bar{\tilde{{\rm s}}}_{\dot{p}\tilde{A}}\bar{\tilde{{\rm s}}}_{\dot{p}\tilde{B}}=0\; , \qquad \\ \nonumber
\tilde{A}, \tilde{B}=1,2,3,4\; , \qquad q,p=1,\ldots ,8 \; , \qquad \dot{q},\dot{p}=1,\ldots ,8\; .
\eea

At this stage it might seem that tilde on $ \tilde{{\rm s}}_{\dot{q}}{}^{\tilde{A}}=( \bar{\tilde{{\rm s}}}_{\dot{q}\tilde{A}})^*$ is redundant as its difference with $ {\rm s}$  is reflected by c-spinor index of $SO(8)$ {\it vs} s-spinor index of ${\rm s}_{{q}}{}^{\tilde{A}}=( \bar{{\rm s}}_{q\tilde{A}})^*$. However, this difference becomes useful when we convert $SO(8) $
indices into $SU(8)$ ones,
the same as  carried by components of the state vector superfield(s),
thus arriving at

\bea
{\rm s}_{B}{}^{\tilde{A}}=\bar{w}_{qB}{\rm s}_{{q}}{}^{\tilde{A}}=(\bar{{\rm s}}^{B}{}_{\tilde{A}})^*=
( w_{q}^B\bar{{\rm s}}_{q\tilde{A}})^*\; , \qquad
{\rm s}{}^{B\tilde{A}}=\bar{w}_{qB}{\rm s}_{{q}}{}^{\tilde{A}}=(\bar{{\rm s}}_{B\tilde{A}})^*=
(\bar{w}_{qB}\bar{{\rm s}}_{q\tilde{A}})^*\; , \qquad
\\
 \tilde{{\rm s}}_{B}{}^{\tilde{A}}=  \bar{w}_{\dot{q}B}
 \tilde{{\rm s}}_{\dot{q}}{}^{\tilde{A}}=( \bar{\tilde{{\rm s}}}{}^B_{\tilde{A}})^*=( w_{\dot{q}}{}^{B}\bar{\tilde{{\rm s}}}_{\dot{q}\tilde{A}})^*
 \; , \qquad
 \tilde{{\rm s}}{}^{B\tilde{A}}= w_{\dot{q}}{}^B\tilde{{\rm s}}_{\dot{q}}{}^{\tilde{A}}=( \bar{\tilde{{\rm s}}}_{B\tilde{A}})^*=( \bar{w}_{\dot{q}B}\bar{\tilde{{\rm s}}}_{\dot{q}\tilde{A}})^*\; .
\eea

Eqs. \eqref{sbs=1} mean that both of these  complex $8\times 4$  matrices, $ {\rm s}_{{q}}{}^{\tilde{A}}=( \bar{{\rm s}}_{q\tilde{A}})^*$ and $ \tilde{{\rm s}}_{\dot{q}}{}^{\tilde{A}}=( \bar{\tilde{{\rm s}}}_{\dot{q}\tilde{A}})^*$, actually parametrize SO(8) group. Furthermore, taking into account that factorization is invariant under $SU(4)=Spin(6)$ gauge transformations, they can be treated as a kind of homogeneous coordinates for $SO(8)/SU(4)\approx SO(8)/SO(6)$ coset. This is a reflection of the fact that the presence of complex null $8$--vector in our three-body (better three-wave) problem breaks the $SO(8)$ symmetry down to $SO(6)$.

Notice that at this point our moving frame based approach becomes different from that of \cite{Boels:2012ie}
in which a set of 3 different matrices with indices of small group $SO(8)$
and tiny group $SU(4)\approx SO(6)$ is introduced. This results in an ambiguity in an invariant tensor, which then has to be fixed.

The basic contractions of helicity and polarization spinors with $SO(8)$ indices are now expressed by universal formula

\bea\label{i-qj-dp=sts}
<{\rm i}^-_q\; |\; {\rm j}^-_{\dot{q}}> = - <{\rm j}^-_q\; |\; {\rm i}^-_{\dot{q}}> = \, \beta_{{\rm ij}}\, |\vec{K}{}_{{\rm i}{\rm j}}^=| \, {\rm s}_q{}^{\tilde{A}}\bar{\tilde{{\rm s}}}_{\dot{p}\tilde{A}}\; , \qquad {\rm i}\, < \, {\rm j}=1,2,3\; ,
\eea
where the last inequality is understood ${\rm mod}( 3)$, i.e. in the sense of cyclic order, with $3<1$.

Passing to variables with $SU(8)$ indices, we find
from \eqref{i-qj-dp=sts}
\bea\label{i-Bj-A=sts}
<{\rm i}^-_B\; |\; {\rm j}^-_{A}> = - <{\rm j}^-_B\; |\; {\rm i}^{-}_{A}> = \, \beta_{{\rm ij}}\, |\vec{K}{}_{{\rm i}{\rm j}}^=| \, {\rm s}_B{}^{\tilde{C}}\bar{\tilde{{\rm s}}}_{A\tilde{C}}\; , \qquad {\rm i}\, < \, {\rm j}=1,2,3\; ,
\eea

In our approach the complex invariant tensor 4-th rank antisymmetric spin--tensor with SO(8) c- and s-spinor indices are defined uniquely by

\bea\label{tsdq4=}
 \tilde{{\rm s}}_{\dot{q}_1\ldots \dot{q}_4} =
\,\frac 1 {4!}\, \epsilon_{\tilde{A}_1 \ldots \tilde{A}_4}   \tilde{{\rm s}}_{\dot{q}_1}{}^{\tilde{A}_1}\ldots  \tilde{{\rm s}}_{\dot{q}_4}{}^{\tilde{A}_4}= (\bar{\tilde{{\rm s}}}_{\dot{q}_1\ldots \dot{q}_4})^* =: \tilde{{\rm s}}{}^{A_1\ldots A_4} \bar{w}_{\dot{q}_1 A_1} \ldots  \bar{w}_{\dot{q}_4 A_4}\; ,  \qquad
%\\ \nonumber \text{ and }
\\ \label{sq4=}
{{\rm s}}_{{q}_1\ldots {q}_4} =
\,\frac 1 {4!}\, \epsilon_{\tilde{A}_1 \ldots \tilde{A}_4}   {{\rm s}}_{{q}_1}{}^{\tilde{A}_1}\ldots {{\rm s}}_{{q}_4}{}^{\tilde{A}_4}= (\bar{{{\rm s}}}_{{q}_1\ldots {q}_4})^*=:
{{\rm s}}{}^{A_1\ldots A_4} \bar{w}_{{q}_1 A_1} \ldots  \bar{w}_{{q}_4 A_4}\;  \; . \qquad
\eea
Notice that
\be \label{epsbs=1} \epsilon_{{q}_1 \ldots {q}_4{p}_1 \ldots {p}_4}
 {{{\rm s}}}{}_{{q}_1 \ldots {q}_4} \bar{{{\rm s}}}{}_{{p}_1 \ldots {p}_4}=1\; , \qquad \hat{K}_{q_1[\dot{q}_1|} \ldots \hat{K}_{q_4|\dot{q}_4]}= 2 \cdot 4!\,  {\rm s}_{q_1\ldots q_4} \bar{\tilde{{\rm s}}}_{\dot{q}_1 \ldots \dot{q}_4}\; . \qquad
\ee

Below, to relate amplitudes and superamplitudes, we will use the relation
\bea\label{KKtij==sqprs}
-\hat{K}^{i} K^{j}t^{ij}_{qprs}= \hat{K}_{[q|\dot{q}}\hat{K}_{|p|\dot{p}}
 {\gamma}{}^{i}_{|r|\dot{q}} {\gamma}{}^{i}_{|s]\dot{p}}
=8 \cdot 4!\,  {\rm s}_{qprs}\;
%\\  \nonumber
\eea
which was presented (in a slightly different notation) and was used essentially in \cite{Boels:2012ie}.
We refer to Appendix  \ref{app:3-point=IIB} for more discussion and for technical details.

The covariant spin--tensor \eqref{tsdq4=} can be used to construct the invariant expression for 12-th power of
supercharge \eqref{bbQ=},
\bea\label{bbQ12=}
{\bb Q}^{{12}} & := & \frac 1 { 12!}\, \epsilon^{\alpha_1\ldots \alpha_{16}} {\tilde{{\rm s}}}_{\dot{q}_1\ldots \dot{q}_4}\,
(\rho^{\#})^{-2} v^+_{\alpha_1\dot{q}_1}\ldots v^+_{\alpha_4\dot{q}_4}\,
{\bb Q}_{\alpha_5}\ldots {\bb Q}_{\alpha_{16}}
\; \qquad \nonumber
%\\ \nonumber
\\
& = & \frac 1 {12!}\, \epsilon^{\alpha_1\ldots \alpha_{16}} {\tilde{{\rm s}}}{}^{A_1\ldots A_4} \, (\rho^{\#})^{-2} v^+_{\alpha_1A_1}\ldots v^+_{\alpha_4A_4}\,  {\bb Q}_{\alpha_5}\ldots {\bb Q}_{\alpha_{16}}\;
%\qquad \nonumber \\ \nonumber \\ &=:& \, 2^{24}\,  Q^{12}
\; .  \qquad
\eea
Here we introduced a compensator $\rho^{\#}$ for the $SO(1,1)$ symmetry of the reference spinor moving frame used in
\eqref{v-i=v-+Kigv+} --
\eqref{u++=u0}.
This compensator can be identified with any of $\rho^{\#}_{{\rm i}}$, e.g. with  $\rho^{\#}_{{\rm 1}}$.

The invariant \eqref{bbQ12=} can be used to write the expression for 3-particle
(3-wave) superamplitude

\bea\label{cA3=Q12cB3}
{\cal A}_3 = {\bb Q}^{{12}} {\cal B}_3\; , \qquad
\bar{{\bb Q}}_\alpha {\cal B}_3=0\; .
\eea
To specify it we, following \cite{Boels:2012ie}, are going to use knowledge of one of particular basic 3-point amplitudes describing the process involving  two scalar fields and a graviton.
This amplitude in its turn can be constructed from the amplitude involving two scalars and one vector field using KLT (Kawai--Lewellen--Tye) relation \cite{Kawai:1985xq}.

The two-scalar one vector amplitude can be written in the form  (see e.g. \cite{Boels:2012ie} for $D=8$ case)

\be\label{Apcip=p1ui2}
A (\bar{\phi}c^i \phi)= p_{{\rm 1}\mu}u^{\mu i}_{{\rm 2}}
\ee
where $p_{{\rm 1}\mu} $ is the momentum of the first (scalar) particle and $u^{\mu i}_{{\rm 2}}$ is the polarization vector of
the
second, vector particle which can be identified with the ''orthogonal'' to momentum vectors of the associated moving frame.
This amplitude is relevant for the case of $10D$ SYM theory, and can be discussed in the context of type IIA theory, but in type IIB case its calculation is an intermediate stage.

In the spinor moving frame approach we can further
specify this amplitude as
\bea\label{Apcip==}
A (\bar{\phi}c^i \phi)= p_{{\rm 1}\mu}u^{\mu i}_{{\rm 2}}= \frac 1 4 \rho^{\#}_{{\rm 1}} <\, {\rm 2}^-_q\, |\, {\rm 1}^-_{\dot{p}}\, > \gamma^i_{q\dot{p}} =  -\frac 1 4 \rho^{\#}_{{\rm 1}}\, |\vec{K}^=_{{\rm 21}}|\, \beta_{{\rm 21}}{\rm s}_q{}^{\tilde{A}}\bar{\tilde{{\rm s}}}_{\dot{p}\tilde{A}}\gamma^i_{q\dot{p}}\;  \qquad
\eea
which implies
\bea\label{Apcqdpp==}
A (\bar{\phi}c^i \phi) \gamma^i_{q\dot{p}} &=&   -2 \rho^{\#}_{{\rm 1}}  <\, {\rm 1}^-_q\, |\, {\rm 2}^-_{\dot{p}}\, >
 = -2 \rho^{\#}_{{\rm 1}}\, |\vec{K}^=_{{\rm 21}}|\, \beta_{{\rm 21}}{\rm s}_q{}^{\tilde{A}}\bar{\tilde{{\rm s}}}_{\dot{p}\tilde{A}}\; . \qquad
\eea

Now, following  \cite{Kawai:1985xq},  we have to write the  amplitude for the process with  two scalars and one graviton as a kind of product of two \eqref{Apcip=p1ui2},
\bea\label{Aphijp==p1u2p3u2}
A (\bar{\phi}h^{ij} \phi)&=& p_{{\rm 1}\mu}u^{\mu i}_{{\rm 2}} p_{{\rm 3}\nu}u^{\nu i}_{{\rm 2}}
=\rho^{\#}_{{\rm 1}} \rho^{\#}_{{\rm 3}}
u^=_{{\rm 1}\mu}u^{\mu i}_{{\rm 2}}  u^=_{{\rm 3}\nu}u^{\nu j}_{{\rm 2}} =- (\rho^{\#}_{{\rm 1}})^2 K^{=i}_{{\rm 21}} K^{=j}_{{\rm 21}} \; .
%\\ \nonumber
\eea
One can immediately check that the {\it r.h.s.} of this expression is traceless on its $SO(8)$ v-indices, as the {\it l.h.s.} should be.

In our approach, the natural combination to be constructed from \eqref{Apcip=p1ui2} is
\bea\label{AphABCDp=}
A(\bar{\phi}h_{ABCD} \phi)&:=& A (\bar{\phi}h^{ij} \phi) t^{ij}_{ABCD} = p_{{\rm 1}\mu}u^{\mu i}_{{\rm 2}} p_{{\rm 3}\nu}u^{\nu j}_{{\rm 2}} t^{ij}_{ABCD} = - (\rho^{\#}_{{\rm 1}})^2 K^{=i}_{{\rm 21}} K^{=j}_{{\rm 21}}  t^{ij}_{ABCD}\; .
%\\ \nonumber
\eea
Notice that this is a ''self-dual'' or real part of the amplitude involving $\phi_{ABCD}$ component of the state vector superfield (which is a complex field  obeying \eqref{phi4=ep-phi4*}).
The 'anti-self dual' or imaginary  part of the amplitude $A(\bar{\phi}\phi_{ABCD} \phi)$, should be expressed through the amplitude of the  self-dual
field antisymmetric 4-rank $SO(8)$ tensor,
$A(\bar{\phi}c^{ijkl} \phi)$. However this vanishes just because it is impossible to reproduce its tensor  structure using two light-like momenta, $ p_{{\rm 1}\mu}$ and $ p_{{\rm 3}\mu}$ and polarization vector $u^{i\mu}_{{\rm 2}}$ only.

Thus, we conclude that
%% CORRECTIONS
\bea\label{App4p=}
A(\bar{\phi}^{[+8]}\phi^{[+4]}_{ABCD} \phi) = \, (\rho^\#_{\rm 1})^4\,
(\rho^\#_{\rm 2})^2\,
p_{{\rm 1}\mu}u^{\mu i}_{{\rm 2}} p_{{\rm 3}\nu}u^{\nu j}_{{\rm 2}} t^{ij}_{ABCD} =  - (\rho^{\#}_{{\rm 1}})^6 \, (\rho^\#_{\rm 2})^2\,   K^{=i}_{{\rm 21}} K^{=j}_{{\rm 21}}  t^{ij}_{ABCD}=  \qquad \nonumber \\  =\, 8\, 4!\,  (\rho^{\#}_{{\rm 1}})^6 \, (\rho^\#_{\rm 2})^2\,  |\vec{K}^=_{{\rm 12}}|^2 \, {{\rm s}}_{ABCD}\; , \qquad
%\\ \nonumber
\eea
where at the last stage Eq. \eqref{KKtij==sqprs} was used. Notice that in this equation we have indicated the SO(1,1) weights of the components of chiral superfield and  restored the multiplier $(\rho^\#_{\rm 2})^2$ which was omitted for simplicity in Eqs.
\eqref{phiqp=phiij}, \eqref{Hij:==}, \eqref{Cijkl:==} and  \eqref{phiq1-q4==} as well as in the above paragraph.  The  multiplier  $(\rho^\#_{\rm 1})^4$  relates the field $(\phi^*)=\bar{\phi}$ conjugate to complex axion--dilaton $\phi$ of type IIB supergrvaity, which we identify with leading component of the superfield \eqref{Phi=phi+}, with the highest component,  $\bar{\phi}^{[+8]}$, of this superfield.
The role of such  multipliers is  to balance the SO(1,1) weights of {\it l.h.s}-s and {\it r.h.s}-s in
\eqref{App4p=} and in some of the  equations below.

To find the coefficient ${\cal B}_3$, which in this case should be assumed to be $\Theta$--independent,
we have to calculate the same amplitude from superamplitude \eqref{cA3=Q12cB3} \footnote{In the notation similar to the one used for amplitudes in the {\it l.h.s.} of  Eq. \eqref{Aphijp==p1u2p3u2}, this can be written as
$ {\cal A}_3 (\Phi_{{\rm 1}}\Phi_{{\rm 2}}\Phi_{{\rm 3}})= {\bf Q}^{{12}} {\cal B}_3\; $,
where under  $\Phi_{{\rm i}}$ the superfields \eqref{Phi=phi+} are understood. The statements that superamplitudes depend on superfields which can be found in some articles (for instance in  \cite{Elvang:2013cua} arround Eq. (4.33) of this paper) should be understood in this sense and not as real dependence of on-shell superamplitudes on the generic superfields
$\Phi_{{\rm i}}= \Phi(\rho^\#_{{\rm i}}, v^-_{{\rm i}}, \Theta^-_{{\rm i}})$. We avoid such type statements as they might produce a confusion.  }.
This can be done by Grassmann integration
or equivalently by applying covariant derivatives and then setting  all the
$\Theta^-_{{\rm i}}$ equal to zero, i.e. from
\be\label{App4p=int}
A(\bar{\phi}\phi_{ABCD} \phi)=
\left(\int d^8\Theta^-_{{\rm 1}}(d^4\Theta^{-}_{{\rm 2}})_{ABCD} \, {\bb Q}^{{12}}\,  {\cal B}_3\right)\vert_{\Theta^-_{{\rm 2}}=0=\Theta^-_{{\rm 3}}}  \equiv
\left(((D^+_{{\rm 1}})^{\wedge 8} D^+_{A\,{\rm 2}}D^+_{B\,{\rm 2}}D^+_{C\,{\rm 2}}D^+_{D\,{\rm 2}} \, {\bb Q}^{{12}})\,  {\cal B}_3\right)\vert_{\Theta^-_{{\rm i}}=0}\; ,
\ee
where $(D^+_{{\rm 1}})^{\wedge 8}:= \frac 1 {8!} \epsilon^{A_1\ldots A_8}D^+_{{\rm 1}A_1}\ldots D^+_{{\rm 1}A_8} $.
% Here proportionality symbol  $\propto$ hides  multipliers
% compensating $SO(1,1)$ weight of the explicitly written expressions.

Grassmann integral in \eqref{App4p=int}  extracts from
$ {\bb Q}^{{12}}$ the coefficient for the terms with highest degree
of $\Theta^-_{{\rm 1}}$, fourth degree of $\Theta^-_{{\rm 2}}$ and  no $\Theta^-_{{\rm 3}}$. Separating the relevant term,
\be\label{Q12=4x8+}
{\bb Q}^{{12}}=\frac {1 } {4!\, 8!} {\cal B}_3 \tilde{s}{}^{\alpha_1\ldots  \alpha_{12}} {\bb Q}_{\alpha_1 {\rm 2}}\ldots {\bb Q}_{\alpha_4 {\rm 2}}\; {\bb Q}_{\alpha_5 {\rm 1}}\ldots  {\bb Q}_{\alpha_{12} {\rm 1}}+... \; , \qquad
\ee
where
\be\label{s1-12=}
\tilde{s}{}^{\alpha_1\ldots  \alpha_{12}} := \epsilon^{\alpha_1\ldots \alpha_{16}} \tilde{{\rm s}}_{\dot{q}_1\ldots \dot{q}_4} \,
(\rho^{\#})^{-2} v^+_{\alpha_{13}\dot{q}_1}\ldots v^+_{\alpha_{16}\dot{q}_4}= \epsilon^{\alpha_1\ldots \alpha_{16}} \tilde{{\rm s}}{}^{A_1\ldots A_4} \,
(\rho^{\#})^{-2} \bar{v}{}^+_{\alpha_{13}A_1}\ldots \bar{v}{}^+_{\alpha_{16}A_4}\, ,
\ee
we find that
\bea\label{D4Q12==}
\left((D^+_{{\rm 1}})^{\wedge 8}D^+_{{\rm 2}A}D^+_{{\rm 2}B}D^+_{{\rm 2}C}D^+_{{\rm 2}D} {\bb Q}^{12}\right)\vert_{\Theta^-_{{\rm i}}=0}
 =2^{24} \, 4!\, (\rho^{\#}_{{\rm 1}})^8  (\rho^{\#}_{{\rm 2}})^4 (\rho^{\#})^{-2} \;  |\vec{K}^=_{{\rm 21}}|^4\;  {\rm s}_{_{ABCD}}  \qquad \nonumber \\ =\frac {2^{21}(\rho^{\#}_{{\rm 1}})^8(\rho^{\#}_{{\rm 2}})^4 } { (\rho^{\#})^{2}}\,   |\vec{K}^=_{{\rm 21}}|^{-2}\;  A(\bar{\phi}^{[+8]}\phi_{_{ABCD}}^{[+4]} \phi)\; , \qquad
\eea
where at the last stage Eq \eqref{App4p=} was used.
See
Appendix \ref{app:3-point=IIB-2} for details on derivation of this result
(which is straightforward but requires some experience in spinor moving frame approach).

Comparing  \eqref{D4Q12==} with  \eqref{App4p=int}, we find
\be
{\cal B}_3= \frac {(\rho^{\#})^2}
{2^{21}(\rho^{\#}_{{\rm 1}})^2\,
(\rho^{\#}_{{\rm 2}})^2
\,
|\vec{K}^=_{{\rm 21}}|^2}
\ee
which results in the following expression for three point superamplitude
\be\label{cA3=Q12r:}
{\cal A}_3 = {\bb Q^{12}}\,(\rho^{\#})^2\,  \frac  {1} {2^{21} \, (\rho^{\#}_{{\rm 1}})^2  (\rho^{\#}_{{\rm 2}})^2   |\vec{K}^=_{{\rm 21}}|^2}\; .
\ee

At first glance, one can be surprised by the presence in the superamplitude of the multiplier $(\rho^{\#})^2$ which is  the Langrange multiplier for $SO(1,1)$ gauge symmetry of the reference moving frame and hence is clearly unphysical. However, a more careful look shows that this multiplier just cancels  $(\rho^{\#})^{-2}$
which was introduced in the definition  \eqref{bbQ12=} of ${\bb Q}^{12}$ to make this gauge invariant.
The quantity independent on unphysical $(\rho^{\#})^2$ has nonvanishing SO(1,1) weight and is given just by the product ${\bb Q^{12}}\,(\rho^{\#})^2\,$, i.e. by
\bea\label{bbQ12=+4}
{\bb Q}^{{12}\; [+4]} & = & \frac 1 {12!}\, \epsilon^{\alpha_1\ldots \alpha_{16}} {\tilde{{\rm s}}}{}^{A_1\ldots A_4} \,  v^+_{\alpha_1A_1}\ldots v^+_{\alpha_4A_4}\,  {\bb Q}_{\alpha_5}\ldots {\bb Q}_{\alpha_{16}}\;
%\qquad \nonumber \\ \nonumber \\ &=:& \, 2^{24}\,  Q^{12}
\; .   \qquad
\eea
In terms of this covariant expression, the 3--point superamplitude \eqref{cA3=Q12r:} reads
\be\label{cA3=}
{\cal A}_3 = {\bb Q^{12\; [+4]}}\,\frac  {1} {2^{21} \, (\rho^{\#}_{{\rm 1}})^2  (\rho^{\#}_{{\rm 2}})^2   |\vec{K}^=_{{\rm 21}}|^2}\; .
\ee
Of course, the numerical coefficient $2^{-21}$ in this expression is purely conventional: it comes from the coefficient 2 in the r.h.s. of supersymmetry algebra
\eqref{QbQ=} which  results in coefficient 4 in the representation \eqref{bbQ=} for supersymmetry generators.

%\bigskip

%\subsection{On four particle superamplitude  of type IIB supergravity}

%******

\subsection{On on-shell amplitudes of type IIA supergravity}
\label{sec=AmplIIA}

The fact that the state vector of type IIA superparticle, when quantized in our formalism, is described by the same analytic on-shell  superfield as the type IIB one, suggests that  the superamplitudes discussed in secs. \ref{sec=IIBsuperA} and  \ref{sec=IIBsuperA3} also describe the processes of scattering of type IIA supergravitons, i.e. of the members of type IIA supergravity multiplet.
In  this section we will discuss the restrictions on such a description.

To begin with, let us notice that, when the unified description of type IIB and type IIA processes is valid, the
spacetime interpretation of type IIA amplitudes is a bit more convoluted and involves the covariantly constant $SO(8)$ vector $k^i$ \eqref{Dki=0} related to the complex St\"uckelberg fields defining the complex structure by \eqref{wAc=wAsk} and \eqref{kg=wbw}.

In particular the amplitude extracted from  superamplitude \eqref{cA3=Q12cB3} as shown in  \eqref{App4p=int}, describes  processes involving, besides a graviton $h^{ij}$ and a dilaton $\phi^{IIA}$, also the a solution for $1$-form field of type IIA supermultiplet in contraction with this $8$-vector, $c^ik^i$ (see \eqref{phiIIB=phi+ck}).
The part of this one-form, orthogonal to $k^i$, namely the combination $k^{[i}c^{j]}$,  contributes, together with the projection of the type IIA $3$-form on $k^i$,
into the amplitude (see \eqref{phiAB=IIA})
\bea\label{App2p=int=IIA}
\left(\int d^8\Theta^-_{{\rm 1}}(d^2\Theta^{-}_{{\rm 2}})_{AB} \, {Q}^{{12}}\,  {\cal B}_3\right)\vert_{\Theta^-_{{\rm 2}}=0=\Theta^-_{{\rm 3}}} =
\left(((D^+_{{\rm 1}})^{\wedge 8} D^+_{A\,{\rm 2}}D^+_{B\,{\rm 2}} \, {Q}^{{12}})\,  {\cal B}_3\right)\vert_{\Theta^-_{{\rm i}}=0}\; .
\eea

Notice that above we have not attributed
$8$-vector $k^i$ to a definite (super)particle
(i.e. did not supply it with some subindex ${\rm i}$, e.g. ${\rm 2}$) and this  requires some explanation.
What is manifest (in particular from the discussion of supercharges below) is that for each particle  ($\forall \; {\rm i}$) we have to use the same spacelike unit 10-vector  $k_\mu$.
All  momenta are orthogonal to the
T-duality direction  defined by this vector,
$k_\mu p^\mu_{{\rm i}}=0$, the fact which already puts a restrictions on the  processes with 8 or more particles which can be described within our present formalism.

Our quantization of type IIA superparticle indicates that, when scattering process of type IIA supergavitons
is considered, the spacelike unit vector $k_\mu$ has similar decomposition on each moving frame:
\be\label{k=kijuij}
k_\mu = -k^j_{{\rm i}}u_{\mu {\rm i}}^j \; .
\ee
This decomposition apparently involves different unit eight-vector vectors $ k^j_{{\rm i}}$ which are transformed by different  $SO(8)$ groups: by $SO(8)_{{\rm i}}$ acting on spinor moving frame of
${\rm i}$-th particle; these $ k^j_{{\rm i}}$ are covariantly constant with respect to this group,
$dk^j_{{\rm i}}+k^k_{{\rm i}}\Omega^{kj}_{{\rm i}}=0$ (see Appendix \eqref{App:kij=kj} for relation between Cartan forms of different particles).

However,  in the gauge \eqref{u--=Kiui}--\eqref{u++=u0} it can be shown  (see Appendix \ref{App:kij=kj}) that actually
all $k^j_{{\rm i}}= k_\mu  u^{\mu j}_{{\rm i}}$ coincide with $k^j= k_\mu u^{\mu j}$, so that
\be\label{k=kjuj}
k_\mu = -k^ju_{\mu }^j \; = \; -k^ju_{\mu {\rm i}}^j  \qquad \forall {\rm i}\; .
\ee

Also the form of complex supercharges used to construct superamplitude in type IIA case requires some comments.
From the definition of the complex coordinate of the analytical basis \eqref{Th-A==IIA}, we find that the rigid supersymmetry transformations act on that as
\bea\label{susyTh-A=IIA}
\delta_{susy}\Theta^{- A}= \frac 1 {\sqrt{2}} {w}_q^A (\epsilon^{1\alpha}v^{\;-}_{\alpha q}+ik\!\!\!/{}_{q\dot{p}}  v^{-\alpha}_{\dot{q}}\epsilon^{2-}_{\alpha})
= \frac 1 {\sqrt{2}}   (\epsilon^{1\alpha}v^{\;-}_{\alpha q}{w}_q^A+i\epsilon^{2-}_{\alpha}v^{-\alpha}_{\dot{q}}{w}_{\dot{q}}^A )\; . \qquad
%\\ \nonumber
\eea
This determines the corresponding derivative terms in the
representation of supersymmetry generators defined by
$\delta_{susy}= \epsilon^{1\alpha}{\bb Q}^1_{\alpha}+\epsilon^{2-}_{\alpha} {\bb Q}^{2\alpha}$ to be
$
{\bb Q}^1_{\alpha}= v^{\;-}_{\alpha q}{w}_{{q}}^A\; \frac {\partial} {\partial\Theta^{-A}}+... $ and $
 {\bb Q}^{2\alpha}= iv^{-\alpha}_{\dot{q}}{w}_{\dot{q}}^A\; \frac {\partial} {\partial\Theta^{-A}}+... $.
The remaining terms linear in $\Theta^{-A}$, can be determined
from type IIA supersymmetry algebra
\bea\label{Q1Q1=P=IIA}
{}\{ {\bb Q}^1_{\alpha}, {\bb Q}^1_{\beta}\}\; &=&2p_\mu\sigma^\mu_{\alpha\beta} =\; 4 \rho^\#  v^{\;-}_{\alpha q}v^{\;-}_{\beta q} =\; 4 \rho^\#  v^{\;-A}_{\alpha }\bar{v}{}^{\;-}_{\beta A } \; , \qquad \\ {}\{ {\bb Q}^1_{\alpha}, {\bb Q}^{2\beta}\}\; &=&\; 0\; , \qquad  \\
\label{Q2Q2=P=IIA}
{}\{ {\bb Q}^{2\alpha}, {\bb Q}^{2\beta}\}&=&2p_\mu\tilde{\sigma}^{\mu\,, \alpha\beta} =4 \rho^\#  v^{-\alpha }_{\dot{q}}v^{-\beta }_{\dot{q}} = 4 \rho^\#  v^{-\alpha  A}\, \bar{v}_A^{-\beta }\; , \qquad
\eea
where at the last stages we have used \eqref{w=inSU8}. In such a way we find
\bea\label{Q1=v-}
{\bb Q}^1_{\alpha}= \; v^{\;-}_{\alpha q} {\bb Q}^1_{q}= \;v^{\;-}_{\alpha q}\left( {w}_{{q}}^A\; \frac {\partial} {\partial\Theta^{-A}}+4\rho^\# \bar{w}_{qA} \Theta^{-A}\right) \; ,
\qquad
%\\ \nonumber
\\
\label{Q2=v-}
{\bb Q}^{2\alpha}= v^{-\alpha}_{\dot{q}} {\bb Q}^{2}_{\dot{q}} =  iv^{-\alpha}_{\dot{q}}\left({w}_{\dot{q}}^A\; \frac {\partial} {\partial\Theta^{-A}}-4\rho^\# \bar{w}_{\dot{q} A} \Theta^{-A} \right)  \; . \qquad
\eea

Now let us contract $ {\bb Q}^{2\alpha}$ with
\be\label{ks=kiuis}
k\!\!\!/{}_{\alpha\beta} = k^i u_\mu^i \sigma^\mu_{\alpha\beta} = k^i\gamma^i_{q\dot{p}} (v^-_{\alpha q } v^+_{\beta \dot{p}}+v^+_{\alpha \dot{p}} v^-_{\beta q})\; ,
\ee
where $k^i$ is the unit covariantly constant eight-vector \eqref{ki} used to define the complex structure in type IIA superparticle model through  Eqs.  \eqref{wAc=wAsk}, \eqref{kg=wbw} and \eqref{Th-A=IIA}. This gives
\be
k\!\!\!/{}_{\alpha\beta}{\bb Q}^{2\beta}=iv_{\alpha q}^-  {\bb Q}^{2}_{{q}} = v_{\alpha q}^- \,i\,  \left({w}_{{q}}^A\; \frac {\partial} {\partial\Theta^{-A}}-4\rho^\# \bar{w}_{{q} A} \Theta^{-A} \right) \; ,
\ee
so that the complex generators
\bea\label{bbbQ=IIA}
\bar{{\bb Q}}_{\alpha {\rm i}}&=&\frac 1 {\sqrt{2}} ({\bb Q}_{\alpha {\rm i}}^1-ik\!\!\!/{}_{\alpha\beta}{\bb Q}_{\rm i}^{2\beta}) = v_{\alpha q\, {\rm i}}^{\; -} w_{q {\rm i}}^A \frac {\partial} {\partial \Theta_{{\rm i}}^{-A}}= v_{\alpha {\rm i}}^{- A} \frac {\partial} {\partial \Theta_{{\rm i}}^{-A}} \; , \qquad    \\ \label{bbQ=IIA}
{\bb Q}_{\alpha {\rm i}}&=&\frac 1 {\sqrt{2}}
({\bb Q}_{\alpha {\rm i}}^1+ik\!\!\!/{}_{\alpha\beta}{\bb Q}_{\rm i}^{2\beta}) = 4\rho^{\#}_{ {\rm i}}v_{\alpha q\, {\rm i}}^{\; -} \bar{w}_{qA {\rm i}}  \Theta_{{\rm i}}^{-A}= 4\rho^{\#}_{ {\rm i}}\bar{v}_{\alpha A {\rm i}}^{\; - }  \Theta_{{\rm i}}^{-A}
%\\ \nonumber \\ \label{bbQi=-4iQiIIA} &=:& 4 {Q}_{\alpha {\rm i}}\;   \qquad
\eea
have exactly the same representation in terms of complex coordinates $\Theta^{-A}$ as those for type IIB case.

Notice that in this last equation we have restored the subindex ${\rm i}$ enumerating points (superparticles) in the scattering process. However, as we have discussed above, for all ${\rm i}$  we can use the same covariantly constant $8$-vector $k^i$ after the gauge \eqref{u--=Kiui}--\eqref{u++=u0}. Then
$w_{q {\rm i}}^A $ with different values of
${\rm i}$ can differ by $SU(8)$ transformations only. Furthermore, we can fix the gauge when all these coincide, $w_{q {\rm i}}^A =w_{q}^A $ and this gauge is preserved by a single $SU(8)$ symmetry acting on the reference bridge variable $w_{q}^A $.

It is also important to notice that in the gauge \eqref{u--=Kiui}--\eqref{u++=u0}  we can also find
\be\label{kKii'=0}
\vec{k} {\vec{{\bb K}}}_{{\rm i}'{\rm i}}:=\vec{k} ({\vec{{\bb K}}}_{{\rm i}'}-{\vec{{\bb K}}}_{{\rm i}})=0\qquad \forall \; {\rm i}', {\rm i}'\; ,
\ee
which follows from
\be\label{kK--=-k--}
\vec{k} {\vec{{\bb K}}}_{{\rm i}}=-k^{=}:=-k_\mu u^{\mu =}\; . \ee
If in our reference frame $k^{=}:=k_\mu u^{\mu =}=0$, we have even more illustrative relations
\be\label{kKi=0}
\vec{k} \vec{{\bb K}}_{{\rm i}}=0 \qquad \forall\; {\rm i}\; .
\ee
This makes clear that in the case of process
which contains
seven or less of type IIA supergravitons, we can always choose the T-duality direction in such a way that the condition \eqref{kKi=0} holds.
Thus, the use of a special T-duality direction to define complex structure can put restrictions on  processes with more than seven type IIA supergravitons which can be described by our formalism.

This is clearly better than the restriction put by using for $10$ or $11$ dimensional processes the D=4 formalism which restricts the possibility of generic studies  to processes with four or less particles (see e.g. \cite{Herderschee:2019ofc}).

Thus, at least the amplitudes for processed with seven or less type IIA supergravitons can be read off  the superamplitudes found for its type IIB cousin with the above correspondence originated in T-duality. Notice that at the level of linearized supergravity the necessity of compactness of the T-duality direction is not manifest.

\subsection{Towards   on-shell amplitudes of type IIA string theory involving D$0$ branes }

In type IIA string theory, besides the scattering  of
supergravitons, there are also  processes involving supergravity multiplets and D$0$--branes, which are massive superparticles in type IIA superspace. An interesting problem is then what are {\it superamplitudes} for such processes.  The issue  is not trivial: to our best knowledge, even in $D=4$ its counterpart was addressed only recently and only for ${\cal N}=1$ case \cite{Herderschee:2019ofc,Delgado:2025oev}.
Below we will discuss some first stages towards such  superamplitudes in $D=10$ type IIA string theory and indicate an issue hampering the straightforward generalization of the known ways to solve the Ward identities in this case.

\subsubsection{Spinor helicity variables,  on-shell superfield and quantum state spectrum of D$0$ brane }

The quantization of spinor moving frame formulation of D$0$--brane was elaborated recently in \cite{Bandos:2025pxv} and resulted in a state vector described by on-shell superfield quite similar to \eqref{Phi=phi+}:

\bea\label{D0:Phi=phi+=D0}
\Xi_{_{{\text{D0}}}}  (x^0, \Theta^A, \bar{\Theta}_A; v, w) &=& e^{-i{\rm m}(x^0+i\Theta^A\bar{\Theta}_A)}   \left(  \phi_{{\rm m}}+ \Theta^A \psi_{A{\rm m}} + \frac 1 2 \Theta^B\Theta^A \phi_{AB {\rm m}} +
\frac 1 {3!} \Theta^C \Theta^B \Theta^A \psi_{ABC  {\rm m}} + \right. \qquad  \nonumber
%\\  \nonumber
\\   && + \frac 1 {4!} \Theta^D  \Theta^C \Theta^B \Theta^A \psi_{ABCD  {\rm m}} +  \nonumber
%\\ \nonumber
\\
&& +\frac 1 {5!\, 3!} \epsilon_{A_1\ldots A_5B_1B_2B_3} \Theta^{A_5}  \ldots \Theta^{A_1} \tilde{\psi}{}^{B_1B_2B_3}_{{\rm m}} +   \frac 1 {6!\, 2!} \epsilon_{A_1\ldots A_4B_1B_2} \Theta^{A_6}  \ldots \Theta^{A_1} \tilde{\phi}{}^{B_1B_2  {\rm m}}_{{\rm m}} + \qquad   \nonumber
%\\ \nonumber
\\
&& \left.  + \frac 1 {7!} \epsilon_{A_1\ldots A_7B} \Theta^{A_7}  \ldots \Theta^{A_1} \tilde{\psi}{}_{ {\rm m}}^{B} + \frac 1 {8!} \epsilon_{A_1\ldots A_8} \Theta^{A_8}  \ldots \Theta^{A_1} \tilde{\phi}_{ {\rm m}} \right) \;   \qquad
%\\ \nonumber
\eea
obeying, besides
\bea \label{D0:bDAXi=0}
\bar{\text{D}}{}^A \Xi_{_{{\text{D0}}}} =0\; ,  \qquad
%\ee with \bea\label{D0:bDA=}
 \bar{\text{D}}{}^A = \bar{\partial}{}^A + i \Theta^A\partial_{x^0}= \bar{\partial}{}^A \,  e^{- i \Theta^A\bar{\Theta}_A\partial_{x^0}} = -({\text{D}}{}_A)^*  \; , \qquad   \eea
also
\bea\label{D=:D4Xi=ebD4bXi}
 \text{D}_{A_1}\ldots \text{D}_{A_4}\Xi_{_{{\text{D0}}}}  =\frac 1 {4!} \epsilon_{A_1\ldots A_4B_1B_2B_3B_4} \bar{\text{D}}{}^{B_1}\ldots \bar{\text{D}}{}^{B_4}\bar{\Xi}_{_{{\text{D0}}}} \; , \qquad  (\Xi_{_{{\text{D0}}}})^* =\bar{\Xi}_{_{{\text{D0}}}}\; .
 %\\ \nonumber
\eea
with $ (\Xi_{_{{\text{D0}}}})^* =\bar{\Xi}_{_{{\text{D0}}}}$ and
\bea \label{D0:DA=}
{\text{D}}_A &=&  {\partial}_A + i \bar{\Theta}_A\partial_{x^0}= {\partial}_A \,  e^{i \Theta^A\bar{\Theta}_A\partial_{x^0}} =- ( \bar{\text{D}}{}^A)^* \; . \qquad
%\\   \nonumber
\eea
In \eqref{D0:Phi=phi+=D0} ${\rm m}$ in the exponent is the mass of D$0$--brane. We also use subindex  ${\rm m}$ to distinguish the component of state vector superfield of D$0$-brane from that of the state vector of IIA supergraviton.

Like in the case of the linearized supergravity discussed above, the complex on-shell superfield essentially depends on
octuplet of complex fermionic  coordinates $\Theta^A$, but not on its c.c. This fermionic coordinate carries the index of fundamental representation of ·$SU(8)$ which was shown to be a hidden symmetry also of the quantum theory of D$0$ brane
\cite{Bandos:2025pxv}.  The complex fermionic coordinate function of D$0$ brane is constructed from the standard real fermionic coordinates $ \theta^{\alpha 1}$ and $\theta_{\alpha}^{2}$ of type IIA superspace by
\bea\label{D0:ThetaA:=}
&& \Theta^A= \Theta^{\underline{q}}w_{\underline{q}}^A=
\frac 1 {\sqrt{2}}( \theta^{\alpha 1}{{\rm v}}_\alpha{}^{\underline{q}} +\theta_{\alpha}^{2}{{\rm v}}_{\underline{q}}{}^\alpha) {{\rm w}}_{\underline{q}}^A\;
=(\bar{\Theta}_A)^* \;  \eea
using spinor moving frame variable of D$0$--brane
\be\label{D0:v=inSpin}
{{\rm v}}_\alpha{}^{\underline{q}} \in \text{Spin}(1,9) \; ,  \qquad \alpha =1,\ldots, 16 \; ,  \qquad {\underline{q}}=1,\ldots, 16
\ee
 and the suitable bridge variables given by
 complex 16$\times 8$ matrix ${{\rm w}}_{\underline{q}}^A = (\bar{{\rm w}}_{\underline{q}A} )^*$ obeying

\begin{eqnarray}
\label{D0:wbw=I}
&& \bar{{{\rm w}}}_{\underline{q}A}\bar{{{\rm w}}}_{\underline{q}B}=0 ={{\rm w}}_{\underline{q}}^A {{\rm w}}_{\underline{q}}^B\; , \qquad \bar{{\rm w}}_{\underline{q}A} {{\rm w}}_{\underline{q}}^B=\delta_A{}^B\; , \qquad \bar{{{\rm w}}}_{\underline{q}A} {{\rm w}}_{\underline{p}}^A +  {{\rm w}}_{\underline{q}}^A \bar{{{\rm w}}}_{\underline{p}A} = \delta_{qp}  \; .  \qquad
\end{eqnarray}
These relations imply that $({\rm w}_{\underline{q}}^A \, ,\, \bar{{\rm w}}_{\underline{q}A} )$ variables parametrize the $\text{SO}(16)$ valued matrix
\be\label{D0:w+bw=inSO16}
\left(\begin{matrix}\Re{\rm e}{{\rm w}}_{\underline{q}}^A \; \vert \; \Im{\rm m}{{\rm w}}_{\underline{q}}^A \;  \end{matrix}\right)  \in \text{SO}(16)\;
\ee

The moving frame vectors of the D0--brane
\be\label{u0uI=inSO}
\left({{\rm u}}_\mu^{0}, {{\rm u}}_\mu^{I}\right) \; \in \; SO(1,9)\qquad \Rightarrow \qquad  {{\rm u}}_\mu^{0} {{\rm u}}^{\mu (0)}=1\; , \qquad  {{\rm u}}_\mu^{0} {{\rm u}}^{\mu I}=0\; , \qquad {{\rm u}}_\mu^{I}{{\rm u}}^{\mu J}= \delta^{IJ}\; , \qquad
\ee
are constructed from the spinor moving frame matrix \eqref{D0:v=inSpin} as indicated by the constraints \cite{Bandos:2000tg}

\begin{eqnarray}\label{D0:u0s=vv}
&&{{\rm u}}_\mu^{0} \sigma^\mu_{\alpha\beta}={{\rm v}}_\alpha{}^{\underline{q}} {{\rm v}}_\beta{}^{\underline{q}} \; ,  \qquad
% \nonumber \\ \nonumber
\\
\label{D0:uIs=vgv} && {{\rm u}}_\mu^{I} \sigma^\mu_{\alpha\beta}={{\rm v}}_\alpha{}^{\underline{q}} \gamma^I_{\underline{q}\underline{p}}{{\rm v}}_\beta{}^{\underline{p}} \; , \qquad
{{\rm v}}_{\alpha}{}^{\underline{q}} \tilde{\sigma}{}_{\mu}^{\alpha\beta}{{\rm v}}_{\beta}{}^{\underline{q}}= {{\rm u}}_\mu^{0} \delta_{\underline{q}\underline{p}}+{{\rm u}}_\mu^{I} \gamma^I_{\underline{q}\underline{p}}\; ,  \qquad I=1,\ldots,9\; . \qquad
%\\ \nonumber
\end{eqnarray}
Here   $\gamma^I_{\underline{q}\underline{p}}$ are 9d Dirac matrices which are symmetric, traceless and obey the Clifford algebra,

\begin{eqnarray}\label{D0:gIgJ=dIJ}
\gamma^I_{\underline{q}\underline{p}} =\gamma^I_{\underline{p}\underline{q}} \; ,  \qquad \gamma^I_{\underline{p}\underline{p}}=0 \; ,  \qquad \gamma^I\gamma^J+\gamma^J\gamma^I=2\delta^{IJ} {\mathbb I}_{16\times 16} \; .  \qquad
\end{eqnarray}

The timelike momentum of the D0--brane is proportional to
${{\rm u}}_\mu^{0}$,

\be\label{pD0=mu0} p_\mu^{{D0}} ={\rm m} {{\rm u}}_\mu^{0}\,  \qquad \Rightarrow \qquad p_\mu^{{D0}}\sigma^\mu_{\alpha\beta} = {\rm m} {\rm v}_\alpha{}^{\underline{q}} {\rm v}_\beta{}^{\underline{q}}
 \ee

While the first of the constraints \eqref{D0:u0s=vv} is invariant under $SO(16)$ symmetry acting on $\underline{q}, \underline{p}$ indices, the remaining constraints \eqref{D0:uIs=vgv} break this down to $Spin(9)\subset SO(16)$ which acts also on
$({w}_{\underline{q}}^{\, A}, \bar{w}_{\underline{q}A})$ variables. These
provide a bridge between $Spin(9)$ and $SU(8)$ groups and can be called internal harmonic variables. The discussion on the degrees of freedom of the superparticle mechanics described in terms of all these variables can be found in \cite{Bandos:2025pxv}.

Like in the case of type II massless superparticles, it is useful to construct the complex version of spinor frame variables/helicity spinors by contraction of spinor frame matrix with internal harmonics,

\be\label{D0:vA=vwA}
{\rm v}_{\alpha}^A = {\rm v}_{\alpha}{}^{\underline{q}} {\rm w}_{\underline{q}}^A\; , \qquad  {\bar{\rm v}}_{\alpha A} = {\rm v}_{\alpha}{}^{\underline{q}} \bar{{\rm w}}_{\underline{q} A}\; . \qquad
\ee

In the case of multiparticle amplitude involving, beside the massive, also massless particles, the above (spinor) moving frame  variables of D$0$--brane can be decomposed on the same reference (spinor) moving frame as the (spinor) moving frame of massless superparicles are decomposed. In this case, in addition to Eqs. \eqref{v-i=v-+Kigv+}, \eqref{v-1+i=v-Kigv},
%and \eqref{u--=Kiui}, \eqref{ui=Kiu++}, \eqref{u++=u0},
we can consider

\begin{eqnarray}\label{v=v-l++v+l-}
{\rm v}_{\alpha}{}^{\underline{q}} &=&
v_{\alpha q}^{\; -} l_q^{+\underline{q}} + v_{\alpha  \dot{p}}^{\; +} l_{\dot{p}}^{-\underline{q}}
\; . \qquad \end{eqnarray}
However, we have not found such a decomposition particularly useful. For the case of processes involving more than one massless superparticles, it turns out
to be more practical to use, following the $D=4$ approach in \cite{Arkani-Hamed:2017jhn}, the decomposition on helicity spinors of two of these, assuming that their momenta are not collinear,
\begin{eqnarray}\label{v=v1l1+v2l2}
{\rm v}_{\alpha}{}^{\underline{q}} &=&
v_{\alpha q{\rm 1}}^{\; -} l_{q {\rm 1} }^{+\underline{q}} + v_{\alpha  {q}{\rm 2}}^{\; -}
l_{{q}{\rm 2}}^{+\underline{q}}
\; . \qquad
\end{eqnarray}
We will use such a decomposition below in our discussion on  three point amplitude which involves
one  D$0$-brane.

The quantum state spectrum of D$0$--brane was shown to be the massive counterpart of type IIA supergravity multiplet \cite{Bandos:2025pxv} formed by a massive graviton, massive gravitino and massive 3-form fields. The solution of (the Fourier image of) the linearized equations can be written in terms of the moving frame and spinor moving frame variables
described above by
 \cite{Bandos:2025pxv}
\bea\label{hmn=hIJ}
h_{\mu\nu} = {\rm u}_{\mu}^I  {\rm u}_{\nu}^J h^{IJ}\; , \qquad h^{IJ} =h^{JI}=h^{(IJ)}\; , \qquad h^{II}=0\; , \qquad
%\\ \nonumber
\\
A_{\mu\nu\rho}= {\rm u}_{\mu}^I  {\rm u}_{\nu}^J {\rm u}_{\rho}^K A^{IJK}\; , \qquad A^{IJK}=A^{[IJK]}\; ,\qquad
%\\ \nonumber
\\
\Psi_{\mu}^{\alpha 1}={\rm v}_{\underline{q}} ^{\alpha }\Psi^I_{\underline{q}}\; ,
\qquad \Psi_{\mu\alpha}^{\; 2}={\rm v}^{\underline{q}} _{\alpha }\Psi^I_{\underline{q}}\; ,
\qquad  \gamma^{I}_{qp}\Psi^I_p=0\; , \qquad
%\\ \nonumber
\eea
where $p_\mu= {\rm m}{\rm u}_{\mu}^{{\rm 0}}$ is assumed.

The expressions for bosonic $h^{IJ}$, $A^{IJK}$ and fermionic $\Psi^I_{\underline{q}}$ in terms of the components of the on-shell superfield can be found in  \cite{Bandos:2025pxv}.
One can also express the components of the on-shell superfield in terms of $h^{IJ}$, $A^{IJK}$ and $\Psi^I_{\underline{q}}$. In particular one can find
\footnote{The relative coefficients for the terms in this expression are not essential for our discussion below and are set to unity for simplicity. This is to say, we did not try to fix them by correspondence with 'inverse' expressions in   \cite{Bandos:2025pxv} which themselves are indicative showing the correspondence between degrees of freedom. }
\bea\label{phimABCD=}
\phi_{ABCD\; {\rm m}} = h^{IJ} \left(U_I^{\check{I}} U_J^{\check{J}}
t^{\check{I}\check{J}}_{ABCD}+ U_J \bar{U}_J \sigma^{\check{I}}{}_{[AB}\sigma^{\check{J}}{}_{CD]}
\right) +  A^{IJK}\left(U_I^{\check{I}} U_J \bar{U}_K\;
it^{\check{I}}_{ABCD}+  U_I^{\check{I}} U_J^{\check{J}}  U_K^{\check{K}}\;
it^{\check{I}\check{J}\check{K}}_{ABCD}\right) \; , \qquad
%\\ \nonumber
\eea
where
$U_I{}^{\check{K}}$ are six real, mutually orthogonal unit vectors vectors  $U_I=(\bar{U}_I)^*$ is complex null vector orthogonal to the above vectors and  normalized by
$U_I\bar{U}_I=2$. Resuming;
 \begin{eqnarray}\label{UU=0} && U_IU_I=0=  \bar{U}_I \bar{U}_I \; , \qquad U_I \bar{U}_I=2 \; , \qquad U_IU_I{}^{\check{J}} =0=  \bar{U}_I U_I{}^{\check{J}} \; , \qquad  U_I{}^{\check{J}}U_I{}^{\check{K}}  =\delta^{\check{J}\check{K}}\;  \qquad
 %\\ \nonumber
 \end{eqnarray}
which implies that these vectors  form the $SO(9)$ valued ''internal frame matrix''

 \begin{eqnarray}
\label{UinSO9} &&
  U_I^{(J)}= \left(U_I{}^{\check{J}}, U_I{}^{(8)}, U_I{}^{(9)}\right)= \left(U_I{}^{\check{J}}, \frac 1 2 \left( U_I+ \bar{U}_I\right), \frac 1 {2i} \left( U_I- \bar{U}_I \right)\right) \; \in \; SO(9)
 \; . \qquad
 %\\ \nonumber
 \end{eqnarray}

Following \cite{Bandos:2017eof}, we can define the square  root of the above spinor frame variables as $Spin(9)$ valued matrix constrained by the condition of the invariance of  the  generalized $SO(9)$ Pauli matrices $\gamma^I_{pq}$. Actually such variables are provided by the above introduced  ${\rm w}_{\underline{q}}^A$ and $\bar{{\rm w}}_{\underline{q}A}=({\rm w}_{\underline{q}}^A)^*$ after we impose on them, besides \eqref{D0:wbw=I} also the above mentioned constraints.
We however prefer to impose on them a bit more general constraints
\begin{eqnarray}\label{bUg=}
&&  U\!\!\!\!/{}_{\underline{q}\underline{p}} := U_I \gamma^I_{\underline{q}\underline{p}} = 2\bar{{\rm w}}_{\underline{q}A} \bar{{\cal U}}{}^{AB} \bar{{\rm w}}_{\underline{p}B}\; , \qquad
\bar{U}{}\!\!\!\!/{}_{\underline{q}\underline{p}} := \bar{U}{}_I \gamma^I_{\underline{q}\underline{p}} = 2{\rm w}_{\underline{q}}^{A}
{{\cal U}}_{AB}{\rm w}_{\underline{p}}^{B} \; , \qquad
%\\ \nonumber
\\  \label{UchJg==} &&  U\!\!\!\!/{}^{\check{J}}_{\underline{q}\underline{p}} := U_I^{\check{J}} \gamma^I_{\underline{q}\underline{p}} =
i {\rm w}_{\underline{q}}^{A} (\sigma^{\check{J}}\bar{{\cal U}}){}_A{}^{B }\bar{{\rm w}}_{\underline{p}B} +i \bar{{\rm w}}_{\underline{q}A} (\tilde{\sigma}{}^{\check{J}}{\cal U})^A{}_B{\rm w}_{\underline{p}}^{B}\; , \qquad
%\\ \nonumber
\\ \label{wgIw=UIcU} &&
  \bar{{\rm w}}_{\underline{q}A}\gamma^{I}_{\underline{q}\underline{p}}\bar{{\rm w}}_{\underline{p}B}= U_I {\cal U}_{AB}\; , \qquad
 {\rm w}_{\underline{q}}^{A}\gamma^{I}_{\underline{q}\underline{p}}{\rm w}_{\underline{p}}^{B}= \bar{U}_I \bar{{\cal U}}^{AB}\; , \qquad
    \bar{{\rm w}}_{\underline{q}A}\gamma^{I}_{\underline{q}\underline{p}}{\rm w}_{\underline{p}}^{B}= iU_I ^{\check{J}}(\sigma^{\check{J}}\bar{{\cal U}})_{A}{}^{B}\; .  \qquad
    %\\ \nonumber
\end{eqnarray}
These include  the symmetric unitary complex matrix ${\cal U}_{AB}=(\bar{{\cal U}}{}^{AB})^*$ parametrizing the coset $SU(8)/SO(7)$ and complex matrices
$\sigma^{\check{J}}_{AB}=(\tilde{\sigma}{}^{\check{J}BA})^*$ which are obtained from $SO(7)$ Clebsch-Gordan coefficients by $SU(8)$ rotation
(see Appendix \ref{SU8-SO7} for details).

%%%%%%%%
\if{}
As we have already noticed, constraints \eqref{D0:wbw=I} imply that the matrix  $({\rm w}_{\underline{q}}^A \, ,\, \bar{{\rm w}}_{\underline{q}A} )$ belongs to $SO(16)$ group (see \eqref{D0:w+bw=inSO16}) and thus carries 120 ''degrees of freedom''.  Then, after imposing the constraints \eqref{wgIw=UIcU}, it becomes a matrix which can be obtained from a $Spin(9)$ valued one by the multiplication on a matrix from $16$ dimensional representation of $SU(8)$ subgroup of $SO(16)$ group. Thus the number of ''degrees of freedom'' it carries is $99=36+63$.
\fi
%%%%%%%

%\bigskip

\subsubsection{Three particle kinematics and on-shell amplitudes of type IIA string theory involving D0 brane }

Let us consider the case of scattering of three (super)particles,
two massless and two massive.
Namely,
we will be interested in amplitudes and superamplitudes for the process involving two type IIA supergravity (${\rm i}={\rm 1,2}$) and one D0--brane supermultiplet  (${\rm i}={\rm 3}$). The basic bra-ket notation for spinor helicity variables and polarization spinors are
\bea\label{bra123=}
&&<\,{\rm 1}^-_q\, |= v_{\alpha q\,{\rm 1}}^- \; , \qquad
<\,{\rm 2}^-_q\, |= v_{\alpha q\,{\rm 2}}^- \; , \qquad
<\,{\rm 3}^{\,\underline{q}}\, |= {\rm v}_{\alpha {\rm 3}}^{\; \;\underline{q}} \; , \qquad
%\\ \nonumber
\\
\label{ket123=}
&& |\, {\rm 1}^-_{\dot{q}}\,>=\, v_{\dot{p}{\rm 1}}^{-\alpha } \; , \qquad  |\, {\rm 2}^-_{\dot{q}}\,>=\,v_{\dot{p}{\rm 2}}^{-\alpha }\; , \qquad   |\, {\rm 3}_{\underline{q}}\,>=\,
{\rm v}_{\underline{q}{\rm 3}}{}^{\alpha} \; .
 \qquad
\eea
The contractions of the helicity spinors corresponding to massless states ${\rm 1}$ and ${\rm 2}$, $<\,{\rm 1}^-_q\, | \, {\rm 2}^-_{\dot{q}}\,>$ and $<\,{\rm 2}^-_q\, | \, {\rm 1}^-_{\dot{q}}\,>$, are determined by \eqref{v-iv-j=}. To specify the contractions involving helicity spinors of a massive state ${\rm 3}$ we have to use  the decomposition \eqref{v=v1l1+v2l2} and the momentum conservation \eqref{Pmu=0} which in this case is solved by
\be\label{u0=u1+u2}
u_{\mu {\rm 3}}^{\, 0}= - \frac 1 m \left(\rho^\#_{{\rm 1}}u_{\mu {\rm 1}}^{\, =}+\rho^\#_{{\rm 2}}u_{\mu {\rm 2}}^{\, =}\right) \qquad \Leftarrow \qquad
0= p_{\mu {\rm 1}}+p_{\mu {\rm 2}}+p_{\mu {\rm 3}}\; .
\ee
Below we will conventionally skip subindex $ {\rm 3}$  as we consider a process with only one massive particle in this section, and also  use the notation
\bea\label{l=-r:m}& l^\#_{{\rm i}}=- \frac {\rho^\#_{{\rm i}}} m \; .
\eea

The decomposition \eqref{v=v1l1+v2l2}
of helicity spinors of the D$0$--brane in terms of helicity spinors of two non-colinear gravitons can always be equivalently  written in the form
\begin{eqnarray}\label{v===v1l1+v2l2}
{\rm v}_{\alpha}{}^{\underline{q}} &=&
\sqrt{2l_{{\rm 1}}^{\#}} v_{\alpha q{\rm 1}}^{\; -} l_{q {\rm 1} }^{\; \underline{q}} + \sqrt{2l_{{\rm 2}}^{\#}} v_{\alpha  {q}{\rm 2}}^{\; -}
l_{{q}{\rm 2}}^{\; \underline{q}}
\;  \qquad
\end{eqnarray}
However, the nontrivial statement is that in this expression
$l_{{\rm 1}q}^{\; \underline{q}}$ and  $l_{{\rm 2}q}^{\; \underline{q}}$  obey
\bea\label{l1l1==I}
&& l_{{\rm 1}q}^{\;\underline{q}}\,l_{{\rm 1}p}^{\;\underline{q}}= \delta_{qp} \; ,
 \qquad {}  \qquad  l_{{\rm 1}q}^{\; \underline{q}}\,l_{{\rm 2}p}^{\;\underline{q}}=0 \; ,
 \qquad {}  \qquad  l_{{\rm 2}q}^{\;\underline{q}}\,l_{{\rm 2}p}^{\; \underline{q}}=\delta_{qp}  \;
 \qquad
\eea
and hence form the $SO(16)$ valued matrix,
$ \left(\, l_{{\rm 1}q}^{\; \underline{q}}\, , \,  l_{{\rm 2}q}^{\; \underline{q}}\right)\;\in\; O(16)$.
This is shown in  Appendix \ref{3=00m}. The expression for the complex version of D$0$--brane spinor moving frame variables
\eqref{D0:vA=vwA}
can be easily obtained from \eqref{v===v1l1+v2l2}.

\subsubsection{On on-shell amplitudes of type IIA string theory involving D0 brane. }

Let us consider a 'massive' counterpart of the three amplitude
\eqref{AphABCDp=}, namely $A(\Re{\rm{e}}{\phi}_{{\rm 1}}\,\Re{\rm{e}}{\phi}_{{\rm 2}}\,\phi_{ABCD {\rm m}} )$ corresponding to the scattering of massive particle described by the intermediate
($\propto \Theta^{\wedge 4}_{{\rm 3}}$)
component of the on--shell superfield of
D$0$--brane and of two real parts of the leading ($\Theta^-_{{\rm 1,2}}$--independent) components of the on-shell superfields that describe type IIA supergravity multiplets. As suggested by Eq. \eqref{phimABCD=}, it should contain some of the contributions proportional to the amplitudes
$A({\phi}^{IIA}_{{\rm 1}}\, {\phi}^{IIA}_{{\rm 2}}\, h^{IJ}_{{\rm m}} )$
and $A({\phi}^{IIA}_{{\rm 1}}\, {\phi}^{IIA}_{{\rm 2}}\,\,A^{IJK}_{{\rm m}} )$.

These amplitudes should be constructed from the light-like momenta of scalar massless particles (dilatons), $p_{\mu{\rm 1}}$ and  $p_{\mu{\rm 2}}$, and the polarization vector of the massive counterpart of graviton and 3-form gauge field  which should be constructed from  the  orthogonal vectors of the D$0$-brane moving frame \eqref{u0uI=inSO}, ${\rm u}_{\mu {\rm 3}}^I={\rm u}_{\mu}^I$. As the index structure of $A^{IJK}_{{\rm m}}$ cannot be reproduced using the contractions of these three ingredients on their 10-vector indices, we conclude that only $A(\Re{\rm{e}}{\phi}_{{\rm 1}}\, \Re{\rm{e}}{\phi}_{{\rm 2}}\, h^{IJ}_{{\rm m}} )$ contributes to
the amplitude   $A(\Re{\rm{e}}{\phi}_{{\rm 1}}\,\Re{\rm{e}}{\phi}_{{\rm 2}}\,\phi_{ABCD {\rm m}} )$ which then reads
({\it cf.} \eqref{App4p=})\footnote{The relative coefficients make the expression in the brackets traceless in its SO(9) vector indices $I$ and $J$ which corresponds to tracelessness of $h^{IJ}$.  }

\bea\label{App4p=m1}
A(\Re{\rm{e}}{\phi}_{{\rm 1}}\, \Re{\rm{e}}{\phi}_{{\rm 2}}\, \phi_{ABCD {\rm m}} ) =
p_{{\rm 1}\mu}{\rm u}^{\mu I}_{{\rm 3}} p_{{\rm 2}\nu}{\rm u}^{\nu J}_{{\rm 3}}\,  \left(U_I^{\check{I}} U_J^{\check{J}}
t^{\check{I}\check{J}}_{ABCD}+ \frac 1 2 U_{(I} \bar{U}_{J)} \sigma^{\check{K}}{}_{[AB}\sigma^{\check{K}}{}_{CD]}
\right)
%\nonumber \\ \nonumber
\eea

Using our result on three particle kinematics, after a straightforward but not short calculations (described in  Appendix \ref{A00m=}) we have obtained the following expression for
this amplitude
\bea\label{App4p==m2}
 A(\Re{\rm{e}}{\phi}_{{\rm 1}}\, \Re{\rm{e}}{\phi}_{{\rm 2}}\, \phi_{ABCD {\rm m}} ) &=& \frac {(\rho^{\#}_{{\rm 1}}\rho^{\#}_{{\rm 2}})^2}{16{\rm m}^2}\, |\vec{K}^=_{{\rm 21}}|^4
 %\times  \nonumber
 %\\  \nonumber \\ &\times &\,
 \left((l_{q{\rm 1}}{\rm w}^E)\, (l_{q{\rm 1}}\bar{{\rm w}}_F)\,
  (\sigma^{(\check{I}|}  {\bar{\cal U}})_E{}^{F} \;
    (l_{p{\rm 1}}{\rm w}^G)\, (l_{p{\rm 1}}\bar{{\rm w}}_H)\,
  (\sigma^{|\check{J})}  {\bar{\cal U}})_G{}^{H} t^{\check{I}\check{J}}_{ABCD}
  - \right.   \nonumber
  %\\   \nonumber
  \\ && {}\hspace{4cm} \left. - \frac 1 8 \left|
  (l_{q{\rm 1}}\bar{{\rm w}}_E){\bar{\cal U}}{}^{EF} (l_{q{\rm 1}}\bar{{\rm w}}_F)\;
  \right|^2\sigma^{\check{K}}{}_{[AB}\sigma^{\check{K}}{}_{CD]}\right)\; .   \eea
 In it the contracted $\underline{q}, \underline{p}=1,..,16$ indices are not written explicitly.

It is natural to try to use this ampitude
as a reference point for finding the superamplitude  in a manner similar to that we followed for the case of process with three type II supergravity multiplets. However, as we are going to show,
there is an issue that
hampers such a straightforward way.

\subsubsection{Problems on the way towards on-shell superamplitudes of type IIA string theory involving D0 brane }

By analogy with the massless case, we can expect that  amplitude \eqref{App4p==m2} can be obtained from superamplitude by applying Grassmann integrations or a product of fermionic derivatives as ({\it cf.} \eqref{App4p=int})
\bea\label{App4p=m3}
 A(\Re{\rm{e}}{\phi}_{{\rm 1}}\, \Re{\rm{e}}{\phi}_{{\rm 2}}\, \phi_{ABCD {\rm m}} ) &=&\propto
\left(\int d^4(\Theta_{{\rm 3}})_{ABCD} \,   {\cal A}_{3\; 00 {\rm m}}\right)\vert_{
\Theta^-_{{\rm 1}}=0=\Theta^-_{{\rm 2}}, \Theta_{{\rm 3}}=0}  \nonumber
%\\ \nonumber
\\ &=&\propto
\left( D_{A\,{\rm 3}}D_{B\,{\rm 3}}D_{C\,{\rm 3}}D_{D\,{\rm 3}} \,  {\cal A}_{3\; 00 {\rm m}}\right)\vert_{\Theta^-_{{\rm 1}}=0=\Theta^-_{{\rm 2}}, \Theta_{{\rm 3}}=0}\; .
%\\ \nonumber
\eea

The problem to be solved is then to find the expression for
the 3-point superamplitude with two type IIA supergravity multiplets and a D$0$-brane supermultiplet (roughly speaking, with two massless and one massive supergravitons),
which we have denoted by ${\cal A}_{3\;  00{\rm m}}$.

The simple prescription which was elaborated for type IIB case and, as we have shown, allowing also for the description of  scattering of type IIA supergravitons, meets in this case some difficulty which can be followed to difference of
 massive and massless representations of the supersymmetry algebra.

If we define the complex supercharges of massive superparticle with momentum \eqref{pD0=mu0} by the first expressions in Eqs. \eqref{bbbQ=IIA} and  \eqref{bbQ=IIA}, their non vanishing anticommutator will be given by
\bea\label{D0:QbQ=}
\{ {\bb Q}_{\alpha {\rm 3}},  \bar{{\bb Q}}_{\beta {\rm 3}}  \} \, &=& \,
-i{\rm m}\, {\rm u}_{\mu {\rm 3}}^0\,  \left(
\sigma^\mu_{\alpha\beta} + k\!\!\!/{}_{\alpha\gamma}\tilde{\sigma}{}^{\mu\gamma\delta} k\!\!\!/{}_{\delta\beta} \right)  \,  = \,
-2i{\rm m}\,  {\rm u}_{\mu {\rm 3}}^0\,
\sigma^\mu_{\alpha\beta}\,
%\\ \nonumber
\\ \label{D0:QbQ==} &=& \,
-i{\rm m}\,  \left({\rm v}_{\alpha {\rm 3}}{}^{\underline{q}}  {\rm v}_{\beta {\rm 3}}{}^{\underline{q}}+ k\!\!\!/{}_{\alpha\gamma}{\rm v}_{\underline{q}}^{\gamma} \, k\!\!\!/{}_{\beta\delta}  {\rm v}_{\underline{q}}^\delta \right) = -2i{\rm m}\,  {\rm v}_{\alpha {\rm 3}}{}^{\underline{q}}  {\rm v}_{\beta {\rm 3}}{}^{\underline{q}} =  -4i{\rm m}\,  {\rm v}_{{\rm 3} (\alpha}{}^{A}  \bar{{\rm v}}_{\beta){\rm 3} A}  \; . %\\ \nonumber
\eea
The first form of this equation, \eqref{D0:QbQ=}, is valid provided $k_\mu$ is spacelike unit vector orthogonal to ${\rm u}_{\mu {\rm 3}}^0$, i.e.
$k^\mu {\rm u}_{\mu {\rm 3}}^0=0$.
Using the momentum conservation conditions
\eqref{u0=u1+u2}, in the gauge \eqref{u--=Kiui}--\eqref{u++=u0}   it is easy to check that this  is indeed the case for $k_\mu$ related with T-duality direction and thus obeying  \eqref{k=kjuj} for  ${\rm i}=1,2$.
(Eq. \eqref{kK--=-k--} and \eqref{k++=0} can  be also used).

The problem comes from the final expression in \eqref{D0:QbQ==}, namely from symmetrization over spinor indices in it. This originates in  the form of the  unity decomposition in the last equation of \eqref{D0:wbw=I}
characteristic for the case of massive superparticle. This term cannot be reproduced if we realize say ${\bb Q}_{\alpha {\rm 3}} $ as multiplication operator
$ \bar{{\rm v}}_{\alpha A}\Theta^A:= {\rm v}_{\alpha}{}^{\underline{q}}\bar{{\rm w}}_{\underline{q}A}\Theta^A$ and $\bar{{\bb Q}}_{\alpha {\rm 3}} $ by differential operator
with respect to the fermionic coordinate,
${\rm v}_{\alpha}{}^A\partial_A := {\rm v}_{\alpha}{}^{\underline{q}}{\rm w}_{\underline{q}}{}^A \frac {\partial}{\partial \Theta^A}$. Indeed, in this case the r.h.s. of the anticommutator would contain $\bar{{\rm v}}_{\alpha A} {\rm v}_{\beta}{}^A $ the antisymmetric part of which,
$\bar{v}_{[\alpha | A} v_{|\beta ]}{}^A =
 {\rm v}_{\alpha}{}^{\underline{q}} {\rm v}_{\beta}{}^{\underline{p}}
 {\rm w}_{[\underline{p}}{}^A \bar{{\rm w}}_{\underline{q}]A}
$, is generically nonvanishing.

It is instructive to take a look on this problem from a bit  different side. To this end,  let us write the representation for original 'real' supersymmetry generators in our complex reduced configuration space. The action of supersymmetry on the original $\theta^{\alpha 1}$ and $\theta_\alpha^2$ coordinates as well as the supersymemtry (super)algebra allows us to find
\bea
{\bb Q}_{\alpha\; {\rm 3}}^{\; 1}=\frac 1 {\sqrt{2}}
\left({\rm v}_{\alpha {\rm 3}}{}^A \frac {\partial}{\partial \Theta^A_{{\rm 3}}} +4{\rm m} \bar{{\rm v}}_{\alpha A {\rm 3}}\Theta^A_{{\rm 3}} \right)\; , \qquad
%\\ \nonumber  \\
{\bb Q}^{\alpha 2}_{{\rm 3}}=\frac 1 {\sqrt{2}}
\left({\rm v}_{\alpha {\rm 3} }{}^A \frac {\partial}{\partial \Theta^A_{{\rm 3}}} +4{\rm m} \bar{{\rm v}}{}^{\alpha }_{A {\rm 3}}\Theta^A_{{\rm 3}}  \right)\; , \qquad
%\\ \nonumber \\
\eea
It is not difficult to observe that \be\label{Q2=u0sQ1} {\bb Q}^{\alpha 2}_{ {\rm 3}}= {\rm u}^0_\mu\tilde{\sigma}^{\mu\alpha\beta}{\bb Q}^1_{\beta  {\rm 3}}\ee  which implies that the action of the
type IIA supersymmetry on the invariant subsuperspace defined through the use of analytical basis (see \cite{Bandos:2025pxv}) is not faithful.
This is not the case for supergravity multiplet described by quantum massless superparicle since no counterparts of
the relation \eqref{Q2=u0sQ1} exists in this case.

The consequence of the above  observation is that the complex supercharge and its conjugate, defined with the use of T-duality related complex structure as in  \eqref{bbQ=IIA} and  \eqref{bbbQ=IIA} in the massive case are expressed algebraically through the same real supercharge,
\bea\label{bbbQ=IIAm}
\bar{{\bb Q}}_{\alpha {\rm 3}}&=&\frac 1 {\sqrt{2}} (Q_{\alpha 1, {\rm 3}}-ik\!\!\!/{}_{\alpha\beta}{\bb Q}_{2, {\rm 3}}^{\beta})  =
\frac 1 {\sqrt{2}} (\delta_{\alpha}{}^{\beta}-i(k\!\!\!/{}\sigma^\mu)_{\alpha}{}^{\beta}) Q_{\beta 1, {\rm 3}}\; ,
 \qquad    \\ \label{bbQ=IIAm}
{\bb Q}_{\alpha {\rm 3}}&=&\frac 1 {\sqrt{2}}(Q_{\alpha 1, {\rm 3}}+ik\!\!\!/{}_{\alpha\beta}{\bb Q}_{2, {\rm 3}}^{\beta})  =
\frac 1 {\sqrt{2}} (\delta_{\alpha}{}^{\beta}+i(k\!\!\!/{}\sigma^\mu)_{\alpha}{}^{\beta}) Q_{\beta 1, {\rm 3}}
 \; , \qquad
\eea
and thus both include the sum of algebraic expression and differential operator.

The above discussed issues  make solution of the supersymmetric Ward identities for superamplitude involving one massive and two massless
superparticles more complicated than it was in the purely massless
case discussed in the previous subsections. Probably this is the reason why superamplitudes for massive supermultiplets have been addressed only recently and elaborated only for $D=4$ ${\cal N}=1$ case \cite{Herderschee:2019ofc,Delgado:2025oev}.
We plan to address this problem for 10D superamplitude in forthcoming paper(s).

\section{Conclusion and discussion}

In this paper we firstly
performed covariant quantization of 10D type IIB superparticle  in such a manner that made manifest the hidden SU(8) symmetry of linearized type IIB supergravity which describes its quantum state spectrum. Then we
quantizatized in a similar way  type IIA superparticle and  discussed  its peculiarities as well as the appearance of the  hidden $SU(8)$ symmetry. Finally we discussed  the generalization of the on-shell superfield formalism obtained as a result of the above mentioned superparticle quantization to superamplitudes, made
some observations on their properties, and pointed
out some problems to be solved.

The hidden $SU(8)$ symmetry of type II supergravities found in this paper  is of interest because $SU(8)$ is R--symmetry of $D=4$ ${\cal N}=8$ supersymmetry algebra and also a local symmetry of the maximal D=4 ${\cal N}=8$ supergravity which can be obtained from $10D$ type IIB one by dimensional reduction.
Actually the   $D=4$  ${\cal N}=8$ supergravity was obtained in \cite{Cremmer:1979up} by dimensional reduction of 11D supergravity \cite{Cremmer:1978km} supplemented by dualization of antisymmetric tensor fields to scalars. After this last stage the scalar fields of the maximal supergravity multiplet was found to parametrize the coset ${\bb E}_{7,+7}/SU(8)$
which implied that $D=4$ ${\cal N}=8$ supergravity is invariant  under nonlinearly realized ${\bb E}_{7,+7}$ symmetry and under local $SU(8)$ gauge symmetry transformations.

The natural question of whether hidden ${\bb E}_{7,+7}$ is present in 11D supergravity was posed and solved
by de Witt and Nicolai in \cite{deWit:1985iy,deWit:1986mz}.
When supergravity is linearized over a flat spacetime,
only  the $SU(8)$ part of the hidden ${\bb E}_{7,+7}$ symmetry can be seen. Our work shows how such  hidden $SU(8)$ symmetry of type IIB supergravity -- and also of its type IIA cousin,-- appears from quantization of superparticle in its spinor moving frame formulation supplemented by auxiliary bridge variables.

These auxiliary bridge variables parametrize the space of possible complex structures which can be introduced in the phase space of type IIB superparticle in its spinor moving frame formulation. This space  is isomorphic to
$SU(8)/SO(8)$ coset. We have shown how to incorporate these bridge variables
into the superparticle Lagrangian in the so--called analytical basis of  type IIB Lorentz harmonic superspace in such a way that the number of physical degrees of freedom of the dynamical system is not changed. This is possible because the bridge variables are St\"uckelberg fields of the type IIB superparticle mechanics. This is in consonance with the fact that the introduction  of St\"uckelberg fields is inevitable to manifest the ${\bb E}_{7,+7}$ symmetry of $11D$ supergravity which is otherwise hidden \cite{deWit:1985iy,deWit:1986mz}.

To observe the  hidden $SU(8)$ symmetry of the linearized type IIA supergravity we have performed similar quantization of type IIA $10D$ superparticle. This resulted in actually the same chiral (analytic) on shell superfield as the one describing quantum state spectrum of type IIB superparticle. However, to obtain this result we needed to introduce a covariantly constant eight-vector $k^i$  carrying the index of v- representation of $SO(8)$. This can be related to the spacelike $10$-vector $k_\mu$ defining the T-duality transformation from type IIA to type IIB superparticle. Notice that the necessity of the compactness of T-duality direction is not seen in the linearized supergravities which appear in the quantum state spectrum of superparticles.

In the final part of the paper we have done first steps towards generalization of the on-shell superfield description of linearized supergravity to on-shell superamplitudes of type IIB supergravity and type IIA theory. We have discussed spinor moving frame derivation of simple superamplitudes of type IIB theory first calculated  in \cite{Boels:2012ie}, and the application of such type IIB formulas and their higher point generalizations to type IIA superamplitudes. The superamplitudes with less or equal to seven IIA supergrvaitons allow for the description within such a formalism. This restriction follows from the necessity to define T-duality vector $k_\mu$
which is
the same
for all scattered particles and orthogonal to all their momenta.

We have also described some stage towards calculation of superamplitudes of type IIA theory involving, besides type IIA supergrvaity multiplets (supergravitons) also D$0$--brane(s) (Dirichlet particle(s)). To this end we have used the results of recent quantization of D$0$--brane in its spinor moving frame formulation \cite{Bandos:2025pxv}.  We pointed out a problem which hampers the straightforward generalization of the approach used to derive superamplitudes of
type II supergravitons to this case. This issue can be followed to the difference of the action of type IIA supersymmetry on the reduced superspaces with complex structures on which the on-shell superfields of massive and massless superparticles are defined.

Probably the 'component amplitude' method of solving the supersymmetric Ward identities generalizing $D=4$ approach of \cite{Bianchi:2008pu} will show itself more productive in this case.
Another alternative is to elaborate a type IIA massless and massive generalizations of the constrained superfield approach to superamplitude proposed for $D=10$ SYM and $D=11$ SUGRA in \cite{Bandos:2017eof}.

%\bigskip

\section*{Acknowledgments}
We would like to thank
Dima Sorokin for useful discussions.
The work of I.B. has been supported in part by the MCI, AEI,
FEDER (UE) grant PID2024-155685NB-C21  (“Gravity, Supergravity and Superstrings” (GRASS)) and by the Basque Government grant IT-1628-22. The work of M.T. has been supported by the Quantum Gravity Unit of the Okinawa Institute of Science and Technology Graduate University (OIST).

\appendix

\setcounter{equation}0

\section{Some useful  relations for 10D spinor moving frame variables}
\label{App=harm}
\def\theequation{A.\arabic{equation}}

\setcounter{equation}0

\subsection{Inverse spinor moving frame matrix}
The rectangular blocks $ v^{-\alpha}_{\dot{q}}$ and $v^{+\alpha}_{{q}}$ of the inverse to the spinor moving frame matrix \eqref{harmV=10},
\begin{eqnarray}\label{harmV-1=10}
V_{(\beta)}^{\;\;\; \alpha}= \left(\begin{matrix}
  v^{-\alpha}_{\dot{q}} \cr v^{+\alpha}_{{q}}\end{matrix} \right) \in Spin(1,9)
 \;  \qquad
\end{eqnarray}
 obey  $ v_{\alpha\dot{q}}^{\; +} v^{-\gamma}_{\dot{q}}
+  v_{\alpha q}^{\; -}v^{+\gamma}_q
=
\delta_{\alpha}{}^\gamma$ and
\begin{eqnarray}\label{v-qv+p=}
&
v^{-\alpha}_{\dot{q}}   v_{\alpha p}^{\; +}=\delta_{\dot{q}\dot{p}}
 \; ,  \qquad & v^{+\alpha}_{{q}}  v_{\alpha \dot{p}}^{\; +}=0\;  , \qquad
 \nonumber
 %\\ \nonumber
 \\
 &
 v^{-\alpha}_{\dot{q}}   v_{\alpha q}^{\; -}=0\;  , \qquad & v^{+\alpha}_{{q}} v_{\alpha {p}}^{\; -} = \delta_{qp}\; .   \qquad
 %\\ \nonumber
\end{eqnarray}
They are related with
moving frame vectors by the constraints similar to \eqref{u--s=v-v-}--\eqref{uIs=v+v-},
 \begin{eqnarray}\label{u--ts=v-v-} && u_a^= \tilde{\sigma}{}^{a\, \alpha\beta}= 2v^{-\alpha}_{\dot q}   v^{-\beta}_{\dot q} \; , \qquad  v^-_{\dot{q}} {\sigma}_{a}v^-_{\dot{p}}= u_a^= \delta_{\dot{q}\dot{p}}
 \; , \qquad
 %\\ \nonumber
 \\
 \label{u++ts=v+v+}
&& \tilde{\sigma}^{ {a} {\alpha} {\beta}} u_{ {a}}^{\# }= 2 v_{{q}}^{+ {\alpha}}v_{{q}}^{+}{}^{ {\beta}}\; , \qquad  v_{{q}}^+ {\Gamma}_{ {a}} v_{{p}}^+ = \; u_{ {a}}^{\# } \delta_{{q}{p}}\; ,  \qquad
%\\ \nonumber
\\  \label{uIts=v-v+}
&& \tilde{\sigma}^{ {a} {\alpha} {\beta}} u_{ {a}}^{i}=-  2 v_{\dot q}^{-( {\alpha}}\gamma^i_{q\dot{q}}v_{{q}}^{+}{}^{ {\beta})}\; , \qquad
  v_{\dot q}^- {\sigma}_{ {a}} v_{{p}}^+ = - u_{ {a}}^{i} \gamma^i_{p\dot{q}}\; . \qquad
  %\\ \nonumber
\end{eqnarray}

\subsection{Cartan forms and admissible variations of spinor moving frame variables}

As spinor moving frame variables $ v_{\alpha q}^{\; -}$ and $v_{\alpha \dot{q}}^{\; +} $ parametrize double covering $Spin(1,9)$ of the Lorentz group $SO(1,9)$, \eqref{harmV=10}, their admissible variations, i.e. variations which preserve the constraints \eqref{harmV=10}, or equivalently, Eqs. \eqref{u--s=v-v-}--\eqref{uIs=v+v-}, should parametrize the space co--target  to the Lorentz group, which is dual to the $so(1,9)$ Lie algebra \footnote{More precisely, this  co-tangent space is dual to the tangent space to the Lorentz group which is the space of vector fields representing $so(1,9)$ Lie algebra.}. The same applies to admissible derivatives of the spinor frame variables, this is to say derivatives which preserve constraints  \eqref{u--s=v-v-}--\eqref{uIs=v+v-} or condition \eqref{harmV=10}.

As the moving frame vectors are composites of the spinor frame variables according to  \eqref{u--s=v-v-}--\eqref{uIs=v+v-}, they both carry the same {\it  local} degrees of freedom.
Then it is convenient to define Cartan forms providing the basis in the space co-tangent to $Spin(1,9)$ group  in terms of moving frame vectors,
\bea\label{Om--i:=}
 \Omega^{= i}&:=&u_a^{=}du^{ai}\; , \qquad
 %\\ \nonumber
 \\ \label{Om++i:=} \Omega^{\# i}&:=&u_a^{\#}du^{ai}\; , \qquad
 %\\ \nonumber
 \\ \label{Om0:=} \Omega^{(0)}&:=&{1\over 4}u_a^{=}du^{a\#} \; , \qquad
 %\\ \nonumber
 \\ \label{Omij:=}\Omega^{ij}&:=&u_a^{i}du^{aj}\; . \qquad
 %\\ \nonumber
\eea
Of these, 1-forms  \eqref{Om--i:=} and  \eqref{Om++i:=} are covariant under local $SO(1,1)\otimes SO(8)$ transformations and form the basis of the (space co-tangent to the) coset $\frac {SO(1,9)} {SO(1,1)\otimes SO(8)}$, while \eqref{Om0:=} and \eqref{Omij:=} transform as $SO(1,1)$ and $SO(8)$ connection,  respectively. Furthermore,  Cartan form  \eqref{Om--i:=},
$\Omega^{= I}:=u_a^{=}du^{aI}$ can be considered as a basis of the   coset $\frac {SO(1,9)} {[SO(1,1)\otimes SO(8)]\subset\!\!\!\!\times  K_{8}}={\bb S}^{8}$,  isomorphic to celestial sphere of a 10D observer \cite{Delduc:1991ir,Galperin:1991gk}.

Using the consequence
\bea\label{I=uu+uu-uiui}
&& \delta_a{}^b= \frac 1 2 u_a^= u^{b\#}+ \frac 1 2 u_a^{\#} u^{b=}-  u_a^i u^{bi} \;  \qquad
%\nonumber \\
\eea
of Eqs. \eqref{u--2=0}, \eqref{u--ui=0},
the (admissible)  derivatives of moving frame vectors can be easily expressed in terms of the above Cartan forms.
We prefer to write these relations as expressions for the $SO(1,1)\times SO(8)$ covariant derivatives of moving frame vectors,

\begin{eqnarray}
\label{Du--}  Du^{=}_a &:=& du^{=}_a + 2u^{=}_a \Omega^{(0)} = u^{i}_a \Omega^{=i}
\; , \qquad
%\\ \nonumber
\\  \label{Du++}  Du^{\#}_a &:=&d u^{\#}_a - 2u^{\#}_a \Omega^{(0)} =
u^{i}_a \Omega^{\# i} \; , \qquad
%\\ \nonumber
\\ \label{Dui}  Du^{i}_a &:=& du^{i}_a + u^{j}_a
\Omega^{ji} = {1\over 2} u^{\#}_a \Omega^{= i} + {1\over 2} u^{=}_a \Omega^{\# i} \; .
%\\ \nonumber
\qquad
\end{eqnarray}

The self consistency conditions of Eqs. (\ref{Du--}), (\ref{Du++}) and (\ref{Dui}), this is to say   Ricci identities
\bea DDu^{=}_a = 2u^{=}_a d\Omega^{(0)}, \qquad DDu^{\#}_a =- 2u^{\#}_a d\Omega^{(0)}\; , \qquad
DDu^{i}_a =u^{j}_a
G^{ji}= u^{j}_a
(d\Omega^{ji}+ \Omega^{jk}\wedge \Omega^{ki}) \; ,
%\nonumber \\
\eea  are equivalent to the Maurer-Cartan equations of the $SO(1,9)$ group
\begin{eqnarray}
\label{DOm--i}  D\Omega^{=i}&:=& d\Omega^{=i} + 2 \Omega^{=i}\wedge  \Omega^{(0)} + \Omega^{ij}\wedge  \Omega^{=j}=0 \; , \qquad
%\\ \nonumber
\\
\label{DOm++i}  D\Omega^{\# i}&:= & d\Omega^{\# i} - 2 \Omega^{\# i}\wedge  \Omega^{(0)}+ \Omega^{ij}\wedge  \Omega^{\# j}=0 \; , \qquad
%\\ \nonumber
\\
\label{dOm0}   && d\Omega^{(0)}= {1\over 4}\Omega^{=i}\wedge \Omega^{\#i}\; , \qquad
%\\  \nonumber
\\
\label{dOmij}   G^{ij}&:= & d\Omega^{ij}+\Omega^{ik}\wedge  \Omega^{kj}= -
\Omega^{=[i}\wedge \Omega^{\# j]}\; . \qquad
%\\ \nonumber
\end{eqnarray}

As the co--tangent space to  $Spin (1,D-1)$ is isomorphic to
the  space co--tangent to  $SO(1,D-1)$, the derivatives of spinor moving frame variables are expressed in terms of the same Cartan forms. These expressions can be derived, after some algebra,  by taking derivatives of the constraints \eqref{u--s=v-v-}--\eqref{uIs=v+v-} and  using  \eqref{Du--}--\eqref{Dui}. In such a way we arrive at the following expression for derivatives (and covariant derivatives) of spinor moving frame variables
\begin{eqnarray}
\label{Dv-q=} &  Dv_{\alpha {q}}^{\; -} := dv_{\alpha {q}}^{\; -}+ \Omega^{(0)} v_{\alpha {q}}^{\; -} + {1\over 4} \Omega^{ij}
v_{\alpha {p}}^{\; -}\gamma_{{p}{q}}^{ij} = {1\over 2} \Omega^{=i}  \gamma_{q\dot{q}}^{i} v_{\alpha \dot{q}}^{\;+} \; , \qquad
%\\ \nonumber
\\
 \label{Dv+q=} & Dv_{\alpha \dot{q}}^{\;+}   :=  dv_{\alpha \dot{q}}^{\;+}    -
\Omega^{(0)} v_{\alpha \dot{q}}^{\;+}   + {1\over 4} \Omega^{ij}  v_{\alpha \dot{p}}^{\;+}  \tilde{\gamma}_{\dot{p}\dot{q}}^{ij} =  {1\over 2}
v_{\alpha {q}}^{\; -}  \Omega^{\# i} \gamma_{q\dot{q}}^{i}\; , \qquad
%\\ \nonumber
\end{eqnarray}
as well as of the blocks of inverse spinor moving frame matrix (which we will also call spinor moving frame variables below),
\begin{eqnarray}
\label{Dv-1-dq} &  Dv_{\dot{q}}^{-\alpha} := dv_{\dot{q}}^{-\alpha} + \Omega^{(0)} v_{\dot{q}}^{-\alpha} +
{1\over 4} \Omega^{ij} \tilde{\gamma}_{\dot{q}\dot{p}}^{ij} v_{\dot{p}}^{-\alpha} = - {1\over 2} \Omega^{=i}
 v_{{q}}^{+\alpha} \gamma_{q\dot{q}}^{i}\; , \qquad
 %\\ \nonumber
 \\
\label{Dv-1+q} &  Dv_{\dot{q}}^{+\alpha} := dv_{\dot{q}}^{+\alpha} - \Omega^{(0)} v_{\dot{q}}^{+\alpha} +
{1\over 4} \Omega^{ij} v_{\dot{p}}^{+\alpha} \gamma_{\dot{p}\dot{q}}^{ij} = - {1\over 2} \Omega^{\# i }
v_p^{-\alpha} \gamma_{p\dot{q}}^{i}\; . \qquad
%\\ \nonumber
\end{eqnarray}
Here  $ \gamma^i_{p\dot{q}}=\tilde{\gamma}^i_{\dot{q}p}$ are $SO(8)$ Clebsch-Gordan coefficients  which obey
\eqref{gitgj+=I}, and $\gamma^{ij}=  \gamma^{[i}\tilde{\gamma}^{j]}$, $\tilde{\gamma}^{ij}=  \tilde{\gamma}^{[i}{\gamma}^{j]}$, so that
\bea
&  \gamma^{i}\tilde{\gamma}^{j}=\delta^{ij}{\bb I}_{8\times 8}+\gamma^{ij}\; , \qquad
  \tilde{\gamma}^{i}{\gamma}^{j}=\delta^{ij}{\bb I}_{8\times 8}+ \tilde{\gamma}^{ij}\; . \qquad
\eea

The admissible variations of the spinor moving frame variables can be obtained formally by contraction of the
above expressions of exterior derivative with variational symbol using the rule
$i_\delta d= \delta$ (which is  the Lie derivative formula $(-)\delta= i_\delta d+di_\delta$ applied to 0-forms).
In this way we obtain, in particular,
\begin{eqnarray}
\label{vu--=} && \delta u^{=}_a =- 2u^{=}_a i_\delta \Omega^{(0)} + u^{i}_a i_\delta \Omega^{=i}
\; , \qquad
%\\ \nonumber
\\ \label{vv-q=} &&  \delta v_{\alpha {q}}^{\; -} =- i_\delta\Omega^{(0)} v_{\alpha {q}}^{\; -} - {1\over 4} i_\delta\Omega^{ij}
v_{\alpha {p}}^{\; -}\gamma_{{p}{q}}^{ij}+  {1\over 2} i_\delta\Omega^{=i}  \gamma_{q\dot{q}}^{i} v_{\alpha \dot{q}}^{\;+} \; .\qquad
%\\ \nonumber \\ && {\text{etc.}} \nonumber
%\\ \nonumber
\qquad
\end{eqnarray}

\subsection{Covariant derivatives for Lorentz harmonic coordinates }

The appropriate differential operators in the harmonic sector of Lorentz harmonic superspace are the so-called covariant harmonic derivatives (their counterparts in $SU(2)/U(1)$ ${\cal N}=2$ and $SU(3)/[U(1)\times U(1)]$ ${\cal N}=3$ harmonic superspaces were introduced in \cite{Galperin:1984av,Galperin:1984bu}),

\bea\label{D0:=}
D^{(0)}&=& {v}_{\alpha \dot{q}}^{\; +}  \frac {\partial } {\partial v_{\alpha \dot{q}}^{\; +} } - v_{\alpha q}^{\; -} \frac {\partial } {\partial v_{\alpha q}^{\; -} } \; , \qquad
%\\ \nonumber
\\ \label{Dij:=} D^{ij} &=& \frac 1 2 {v}_{\alpha \dot{q}}^{\; +}\tilde{\gamma}{}^{ij}_{\dot{q}\dot{p}}  \frac {\partial } {\partial v_{\alpha \dot{p}}^{\; +} } + \frac 1 2 v_{\alpha q}^{\; -} \gamma^{ij}_{qp}\frac {\partial } {\partial v_{\alpha p}^{\; -} } \; , \qquad
%\\ \nonumber
\\ \label{D++i:=} D^{\# i }&=&  \frac 1 2 {v}_{\alpha \dot{q}}^{\; +}\tilde{\gamma}{}^{i}_{\dot{q}p}  \frac {\partial } {\partial v_{\alpha p}^{\; -} }  \; , \qquad
%\\ \nonumber
\\ \label{D--i:=} D^{=i} &=&  \frac 1 2 v_{\alpha q}^{\; -}{\gamma}{}^{i}_{q\dot{p}}  \frac {\partial } {\partial v_{\alpha \dot{p}}^{\; +} } \; . \qquad
%\\ \nonumber
\eea
Their characteristic property is that they commute with all the defining constraints \eqref{u--s=v-v-}--\eqref{uIs=v+v-}. Their form can be derived algorithmically  by decomposing the exterior differential acting in harmonic sector of Lorentz harmonic superspace, this is to say on $Spin(1,9)$ group, on the Cartan forms \eqref{Om--i:=}--\eqref{Omij:=}
\bea\label{dSpin=}
& d^{Spin(1,9)}:=dv_{\alpha q}^{\; -} \frac {\partial } {\partial v_{\alpha q}^{\; -} }+d{v}_{\alpha \dot{q}}^{\; +}  \frac {\partial } {\partial v_{\alpha \dot{q}}^{\; +} }= \Omega^{(0)}D^{(0)} -\frac 1 2 \Omega^{ij}D^{ij} +\Omega ^{=i}D^{\# i }+\Omega ^{\# i }D^{=i}\; . \qquad
%\\ \nonumber
\eea
The r.h.s. is obtained from the l.h.s. by expressing $dv_{\alpha q}^{\; -}$ and $d{v}_{\alpha \dot{q}}^{\; +}$ through Cartan forms using  Eqs.
\eqref{Dv-q=} and  \eqref{Dv+q=}.

This fact also manifests the duality of
the basis of cotangent space to
(the double covering $Spin(1,9)$ of) the Lorentz group SO(1,9), given by Cartan forms \eqref{Om--i:=}--\eqref{Omij:=}, to the basis of the space tangent to (the double covering $Spin(1,9)$ of) SO(1,9) provided by the covariant harmonic derivatives \eqref{D0:=}--\eqref{D--i:=}.

The commutators of these covariant derivatives represent the algebra $so(1,9)$,
\bea
&& {}[D^{(0)}, D^{ij}]=0\; , \qquad {}[D^{(0)}, D^{\# i}]=2D^{\# i}\; , \qquad  {}[D^{(0)}, D^{= i}]=-2D^{= i}\; , \qquad
%\\ \nonumber
\\
&& {}[D^{ij}, D^{kl}]=4\delta^{[i|[k} D^{l]|j]}\; , \qquad  {}[D^{ij},  D^{\# k}]=2D^{\# [i}\delta^{j]k}\; , \qquad   {}[D^{ij},  D^{= k}]=2D^{= [i}\delta^{j]k}\; , \qquad
%\\ \nonumber
\\
&& {}[D^{\# i}, D^{\# j}]=0\; , \qquad   {}[D^{\# i}, D^{= j}]=\frac 1 4 \delta^{ij} D^{(0)} + \frac 1 2 D^{ij} \; , \qquad  {}[D^{= i}, D^{= j}]=0\; , \qquad
%\\ \nonumber
\eea
which is natural as the target space of the double covering $Spin(1,9)$ of the Lorentz group $SO(1,9)$ is isomorphic to the Lorentz algebra  $so(1,9)$.

The action of these covariant derivatives on the vector harmonics can of course be calculated taking into account its composite nature reflected by the constraints \eqref{u--s=v-v-}--\eqref{uIs=v+v-}. However it is much more practical to obtain this by decomposing the differential in the group manifold of $SO(1,9)$ parametrized by vector harmonics
\bea
d^{SO(1,9)}:=du_a^{=} \frac {\partial } {\partial u_a^{=} }+du_a^{\#} \frac {\partial } {\partial u_a^{\#} }+du_a^{i} \frac {\partial } {\partial u_a^{i} }= \Omega^{(0)}D^{(0)} -\frac 1 2 \Omega^{ij}D^{ij} +\Omega ^{=i}D^{\# i }+\Omega ^{\# i }D^{=i}\; . \qquad \\ \nonumber
\eea
On this way
we find the following expressions for covariant harmonic derivatives
\bea
D^{(0)}&=&2u_a^{\#} \frac {\partial } {\partial u_a^{\#} }- 2u_a^{=} \frac {\partial } {\partial u_a^{=} }\; , \qquad
%\\ \nonumber
\\  D^{ij} &=& u_a^{i}\frac {\partial } {\partial u_a^{j} } -u_a^{j}\frac {\partial } {\partial u_a^{i} }\; , \qquad
%\\ \nonumber
\\ D^{\# i }&=&  \frac 1 2 u_a^{\#}\frac {\partial } {\partial u_a^{i} }+ u_a^{i} \frac {\partial } {\partial u_a^{=} } \; , \qquad
%\\ \nonumber
\\ D^{=i} &=&  \frac 1 2 u_a^{=}  \frac {\partial } {\partial u_a^{i} }+ u_a^{i} \frac {\partial } {\partial u_a^{\#} }\;  \qquad
%\\ \nonumber
\eea
which can be used in the cases where only vector Lorentz harmonics \eqref{Uab=in10}, but not spinorial ones \eqref{harmV=10}, are present.

Similarly one can find how covariant harmonic derivatives act on the element of the inverse harmonic variables - counterparts of the blocks of inverse spinor moving frame matrix \eqref{harmV-1=10}.

The superfield of Lorentz harmonic superspace including coset $SO(1,9)/[SO(1,1)\otimes SO(8)]$ are obtained by restricting the class of superfields defined on superspace with $SO(1,9)$ group manifold by conditions of carrying definite representations of $SO(1,1)$ and $SO(8)$ group. This can be formulated by defining the result of the action of the differential operators
 $D^{(0)}$ and $D^{ij}$ on this class of superfields. Below we will find such type conditions as the result of quantization of massless superparticle mechanics.

\subsection{Curvature and variations of the composite $SU(8)$ connection}
\label{App=dmho}
The curvature of the composite $SU(8)$ connection \eqref{mho=wDw} is expressed in terms of $SO(1,9)/[SO(1,1)\otimes SO(8)]$ Cartan forms \eqref{Om--i:=},  \eqref{Om++i:=} as
\bea\label{dmho=IIB}
d{\mho}_{B}{}^{A}- {\mho}_{B}{}^{C}\wedge {\mho}_{C}{}^{A}= \frac 1 4 \Omega^{=i} \wedge \Omega^{\# j}  \gamma^{ij}{}_{B}{}^A  \; , \qquad
%\\ \nonumber
\eea
where
\bea\label{gij=SU8}
& \frac 1 4  \gamma^{ij}{}_{B}{}^A   = \frac 1 4  \bar{{\rm w}}_{qB}\gamma^{ij}_{qp} {\rm w}_p{}^A \; . \eea
This matrix, besides not
being constant, provides a representation of the $so(9)$ generator.

Eq. \eqref{dmho=IIB} can be used to obtain the expression for  admissible variations of the generalized $SU(8)$ Cartan forms,
\bea\label{vmho=IIB}
& \delta {\mho}_{B}{}^{A}= {\mho}_{B}{}^{C}\, i_\delta {\mho}_{C}{}^{A} -  i_\delta {\mho}_{B}{}^{C}\,  {\mho}_{C}{}^{A}+ (\Omega^{=i} \,  i_\delta \Omega^{\# j} - i_\delta\Omega^{=i} \,   \Omega^{\# j} )\frac 1 4  \gamma^{ij}{}_{B}{}^A \; .
%\\ \nonumber
\eea
In it $i_\delta {\mho}_{B}{}^{A}$ parametrize local $SU(8)$ transformation which is the manifest gauge symmetry of the Lagrangian 1--form \eqref{cL1=IIB-An=c}. Notice that  ${\mho}_{B}{}^{A}$ is invariant under $SO(1,1)\otimes SO(8)$ symmetry acting on the spinor frame variables, but its variations  transformed by ${\bb K}_8$ symmetry ($i_\delta \Omega^{\# j}$) and under the most essential variations of spinor frame variables ($i_\delta \Omega^{=i}$) are nontrivial.

\section{SO(8) gamma matrices}
\label{SO8gammas}

\def\theequation{B.\arabic{equation}}

\setcounter{equation}0

The exhaustive list of the useful relations  for $SO(8)$ gamma matrices

\be \gamma^i_{q\dot{p}}=:\tilde{ \gamma}{}^i_{q\dot{p}}\; , \qquad i=1,...,8\; , \qquad q,p=1,...,8\; , \qquad \dot{q}, \dot{p}=1,...,8\;   \ee
can be found in Appendix A of \cite{Green:1983hw}. Here we reproduce some of these relations for readers convenience.

First of all, they obey

\bea\label{gtg=}
\gamma^i_{q\dot{p}}\gamma^j_{p\dot{p}} & \equiv & (\gamma^i\tilde{ \gamma}{}^j)_{q{p}}= \delta^{ij} \delta_{q{p}} + \gamma^{ij}_{q{p}}\; , \qquad \\
\label{tgg=} \gamma^i_{p\dot{q}}\gamma^j_{p\dot{p}}& \equiv & (\tilde{ \gamma}{}^i{ \gamma}{}^j)_{\dot{q}\dot{p}}= \delta^{ij} \delta_{\dot{q}\dot{p}} + \tilde{\gamma}{}^{ij}_{\dot{q}\dot{p}}\; , \qquad \\ \label{gigi=}
\gamma^i_{q\dot{q}}\gamma^i_{p\dot{p}}&= &  \delta_{q{p}}\delta_{\dot{q}\dot{p}}  + \frac 1 4 \gamma^{ij}_{q{p}} \tilde{\gamma}{}^{ij}_{\dot{q}\dot{p}} =:    \delta_{q{p}}\delta_{\dot{q}\dot{p}}  +\gamma^{qp}_{\dot{q}\dot{p}}\; ,
\eea
where, following  \cite{Green:1983hw}, we introduced $\gamma^{qp}_{\dot{q}\dot{p}}= \frac 1 4 \gamma^{ij}_{q{p}} \tilde{\gamma}{}^{ij}_{\dot{q}\dot{p}} $ to make manifest that these relations respect the well known triality property of $SO(8)$ group.
The triality properties implies equivalence of the vector {\bf 8}$_v$ s-spinor  {\bf 8}$_s$ and c-spinor {\bf 8}$_c$ representations of
$SO(8)$  and imply that in any relation one can interchange $i, q$ and $\dot{q}$ indices.

The Fierz identity for the matrix with two s-spinor indices
\be
M_{qp}= \frac 1 8 \delta_{qp}{\rm tr} (M) - \frac 1 {16}  \gamma^{ij}_{q{p}}{\rm tr} ( \gamma^{ij}M) + \frac 1 {2\cdot 8\cdot 4!} \gamma^{ijkl}_{q{p}}{\rm tr} ( \gamma^{ijkl}M)\;  \qquad
\ee
uses
\be\label{gijgij=}
\delta_{q[r}\delta_{s]p}=-\frac 1 {16}  \gamma^{ij}_{qp}\gamma^{ij}_{rs} \; , \qquad
\ee
\be\label{g4g4=}
 \delta_{q(r}\delta_{s)p}= \frac 1 8 \delta_{qp}\delta_{rs} + \frac 1{384}\gamma^{ijkl}_{qp}\gamma^{ijkl}_{rs}\; , \qquad
\ee
and/or
\be\label{tijtij=}
 \text{tr} (\gamma^{ij}\gamma^{kl})=-16  \delta_{[i}{}^{k} \delta_{j]}{}^{l} \; , \qquad \text{tr} (\gamma^{ijkl}\gamma^{i'j'k'l'})= 8\,  (4!\, \delta_{[i}{}^{i'} \ldots \delta_{l]}{}^{l'}+ \epsilon^{ijkli'j'k'l'}) \; . \qquad
\ee
Notice that the last of this equation reflects self-duality of $\gamma^{ijkl}$,
\be
\gamma^{ijkl}= \frac 1 {4!} \epsilon^{ijkli'j'k'l'} \gamma^{i'j'k'l'}\qquad \Leftrightarrow  \qquad \tilde{\gamma}{}^{ijkl}=- \frac 1 {4!} \epsilon^{ijkli'j'k'l'} \tilde{\gamma}{}^{i'j'k'l'}\qquad
\ee
which then implies
\bea
\gamma^{ijkli'}= -\frac 1 {3!} \epsilon^{ijkli'j'k'l'} \gamma^{j'k'l'} \qquad & \Leftrightarrow &\qquad \tilde{\gamma}{}^{ijkli'}= +\frac 1 {3!} \epsilon^{ijkli'j'k'l'} \tilde{\gamma}{}^{j'k'l'} \qquad \\
\gamma^{ijkli'j'}= -\frac 1 {2} \epsilon^{ijkli'j'k'l'} \gamma^{k'l'}\qquad &\Leftrightarrow &\qquad \tilde{\gamma}{}^{ijkli'j'}= +\frac 1 {2} \epsilon^{ijkli'j'k'l'} \tilde{\gamma}{}^{k'l'}\; , \qquad \\
\gamma^{ijkli'j'k'}= +\epsilon^{ijkli'j'k'l'} \gamma^{l'}\qquad &\Leftrightarrow &\qquad  \tilde{\gamma}{}^{ijkli'j'k'}= -\epsilon^{ijkli'j'k'l'} \tilde{\gamma}{}^{l'}\; , \qquad \\
\gamma^{ijkli'j'k'l'}=  \epsilon^{ijkli'j'k'l'} {\bb I} \qquad &\Leftrightarrow &\qquad \tilde{\gamma}{}^{ijkli'j'k'l'}=  -\epsilon^{ijkli'j'k'l'} {\bb I} \; , \qquad
\eea

To proceed, we introduce
\bea
t^i_{qpr\dot{s}} &:=& - \gamma^{ij}_{[qp}\gamma^j_{r]\dot{s}}\; , \qquad \\ \label{t24:=}
t^{ij}_{qprs} &:=&  \gamma^{ik}_{[qp}\gamma^{jk}_{rs]} = t^{ji}_{qprs}\; , \qquad \\  \label{t44:=}
t^{ijkl}_{qprs} &:=& \gamma^{[ij}_{[qp}\gamma^{kl]}_{rs]}\; , \qquad
%\\ \nonumber
\eea
which obey
\bea\label{t24=t2*4}
 t^{ij}_{qprs} & =& \; \frac 1{4!} \epsilon^{qprsq'p'r's'}t^{ij}_{q'p'r's'} \; , \qquad \\
\label{t44=t4*4} t^{ijkl}_{qprs}  &= &-\frac 1{4!} \epsilon^{qprsq'p'r's'}t^{ijkl}_{q'p'r's'} \; , \qquad \\
\label{t44=t*44} t^{ijkl}_{qprs}  &= &-\frac 1{4!} \epsilon^{ijkli'j'k'l'}t^{i'j'k'l'}_{q'p'r's'} \; . \qquad
\eea
Furthermore \eqref{tijtij=} implies that $t^{ij}_{qprs}=t^{ji}_{qprs}$ is traceless on its v-indices
\be\label{tii4=0}
t^{ii}_{qprs}=0\; .
\ee

The signs in \eqref{t24=t2*4}--\eqref{t44=t*44} correspond to the sign plus in
\bea\label{gamma4=*}
\gamma_{qp}^{ijkl}  =\frac 1{4!} \epsilon^{ijkli'j'k'l'}\gamma_{qp}^{i'j'k'l'} \; , \qquad
\eea
which completes the description of the conventions for gamma matrices.

Notice that \eqref{t24=t2*4} and  \eqref{t44=t4*4} imply that
\be\label{t44t24=0}
t^{ijkl}_{q_1\ldots q_4}t^{i'j'}_{q_1\ldots q_4} =0\; . \qquad
\ee
As far as the other traces over four antisymmetrized blocks of SO(8) s-spinor indices are concerned, it is not difficult to find that
\bea\label{t24t24=} t^{ij}_{q_1\ldots q_4}t^{i'j'}_{q_1\ldots q_4} = 384\,  \left(\delta^{i(i'}\delta^{j')j}-\frac 1 8 \delta^{ij}\delta^{i'j'} \right) \;  ,
\qquad
\\ \nonumber {\text{and}} \\
\label{t44t44=}
t^{ijkl}_{q_1\ldots q_4}t^{i'j'k'l'}_{q_1\ldots q_4} =  128\;  \left( \delta_{i'}^{[i}\delta_{j'}^{j}\delta_{k'}^{k}  \delta_{l'}^{l]}- \frac 1 {4!}\epsilon^{ijkli'j'k'l'}\right) \; . \qquad
\eea
Some other very important equations are
\bea\label{gijgkl=t+t}
& \gamma^{ij}_{[qp}\gamma^{kl}_{rs]}= t^{ijkl}_{qprs}-\frac 2 3 \delta^{[i|[k}  t^{l]|j]}_{qprs}\; ,   \qquad
\\
% \nonumber {\text{and}} \\
\label{titj=I+g2}
& t^{i}_{q_1q_2q_3\dot{q}} t^{j}_{q_1q_2q_3\dot{p}}= 48 \delta^{ij} \delta_{\dot{q}\dot{p}}-6(\tilde{\gamma}{}^{i}\gamma^{j})_{\dot{q}\dot{p}}\; . \qquad
\eea
It is easy to check that these are   consistent with \eqref{t24:=}, \eqref{t44:=} and the following consequence of \eqref{gijgij=}:
\be \label{tigi=0} t^{i}_{q_1q_2q_3\dot{q}}\tilde{\gamma}{}^{i}_{\dot{q}p}=0 \; .\ee

More useful identities are
\bea\label{t44t131=}
t^{ijkl}_{qp_1p_2p_3}t^{i'}_{p_1p_2p_3\dot{q}} = \frac 1 3 (\gamma^{[ij|}\gamma^{j'})_{q\dot{q}}\; tr(\gamma^{|kl]}\gamma^{i'j'})- \frac 2 3 (\gamma^{[ij|}\gamma^{i'j'}\gamma^{|kl]} \gamma^{j'})_{q\dot{q}}= \nonumber
\\ =-\frac {16} 3 (\gamma^{[ijk})_{q\dot{q}}\delta^{l]i'}+ \frac {32} 3 \delta^{i'[i } \gamma^{jkl]}_{q\dot{q}} -2 \gamma^{ijkl} \gamma^{i'} \gamma^{j'k'l'}_{q\dot{q}}= \nonumber
\\ =-8 \left(\delta^{i'[i } \gamma^{jkl]} - \frac 1 {4!} \epsilon^{i'ijklj'k'l'} \gamma^{j'k'l'}\right)_{q\dot{q}}
\eea
(which is in consonance with anti-self duality of $t^{ijkl}_{p_1p_2p_3p_4}$ \eqref{t44=t*44}) and
\bea\label{t24t131=}
t^{ij}_{qp_1p_2p_3}t^{i'}_{p_1p_2p_3\dot{q}} =\frac 1 3 (\gamma^{ik}\gamma^{j'})_{q\dot{q}}\; tr(\gamma^{jk}\gamma^{i'j'})- \frac 2 3 (\gamma^{ik}\gamma^{i'j'}\gamma^{jk} \gamma^{j'})_{q\dot{q}}= \nonumber
\\ = -48 \delta^{i'(i} \gamma^{j)}_{q\dot{q}} + 6 \delta^{ij} \gamma^{i'}_{q\dot{q}}
\eea
(which is in consonance with symmetricity and tracelessness of $t^{ij}_{p_1p_2p_3p_4}$ \eqref{tii4=0}).

\setcounter{equation}0

\section{SU(8) and SO(7)}

\label{SU8-SO7}
\def\theequation{C.\arabic{equation}}

As in \cite{Bandos:2025pxv}, we will use the set of 7 complex $8\times 8$  matrices obtained by $SU(8)$ rotation (on $Spin(7)$ indices) of the     $SO(7)$ Clebsch-Gordan coefficients (generalized Pauli matrices),
\be\label{s-chI}
\sigma^{\check{I}}_{AB}= -\sigma^{\check{I}}_{BA}= \sigma^{\check{I}}_{[AB]}\;
\ee
and their complex conjugate
\be\label{ts-chI}
\tilde{\sigma}{}^{\check{I}\; AB}=-\tilde{\sigma}{}^{\check{I}\; BA}\dot{=} \tilde{\sigma}{}^{\check{I}\; [AB]}
= - (\sigma^{\check{I}}_{AB})^*=  (\sigma^{\check{I}}_{BA})^*\;
\ee
which obey

\be
\sigma^{\check{I}}_{AC}\tilde{\sigma}{}^{\check{J}\; CB} + \sigma^{\check{J}}_{AC}\tilde{\sigma}{}^{\check{I}\; CB}:= 2\sigma^{(\check{I}}_{AC}\tilde{\sigma}{}^{\check{J})\; CB}= \delta_A{}^B\; .
\ee

The complex symmetric matrix
\be\label{s-chI=sym}
{\cal U}_{AB} =\frac 1 {7!}  \epsilon^{\check{I}_1\ldots \check{I}_7}  \sigma^{\check{I}_1\ldots \check{I}_7}_{AB}\;  \qquad
\ee
is unitary,
\be\label{cUbcU=1}
  {\cal U}_{AC}\bar{{\cal U}}{}^{CB}=\delta_A{}^B\;  \qquad
\ee
and parametrizes the coset $SU(8)/SO(7)$. It is not constant but is obtained from 8$\times 8$ unity matrix (invariant tensor of $SO(8)$) by $SU(8)$ transformations (and thus represents these transformations modulo their $SO(8)$ part).
Keeping this in mind the matrix ${\cal U}_{AB} $ and its c.c.
$\bar{{\cal U}}{}^{AB}$ can be used to rise and to lower the $SU(8)$ indices, in particular

\be\label{ts=-UsbU} ({\cal U}\tilde{\sigma}^{\check{J}})_{A}{}^{B}=-  (\sigma^{\check{J}}\bar{{\cal U}})_{A}{}^{B} \qquad \Rightarrow \qquad  \tilde{\sigma}^{\check{J}AB}=   -(\bar{{\cal U}}{\sigma}^{\check{J}}\bar{{\cal U}})^{AB}\; .  \qquad
\ee
See Appendix A of  \cite{Bandos:2025pxv} for more discussion and further details.

Finally, let us write down
an explicit representation for the basic matrices using SO(8) matrices  $ \gamma^i_{p\dot{q}}=\tilde{\gamma}^i_{\dot{q}p}= (\gamma^{\check{I}}_{p\dot{q}}, \gamma^8_{p\dot{q}})$, obeying \eqref{gitgj+=I}, and
 \eqref{wbw=inSU8} as SU(8) element:
\be\label{sSO7=gtg8}
\sigma^{\check{J}}_{AB}= (\gamma^{\check{J}}\tilde{\gamma}^8)_{pq}\bar{w}_{pA}\bar{w}_{qB}\; , \qquad
\tilde{\sigma}{}^{\check{J}AB}=  ({\gamma}^8\tilde{\gamma}{}^{\check{J}})_{pq}w_p^Aw_q^B\; .
\ee

Also notice that the identity for $SO(8)$ matrices give rise to the following identities
\bea
\sigma^{\check{J}}_{AB} \sigma^{\check{J}}_{CD}+{\cal U}_{AB}{\cal U}_{CD} ={\cal U}_{AC}{\cal U}_{BD} - \frac 1 2 \sigma^{\check{J}}_{AC} \sigma^{\check{J}}_{BD}+ \frac 1 4( \sigma^{\check{J}\check{K}}{\cal U})_{AC} ( \sigma^{\check{J}\check{K}}{\cal U})_{BD}
%\\ \nonumber
\\ \Rightarrow \qquad  \sigma^{\check{J}}_{A[B} \sigma^{\check{J}}_{CD]} = - \frac 1 2( \sigma^{\check{J}\check{K}}{\cal U})_{A[B} ( \sigma^{\check{J}\check{K}}{\cal U})_{CD]}
\;  \qquad
\eea
as well as to the anti--duality relation between complex conjugate matrices
\bea\label{t44:=SO7}
t^{\check{I}\check{J}\check{K}\check{L}}_{ABCD}&=&
(\sigma^{[\check{I}\check{J}}{\cal U})_{[AB}(\sigma^{\check{K}\check{L}]}{\cal U})_{CD]} \; , \qquad
%\\ \nonumber
\\ \text{and}\nonumber
%\\ \nonumber
\\ \label{tt44:=SO7} \tilde{t}{}_{\check{I}\check{J}\check{K}\check{L}}^{ABCD}&=& (\tilde{\sigma}^{[\check{I}\check{J}|}\bar{{\cal U}})^{[AB}(\tilde{\sigma}^{|\check{K}\check{L}]}\bar{{\cal U}})^{CD]}=(t^{\check{I}\check{J}\check{K}\check{L}}_{ABCD})^*\; , \qquad
%\\ \nonumber
\eea
and duality between
\bea  \label{t24:=SO7} t^{\check{I}\check{J}}_{ABCD}&=&
(\sigma^{\check{I}\check{K}}{\cal U})_{[AB}(\sigma^{\check{J}\check{K}}{\cal U})_{CD]} + \sigma^{\check{I}}_{[AB}\sigma^{\check{J}}_{CD]} \; , \qquad
%\\ \nonumber
\\ \label{t14:=SO7} t^{\check{I}}_{ABCD}&=&
-(\sigma^{[\check{I}\check{K}}{\cal U})_{[AB}\sigma^{\check{K}]}{}_{CD]}\; , \qquad %\\ \nonumber
\\ \text{and}\nonumber
%\\ \nonumber
\\  \label{tt24:=SO7} \tilde{t}{}_{\check{I}\check{J}}^{ABCD}&=& (\tilde{\sigma}^{\check{I}\check{K}}\bar{{\cal U}})^{[AB}(\tilde{\sigma}^{\check{J}\check{K}]}\bar{{\cal U}})^{CD]}+\tilde{\sigma}^{\check{I}[AB}\tilde{\sigma}^{CD]\check{J}}=(t^{\check{I}\check{J}}_{ABCD})^*\; , \qquad
%\\ \nonumber
\\  \label{tt14:=SO7} \tilde{t}{}_{\check{I}}^{ABCD}&=&\;  (\tilde{\sigma}^{\check{I}\check{K}}\bar{{\cal U}})^{[AB}\tilde{\sigma}^{}{}^{CD]\, \check{K}}=(t^{\check{I}}_{ABCD})^*\; ,
\eea
i.e.
\bea\label{t44=-ebt44}
\tilde{t}{}_{\check{I}\check{J}\check{K}\check{L}}^{ABCD}= -\frac 1 {4!}  \epsilon^{ABCDEFGH}
t^{\check{I}\check{J}\check{K}\check{L}}_{EFGH} \; , \qquad
%\\ \nonumber
\\ \label{t24=ebt24}
\tilde{t}{}_{\check{I}\check{J}}^{ABCD}= +\frac 1 {4!}  \epsilon^{ABCDEFGH}
t^{\check{I}\check{J}}_{EFGH} \; , \qquad
%\\ \nonumber
\\ \label{t14=ebt14}
\tilde{t}{}_{\check{I}}^{ABCD}= +\frac 1 {4!}  \epsilon^{ABCDEFGH}
t^{\check{I}}_{EFGH} \; . \qquad
%\\ \nonumber
\eea

Actually, instead of \eqref{t44:=SO7} and \eqref{tt44:=SO7} it is more convenient to use their anti-duals
with three $SO(7)$ vector indices
\bea\label{t34:=SO7}
t^{\check{I}\check{J}\check{K}}_{ABCD}&=&
(\sigma^{[\check{I}\check{J}}{\cal U})_{[AB}\sigma^{\check{K}]}{}_{CD]} =  -\frac 1 {4!}  \epsilon^{\check{I}\check{J}\check{K}\check{L}\check{P}\check{Q}\check{R}}
\tilde{t}^{\check{L}\check{P}\check{Q}\check{R}}_{EFGH}
\; , \qquad
%\\ \nonumber
\\  \label{tt44=SO7=} \tilde{t}{}_{\check{I}\check{J}\check{K}}^{ABCD}&=& (\tilde{\sigma}^{[\check{I}\check{J}|}\bar{{\cal U}})^{[AB}\tilde{\sigma}^{CD]\, |\check{K}]}=(t^{\check{I}\check{J}\check{K}}_{ABCD})^*= -\frac 1 {4!}  \epsilon_{\check{I}\check{J}\check{K}\check{L}\check{P}\check{Q}\check{R}}
\tilde{t}_{\check{L}\check{P}\check{Q}\check{R}}^{ABCD}\; , \qquad
%\\ \nonumber
\eea
which are also anti-dual
\bea\label{t34=-ebt34}
\tilde{t}{}_{\check{I}\check{J}\check{K}}^{ABCD}= -\frac 1 {4!}  \epsilon^{ABCDEFGH}
t^{\check{I}\check{J}\check{K}}_{EFGH} \; . \qquad
%\\ \nonumber
\eea

With our representation \eqref{sSO7=gtg8}, the identities for $SO(8)$ gamma (actually sigma) matrices (first presented, to our best knowledge in \cite{Green:1982tk}, and collected in Appendix \ref{SO8gammas})  also imply the duality relation for contraction of sigma matrices on their 7-vector indices
\be
\tilde{\sigma}^{\check{I}[AB}\tilde{\sigma}^{CD]\check{I}}= +\frac 1 {4!}  \epsilon^{ABCDEFGH}
\sigma^{\check{J}}_{EF} \sigma^{\check{J}}_{GH}\; .
\ee
We did not include these matrices in the above list of basic elements for decomposition of antisymmetric tensors
because it appears as a trace of symmetric $SO(7)$ tensor \eqref{t24:=SO7},
\bea
t^{\check{J}\check{J}}_{ABCD}=- \sigma^{\check{J}}_{[AB} \sigma^{\check{J}}_{CD]} \; , \qquad \tilde{t}_{\check{J}\check{J}}^{ABCD}=- \tilde{\sigma}^{\check{J}[AB} \tilde{\sigma}^{CD]\check{J}} \; . \qquad
%\\ \nonumber
\eea

\section{Some technical details for on spinor moving frame
derivation of $3$-point type IIB superamplitude and on related issues relevant to discussion of type IIA}

\def\theequation{D.\arabic{equation}}

\label{app:3-point=IIB}

\subsection{Some properties of $\hat{K}_{q[\dot{q}|}$ as well as of $ {\rm s}_{q_1\ldots q_4}$ \eqref{sq4=} and its cousins }

Notice the useful relations  involving the unitary matrix
\eqref{hK=sbs}
\bea
\hat{K}_{q_1[\dot{q}_1|} \ldots \hat{K}_{q_4|\dot{q}_4]}= 2\, 4!\,  {\rm s}_{q_1\ldots q_4} \bar{\tilde{{\rm s}}}_{\dot{q}_1 \ldots \dot{q}_4}
\eea
and
\bea
\hat{K}_{q[\dot{q}|} \hat{K}_{p|\dot{p}}:= 4{\rm s}_{q}{}^{\tilde{A}} \bar{\tilde{{\rm s}}}_{[\dot{q}|\tilde{A}}  {\rm s}_{p}{}^{\tilde{B}} \bar{\tilde{{\rm s}}}_{|\dot{p}]\tilde{B}} =  \frac 1 {16}
\hat{K}_{q} \tilde{\gamma}{}^{ij}\tilde{\hat{K}}_{p}   \tilde{\gamma}{}^{kl}_{\dot{q}\dot{p}}   =  \frac 1 {16}
\tilde{\hat{K}}_{\dot{q}} {\gamma}{}^{ij}\hat{K}_{\dot{p}}  {\gamma}{}^{ij}_{qp} =\qquad \nonumber \\ \nonumber \\ =\frac 1 {256}
{\rm tr} \left(\tilde{\hat{K}}{\gamma}{}^{ij}\hat{K}\tilde{\gamma}{}^{kl}\right)  {\gamma}{}^{ij}_{qp}  \tilde{\gamma}{}^{kl}_{\dot{q}\dot{p}}=
\frac 1 {256}
{\rm tr} \left(\tilde{\hat{K}}{\gamma}{}^{ij}\hat{K}\tilde{\gamma}{}^{ij}\right)  {\gamma}{}^{ij}_{qp}  {\gamma}{}^{kl}_{rs}  {\gamma}{}^{i'}_{r\dot{q}} {\gamma}{}^{i'}_{s\dot{p}}\; . \qquad
%\\ \nonumber
\eea
These equations can be considered as an evidence (but not the proof) of the existence of relations

%\bea \hat{K}_{q[\dot{q}|} \hat{K}_{p|\dot{p}}=
%\propto {\rm s}_{qprs}  {\gamma}{}^{i}_{r\dot{q}} {\gamma}{}^{i}_{s\dot{p}}
%= \propto  \bar{\tilde{{\rm s}}}_{\dot{q}\dot{p}\dot{r}\dot{s}}  {\gamma}{}^{i}_{q\dot{r}} {\gamma}{}^{i}_{p\dot{s}}\;  \qquad
%\\ \nonumber\eea
%which imply (after contracting with $ {\gamma}{}^{i}_{r\dot{q}} {\gamma}{}^{i}_{s\dot{p}}$)

\bea\label{KKtij=sqprs}
-\hat{K}^{i} K^{j}t^{ij}_{qprs}= \hat{K}_{[q|\dot{q}}\hat{K}_{|p|\dot{p}}
 {\gamma}{}^{i}_{|r|\dot{q}} {\gamma}{}^{i}_{|s]\dot{p}}
= 4 s_{[q}^{\tilde{A}} s_{p}^{\tilde{B}}{\gamma}{}^{i}_{r|\dot{q}}\bar{\tilde{s}}_{\dot{q}\tilde{A}}{\gamma}{}^{i}_{|s]\dot{p}}\bar{\tilde{s}}_{\dot{p}\tilde{B}}=\qquad
\nonumber
%\\ \nonumber
\\  = 8\epsilon_{\tilde{A}\tilde{B}\tilde{C}\tilde{D}} {\rm s}_q^{\tilde{A}}{\rm s}_p^{\tilde{B}}{\rm s}_r^{\tilde{C}}{\rm s}_s^{\tilde{D}}
=8\, 4!\,  {\rm s}_{qprs}\; .
%\\  \nonumber
\eea
Such equation was presented (in a bit different notation) and essentially used in \cite{Boels:2012ie}
(see Eqs. (4.27)--(4.30) of this paper). We have used it  for relating amplitudes and superamplitudes in our approach.
Eqs. \eqref{KKtij=sqprs} can be obtained from
\be
{\gamma}{}^{i}_{[q|\dot{q}}\bar{\tilde{s}}_{\dot{q}\tilde{A}}{\gamma}{}^{i}_{|p]\dot{p}}\bar{\tilde{s}}_{\dot{p}\tilde{B}}=
2\epsilon_{\tilde{A}\tilde{B}\tilde{C}\tilde{D}}{\rm s}_q^{\tilde{C}}{\rm s}_p^{\tilde{D}}\qquad \Longleftrightarrow \qquad {\rm s}_q^{\tilde{A}}{\gamma}{}^{i}_{q[\dot{q}|}
{\rm s}_q^{\tilde{B}}{\gamma}{}^{i}_{p|\dot{p}]}= 2\epsilon^{\tilde{A}\tilde{B}\tilde{C}\tilde{D}}
\bar{\tilde{s}}_{\dot{q}\tilde{C}}\bar{\tilde{s}}_{\dot{p}\tilde{D}} \; . \qquad
\ee
However, the consistency of imposing such conditions on our
$\bar{\tilde{\rm s}}$ variables has to be proven.

The 4-th rank covariant tensors with indices of fundamental and anti-fundamental representations of SU(8) associated  to
\eqref{tsdq4=} and \eqref{sq4=}
are
\bea\label{tsA4=}
 \tilde{{\rm s}}{}^{A_1\ldots A_4} =  \tilde{{\rm s}}_{\dot{q}_1\ldots \dot{q}_4} w^{A_1}_{\dot{q}_1}\ldots w^{A_4}_{\dot{q}_4}  =  \, \frac 1 {4!}\,
\epsilon_{\tilde{B}_1 \ldots \tilde{B}_4}   \tilde{{\rm s}}_{A_1}{}^{\tilde{B}_1}\ldots  \tilde{{\rm s}}_{A_4}{}^{\tilde{B}_4}= (\bar{\tilde{{\rm s}}}_{A_1\ldots A_4})^*\; , \qquad
%\\ \nonumber
\\ \label{sA4=}
{{\rm s}}^{A_1\ldots A_4}  =  {{\rm s}}_{{q}_1\ldots {q}_4} w^{A_1}_{\dot{q}_1}\ldots w^{A_4}_{\dot{q}_4} =
 \, \frac 1 {4!}\, \epsilon_{\tilde{B}_1 \ldots \tilde{B}_4}   {{\rm s}}_{A_1}{}^{\tilde{B}_1}\ldots {{\rm s}}_{A_4}{}^{\tilde{B}_4}= (\bar{{{\rm s}}}_{A_1\ldots A_4})^*\; . \qquad
\eea

%\bea \hat{K}_{q[\dot{q}|} \hat{K}_{p|\dot{p}}=
%\propto {\rm s}_{qprs}  {\gamma}{}^{i}_{r\dot{q}} {\gamma}{}^{i}_{s\dot{p}}
%= \propto  \bar{\tilde{{\rm s}}}_{\dot{q}\dot{p}\dot{r}\dot{s}}  {\gamma}{}^{i}_{q\dot{r}} {\gamma}{}^{i}_{p\dot{s}}\;  \qquad
%\\ \nonumber\eea
%which imply (after contracting with $ {\gamma}{}^{i}_{r\dot{q}} {\gamma}{}^{i}_{s\dot{p}}$)

\subsection{On  derivation of some amplitudes and of superamplitudes}

Some details of the  derivation of the amplitude \eqref{Apcip==} are
\bea\label{Apcip=}
A (\bar{\phi}c^i \phi)= p_{{\rm 1}\mu}u^{\mu i}_{{\rm 2}}= \frac 1 8 \rho^{\#}_{{\rm 1}} v_{\dot{q}{\rm 1}}^{-\alpha}  v_{\dot{q}{\rm 1}}^{-\beta }\, {\sigma}{}^{\mu}_{ \alpha \beta}u{_\mu 2}^i= \frac 1 4 \rho^{\#}_{{\rm 1}}
v_{\dot{q}{\rm 1}}^{-\alpha}  v_{\dot{q}{\rm 1}}^{-\beta }\, v_{(\alpha|{q}{\rm 2}}^{\; -}\gamma^i_{q\dot{p}}  v_{|\beta )\dot{p}{\rm 2}}^{\; +}\,
 = \frac 1 4 \rho^{\#}_{{\rm 1}} <\, {\rm 2}^-_q\, |\, {\rm 1}^-_{\dot{p}}\, > \gamma^i_{q\dot{p}} = \qquad \nonumber % \\ \nonumber
\\ = -\frac 1 4 \rho^{\#}_{{\rm 1}}\, |\vec{K}^=_{{\rm 21}}|\, \beta_{{\rm 21}}{\rm s}_q{}^{\tilde{A}}\bar{\tilde{{\rm s}}}_{\dot{p}\tilde{A}}\gamma^i_{q\dot{p}}= -\frac 1 4 \rho^{\#}_{{\rm 1}} \, |\vec{K}^=_{{\rm 21}}|\,  \beta_{{\rm 21}}{\rm s}{}^{A\tilde{C}}\bar{\tilde{{\rm s}}}{}^B{}_{\tilde{C}}\gamma^i_{AB}\; . \qquad
\eea
To obtain this result, the Eqs.  \eqref{p=rv-sv-}, \eqref{uIs=v+v-}, \eqref{bv-iAv-jB=}, \eqref{i-qj-dp=sts},  \eqref{i-Bj-A=sts} and \eqref{v-qv+p=} have been used.

To make our expression closer to the one written
(for $D=8$) in \cite{Boels:2012ie} we should consider
$A (\bar{\phi}c_{q\dot{p}} \phi)=A (\bar{\phi}c^i \phi) \gamma^i_{q\dot{p}}$. The calculations within spinor moving frame approach on the line described above give

\bea\label{Apcqdpp=}
A (\bar{\phi}c^i \phi) \gamma^i_{q\dot{p}} &=& p_{{\rm 1}\mu}u^{\mu i}_{{\rm 2}}= -p_{{\rm 1}\mu} {\sigma}{}^{\mu}_{\alpha \beta} v_{\dot{p}{\rm 2}}^{-\alpha} v_{{q}{\rm 2}}^{+\beta} = -2 \rho^{\#}_{{\rm 1}}  <\, {\rm 1}^-_q\, |\, {\rm 2}^-_{\dot{p}}\, >
 = -2 \rho^{\#}_{{\rm 1}}\, |\vec{K}^=_{{\rm 21}}|\, \beta_{{\rm 21}}{\rm s}_q{}^{\tilde{A}}\bar{\tilde{{\rm s}}}_{\dot{p}\tilde{A}}\; , \qquad
 \\ \nonumber \\
 A (\bar{\phi}c^i \phi) \gamma^i_{AB}&=&
  -2 \rho^{\#}_{{\rm 1}} \, |\vec{K}^=_{{\rm 21}}|\,  \beta_{{\rm 21}}{\rm s}{}_A{}^{\tilde{C}}\bar{\tilde{{\rm s}}}{}_{B\tilde{C}}\; .
\eea
It is easy to check that, tracing these  equations with
$\gamma^i_{q\dot{p}}$ and $\tilde{\bar{\gamma}}^{AB}$, respectively, we obtain  \eqref{Apcip=}.
Eq. \eqref{Apcqdpp=} is close, although not identical to the one in \cite{Boels:2012ie}.

The details on derivation of the final form of Eq.
\eqref{Aphijp==p1u2p3u2} are
\bea\label{Aphijp=p1u2p3u2}
A (\bar{\phi}h^{ij} \phi)&=& p_{{\rm 1}\mu}u^{\mu i}_{{\rm 2}} p_{{\rm 3}\nu}u^{\nu j}_{{\rm 2}} =\rho^{\#}_{{\rm 1}} \rho^{\#}_{{\rm 3}} u^=_{{\rm 1}\mu}u^{\mu i}_{{\rm 2}}  u^=_{{\rm 3}\nu}u^{\nu j}_{{\rm 2}}\nonumber
%\\ \nonumber
\\ &=&  \rho^{\#}_{{\rm 1}} \rho^{\#}_{{\rm 3}} K^{=i}_{{\rm 21}} K^{=j}_{{\rm 23}}=-(\rho^{\#}_{{\rm 1}})^2(\rho^{\#}_{{\rm 3}})^2\; K^{=i}_{{\rm 31}} K^{=j}_{{\rm 31}}/(\rho^{\#}_{{\rm 2}})^2 =- (\rho^{\#}_{{\rm 1}})^2 K^{=i}_{{\rm 21}} K^{=j}_{{\rm 21}} \; .
%\\ \nonumber
\eea

One can immediately check that the {\it r.h.s. } of this expression is traceless on its $SO(8)$ v-indices, as the {\it l.h.s.} should be.

\subsection{Details of derivation of \eqref{D4Q12==} }
\label{app:3-point=IIB-2}
%\ref{app:3-point=IIB-2}

\bea
2^{-24}\left((D^+_{{\rm 1}})^{\wedge 8}D^+_{{\rm 2}A}D^+_{{\rm 2}B}D^+_{{\rm 2}C}D^+_{{\rm 2}D} {\bb Q}^{12}\right)\vert_{\Theta^-_{{\rm i}}=0}
=\frac 1 {8!} (\rho^{\#}_{{\rm 1}})^8  (\rho^{\#}_{{\rm 2}})^4 \tilde{s}{}^{\alpha_1\ldots  \alpha_{12}} \bar{v}{}^-_{\alpha_{1}A {\rm 2}}\ldots \bar{v}{}^-_{\alpha_{4}D {\rm 2}}
\epsilon^{B_1\ldots B_8} \bar{v}{}^-_{\alpha_{5}B_1 {\rm 1}}\ldots \bar{v}{}^-_{\alpha_{12}B_8 {\rm 1}} \qquad
%\\ \nonumber
\\ = \frac 1 {8!} (\rho^{\#}_{{\rm 1}})^8  (\rho^{\#}_{{\rm 2}})^4 \tilde{s}{}^{\alpha_1\ldots  \alpha_{12}}\bar{w}_{q_1A}\ldots
\bar{w}_{q_4D}{v}{}^-_{\alpha_{1}q_1 {\rm 2}}\ldots {v}{}^-_{\alpha_{4}q_4 {\rm 2}}
\epsilon_{p_1\ldots p_8} {v}{}^-_{\alpha_{5}p_1 {\rm 1}}\ldots {v}{}^-_{\alpha_{12}p_8 {\rm 1}} \qquad
%\\ \nonumber
\\ \label{l3} =  \frac { (\rho^{\#}_{{\rm 1}})^8  (\rho^{\#}_{{\rm 2}})^4}{(\rho^{\#})^{2}} \bar{w}_{q_1A}\ldots
\bar{w}_{q_4D}
 \tilde{{\rm s}}{}_{\dot{q}_1 \ldots \dot{q}_4}
\epsilon_{\dot{q}_1 \ldots \dot{q}_4\dot{p}_1 \ldots \dot{p}_4} {v}{}^{-\alpha_{1}}_{\dot{p}_1 {\rm 1}}\ldots {v}{}^{-\alpha_{4}}_{\dot{p}_4 {\rm 1}}
{v}{}^-_{\alpha_{1}q_1 {\rm 2}}\ldots {v}{}^-_{\alpha_{4}q_4 {\rm 2}}  \qquad
%\\ \nonumber
\\
= \frac{(\rho^{\#}_{{\rm 1}})^8  (\rho^{\#}_{{\rm 2}})^4}{(\rho^{\#})^{2}}
 \tilde{{\rm s}}{}_{\dot{q}_1 \ldots \dot{q}_4}
\epsilon_{\dot{q}_1 \ldots \dot{q}_4\dot{p}_1 \ldots \dot{p}_4} \bar{w}_{q_1A} <\, 2^-_{q_1}\, |\, 1^{-}_{\dot{p}_1}> \, \ldots
\bar{w}_{q_4D} \,<\, 2^-_{q_4}\, |\, 1^{-}_{\dot{p}_4}> \qquad %\\ \nonumber
\\
=\frac { (\rho^{\#}_{{\rm 1}})^8  (\rho^{\#}_{{\rm 2}})^4}{16 (\rho^{\#})^{2}}
 \tilde{{\rm s}}{}_{\dot{q}_1 \ldots \dot{q}_4}
\epsilon_{\dot{q}_1 \ldots \dot{q}_4\dot{p}_1 \ldots \dot{p}_4} \bar{w}_{q_1A} K^{= k_1}_{{\rm 21}} \gamma^{k_1}_{q_1\dot{p}_1} \, \ldots
\bar{w}_{q_4D} \,
K^{= k_4}_{{\rm 21}} \gamma^{k_4}_{q_4\dot{p}_4} \,\qquad
%\\ \nonumber
%\\
%=\frac { (\rho^{\#}_{{\rm 1}})^8  (\rho^{\#}_{{\rm 2}})^4}{ (\rho^{\#})^{2}}  \;  |\vec{K}^=_{{\rm 21}}|^4\;
% \tilde{{\rm s}}{}_{\dot{q}_1 \ldots \dot{q}_4}
%\epsilon_{\dot{q}_1 \ldots \dot{q}_4\dot{p}_1 \ldots \dot{p}_4} \bar{w}_{q_1A}  {{\rm s}}{}_{q_1}{}^{\tilde{A_1}} \bar{\tilde{{\rm s}}}{}_{\dot{p}_1\tilde{A_1}} \, \ldots
%\bar{w}_{q_4D} \,
 % {{\rm s}}{}_{q_4}{}^{\tilde{A_4}} \bar{\tilde{{\rm s}}}{}_{\dot{p}_4\tilde{A_4}}  \,\qquad
%\\ \nonumber
\\
=\frac { (\rho^{\#}_{{\rm 1}})^8  (\rho^{\#}_{{\rm 2}})^4}{ (\rho^{\#})^{2}}  \;  |\vec{K}^=_{{\rm 21}}|^4\;
 \tilde{{\rm s}}{}_{\dot{q}_1 \ldots \dot{q}_4}
\epsilon_{\dot{q}_1 \ldots \dot{q}_4\dot{p}_1 \ldots \dot{p}_4} \bar{w}_{q_1A}   {{\rm s}}{}_{q_1}{}^{\tilde{A_1}} \bar{\tilde{{\rm s}}}{}_{\dot{p}_1\tilde{A_1}} \, \ldots
\bar{w}_{q_4D} \,
  {{\rm s}}{}_{q_4}{}^{\tilde{A_4}} \bar{\tilde{{\rm s}}}{}_{\dot{p}_4\tilde{A_4}}  \,\qquad
%\\ \nonumber
\\
=\frac { (\rho^{\#}_{{\rm 1}})^8  (\rho^{\#}_{{\rm 2}})^4}{ (\rho^{\#})^{2}}  \;  |\vec{K}^=_{{\rm 12}}|^4\;
\tilde{{\rm s}}{}_{\dot{q}_1 \ldots \dot{q}_4}
\epsilon_{\dot{q}_1 \ldots \dot{q}_4\dot{p}_1 \ldots \dot{p}_4} {\rm s}_{A}{}^{\tilde{A_1}} \bar{\tilde{{\rm s}}}{}_{\dot{p}_1\tilde{A_1}} \, \ldots
\,
 {\tilde{{\rm s}}}{}_{q_4}{}^{\tilde{A_4}} \bar{\tilde{{\rm s}}}{}_{\dot{p}_4\tilde{A_4}}  \,\qquad
%\\ \nonumber
\\
=\frac { (\rho^{\#}_{{\rm 1}})^8  (\rho^{\#}_{{\rm 2}})^4}{ (\rho^{\#})^{2}} \;  |\vec{K}^=_{{\rm 21}}|^4\;
\epsilon_{\tilde{A_1} \ldots \tilde{A_4}} \bar{s}_{A}{}^{\tilde{A_1}} \ldots \bar{s}_{D}{}^{\tilde{A_4}}
 \epsilon_{\dot{q}_1 \ldots \dot{q}_4\dot{p}_1 \ldots \dot{p}_4}
 {\tilde{{\rm s}}}{}_{\dot{q}_1 \ldots \dot{q}_4} \bar{\tilde{{\rm s}}}{}_{\dot{p}_1 \ldots \dot{p}_4}  \,\qquad
%\\  \nonumber
\\
=\frac { (\rho^{\#}_{{\rm 1}})^8  (\rho^{\#}_{{\rm 2}})^4}{ (\rho^{\#})^{2}} \;  |\vec{K}^=_{{\rm 21}}|^4\;
\epsilon_{\tilde{A_1} \ldots \tilde{A_4}} {\rm s}_{A}{}^{\tilde{A_1}} \ldots {\rm s}_{D}{}^{\tilde{A_4}}
 = 4!\, \frac { (\rho^{\#}_{{\rm 1}})^8  (\rho^{\#}_{{\rm 2}})^4}{ (\rho^{\#})^{2}} \;  |\vec{K}^=_{{\rm 21}}|^4\;  {\rm s}_{_{ABCD}} \,\qquad
%\\ \nonumber
\\ \label{l=pre-l}
=4! \,\frac { (\rho^{\#}_{{\rm 1}})^8  (\rho^{\#}_{{\rm 2}})^4}{ (\rho^{\#})^{2}} \;  |\vec{K}^=_{{\rm 21}}|^4\;  \bar{w}_{q_1A}\ldots \bar{w}_{q_4D}
{\rm s}_{q_1q_2q_3q_4}
  \;  \qquad
%\\ \nonumber
\\  \label{=App4p} =\frac {(\rho^{\#}_{{\rm 1}})^2(\rho^{\#}_{{\rm 2}})^2 } {8 \, (\rho^{\#})^{2} }\,   |\vec{K}^=_{{\rm 21}}|^{2}\;  A(\bar{\phi}^{[+8]}\phi_{_{ABCD}}^{[+4]}\phi) =\frac {(\rho^{\#}_{{\rm 1}})^2(\rho^{\#}_{{\rm 3}})^2 } {8 \, (\rho^{\#})^{2} }\,   |\vec{K}^=_{{\rm 31}}|^{2}\;  A(\bar{\phi}^{[+8]}\phi_{_{ABCD}}^{[+4]}\phi)\; . \qquad
%\\ \nonumber
\eea

To obtain the third line
of this series of equations, \eqref{l3}, we have used
the identity
\bea
\tilde{s}^{\alpha_1\ldots  \alpha_{12}}\epsilon^{B_1\ldots B_8} \bar{v}{}^-_{\alpha_{5}B_1 {\rm 1}}\ldots \bar{v}{}^-_{\alpha_{12}B_8 {\rm 1}} = 8!\,(\rho^\#)^{-2} \epsilon_{A_1\ldots A_8}{v}{}^{-\alpha_{1}A_1}_{{\rm 1}}\ldots {v}{}^{-\alpha_{4}A_4}_{ {\rm 1}}\, s^{A_5\ldots A_8} \; , \qquad \\  \Leftrightarrow \qquad \tilde{s}^{\alpha_1\ldots  \alpha_{12}}\epsilon^{q_1\ldots q_8} {v}{}^-_{\alpha_{5}q_1 {\rm 1}}\ldots {v}{}^-_{\alpha_{12}q_8 {\rm 1}} = 8!\,(\rho^\#)^{-2} \epsilon_{\dot{p}_1\ldots \dot{p}_8}\bar{v}{}^{-\alpha_{1}}_{\dot{p}_1{\rm 1}}\ldots \bar{v}{}^{-\alpha_{4}}_{\dot{p}_4{\rm 1}}\, \tilde{s}_{\dot{p}_5\ldots \dot{p}_8} \; ,
\eea
where $s^{\alpha_1\ldots  \alpha_{12}}$ is defined in \eqref{s1-12=}.
%$$
%s^{\alpha_1\ldots  \alpha_{12}}
%\epsilon_{p_1, \ldots p_8}
%{v}{}^-_{\alpha_{5}p_1 }\ldots {v}{}^-_{\alpha_{12}p_8 }
%= s_{\dot{q}_1 \ldots \dot{q}_4}
%\epsilon_{\dot{q}_1 \ldots \dot{q}_4, %\dot{p}_1 \ldots \dot{p}_4}v_{\dot{p}_1}^{-%\alpha_1}\ldots  v_{\dot{p}_4}^{-\alpha_{4}}
%$$
This is the consequence of
\bea\label{ev-8=v-1-8}
&& \epsilon^{\alpha_1\ldots  \alpha_{16}}
{v}{}^-_{\alpha_{9}q_1 }\ldots {v}{}^-_{\alpha_{16}q_8 } \epsilon_{\dot{q}_1\ldots \dot{q}_8}=8!\,\epsilon_{{q}_1\ldots {q}_8}v_{\dot{q}_1}^{-\alpha_1}\ldots  v_{\dot{q}_1}^{\alpha_{8}}\qquad
\nonumber
%\\ \nonumber
\\ && \Leftrightarrow \qquad
\epsilon_{{q}_1\ldots {q}_8} \epsilon^{\alpha_1\ldots  \alpha_{16}}
{v}{}^-_{\alpha_{9}q_1 }\ldots {v}{}^-_{\alpha_{16}q_8 } = 8!\,\epsilon_{{q}_1\ldots {q}_8} v_{\dot{q}_1}^{-\alpha_1}\ldots  v_{\dot{q}_8}^{\alpha_{8}}\qquad  \nonumber
\\ && \Leftrightarrow \qquad
\epsilon^{{B}_1\ldots {B}_8} \epsilon^{\alpha_1\ldots  \alpha_{16}}
\bar{v}{}^-_{\alpha_{9}B_1 }\ldots \bar{v}{}^-_{\alpha_{16}B_8 } = 8!\,\epsilon_{{B}_1\ldots {B}_8} v^{-\alpha_1B_1}\ldots  v^{-\alpha_{8}B_8}\qquad
%\\ \nonumber
\eea
which, in its turn, follows from
%\footnote{Similar relation is presented and used
%in \cite{Boels:2012ie}, without use of spinor moving frame method of \cite{Bandos:1992np,Bandos:1992ze,Bandos:1996ju,Bandos:2007mi,Bandos:2007wm}.}
\be\label{detv-v+=1}
{\rm det} (v_{\alpha q}^{\; -}, v_{\alpha \dot{q}}^{\; +})=1 \qquad \Leftrightarrow \qquad
\epsilon^{\alpha_1\ldots  \alpha_{16}} v_{\alpha_1 q_1}^{\; -} \ldots v_{\alpha_8 q_8}^{\; -} v_{\alpha_{9} \dot{q}_1}^{\; +}\ldots v_{\alpha_{16} \dot{q}_8}^{\; +}= \epsilon_{{q}_1\ldots {q}_8} \epsilon_{\dot{q}_1\ldots \dot{q}_8}\; .
\ee
Line \eqref{l=pre-l} is obtained by using the identity
\be \label{eptsbts=1} \epsilon_{\dot{q}_1 \ldots \dot{q}_4\dot{p}_1 \ldots \dot{p}_4}
 {\tilde{{\rm s}}}{}_{\dot{q}_1 \ldots \dot{q}_4} \bar{\tilde{{\rm s}}}{}_{\dot{p}_1 \ldots \dot{p}_4}=1\; ,
\ee
and passing to the last line we use the definition \eqref{sA4=}.

Finally the last form, \eqref{=App4p},  is obtained by using \eqref{App4p=}.
%\newpage

\subsection{Some technicalities on type IIA scattering description with T-duality directions and Cartan forms of spinor moving frame of scattering particles }

\label{App:kij=kj}

The condition \eqref{k=kijuij} is equivalent to
\be\label{k=kijuij-}
k_\mu u^{\mu =}_{{\rm i}}=0= k_\mu u^{\mu \# }_{{\rm i}}\qquad \forall\,  {\rm i} \; .
\ee
Considering these equations
in the gauge \eqref{u--=Kiui}--\eqref{u++=u0} we find that it results in the following
conditions for contractions of $k_\mu$ with vectors from reference frame
\bea\label{k++=0}
k^\# = k_\mu u^{\mu \# }=0 \; , \qquad \\
\label{k--=-kjKij--}
{\bb K}_{{\rm i}}^{=j} k^j +k^= :=
k_\mu u^{\mu j }{\bb K}_{{\rm i}}^{=j} +k_\mu u^{\mu =} =0 \; . \qquad
\eea
Eq. \eqref{k++=0} results in identification of all $k^j_{{\rm i}}= k_\mu  u^{\mu j}_{{\rm i}}$ with $k^j= k_\mu u^{\mu j}$
\be
k^j_{{\rm i}}=k^j:= k_\mu u^{\mu j}\; , \qquad \forall \; {\rm i}\; . \qquad
\ee
Furthermore, \eqref{k--=-kjKij--} implies
\be\label{ki'Ki'-Ki=0}
\vec{k} {\vec{{\bb K}}}_{{\rm i}'{\rm i}}:=\vec{k} ({\vec{{\bb K}}}_{{\rm i}'}-{\vec{{\bb K}}}_{{\rm i}})=0
\ee
for any ${\rm i}$ and ${\rm i}'$.

In the gauge
\eqref{u--=Kiui}--\eqref{u++=u0}, the Cartan form of the reference spinor moving frame are related by
\bea\label{Om--i=gauge}
\Omega^{=j}_{{\rm i}}&=& \Omega^{=j} +D {\bb K}^{=j}_{{\rm i}} -\frac 1 2
 {\bb K}^{=j}_{{\rm i}}\,
\;  \Omega^{\# j'}
 {\bb K}^{=j'}_{{\rm i}}
 -\frac 1 4
 \Omega^{\# j} \,
 |\vec{{\bb K}}{}^{=}_{{\rm i}}|^2
\; , \qquad
 \\
\label{Om++i=gauge} \Omega^{\#j}_{{\rm i}}&=&\Omega^{\#j}\; , \qquad  \\ \label{Om0=gauge} \Omega^{(0)}_{{\rm i}}&=&\Omega^{(0)}+ \frac 1 4  {\bb K}^{=j}_{{\rm i}} \Omega^{\# j}=0\; ,   \qquad\\
\label{Omij=gauge} \Omega^{jk}_{{\rm i}}&=&\Omega^{jk}+ {\bb K}^{=[j}_{{\rm i}} \Omega^{\# k]}  \; , \qquad
\eea
where
\bea
D {\bb K}^{=j}_{{\rm i}}=d {\bb K}^{=j}_{{\rm i}} -2
 {\bb K}^{=j}_{{\rm i}}\,
\;  \Omega^{(0)} +
 {\bb K}^{=j'}_{{\rm i}}
\;  \Omega^{j'j}\; \qquad
\eea
is covariant derivative with connection composed from reference spinor moving frame.

\section{Spinor moving frame approach to three point type IIA amplitudes involving D$0$--brane }

\label{3=00m}

\def\theequation{E.\arabic{equation}}

\subsection{Kinematics of three particle type IIA process involving D$0$-brane}

Using the gauge fixing decomposition of  spinor moving frame variables of massless superparticles on the reference  spinor moving frame, Eqs.
\eqref{v-i=v-+Kigv+},  and also its consequence \eqref{u--=Kiui},  we can split the above momentum conservation condition \eqref{u0=u1+u2} on three equations. These appear as coefficients
for $v_{\alpha q}^-v_{\beta p}^-$, $v_{(\alpha| q}^-v_{|\beta ) \dot{p}}^+ \tilde{\gamma}{}^i_{q\dot{p}}$ and $v_{(\alpha|\dot{q}}^+v_{|\beta ) \dot{q}}^+$
in the contraction of Eq. \eqref{u0=u1+u2} with  $\sigma^\mu_{\alpha\beta}$ and read
\bea\label{(l1+l2)2=}
&& (l_{{\rm 1}q}^{+\underline{q}}+l_{{\rm 2}q}^{+\underline{q}})\, (l_{{\rm 1}p}^{+\underline{q}}+l_{{\rm 2}p}^{+\underline{q}})=2 (l_{{\rm 1}}^{\#}+l_{{\rm 2}}^{\#})
\delta_{qp} \; ,
%\\ \nonumber
\\ \label{Kl+l+1+2+12=}
&& K^{=i}_{{\rm 1}} \, (l_{{\rm 1}q}^{+\underline{q}}+l_{{\rm 2}q}^{+\underline{q}})\, l_{{\rm 1}p}^{+\underline{q}} + K^{=i}_{{\rm 2}} \, (l_{{\rm 1}q}^{+\underline{q}}+l_{{\rm 2}q}^{+\underline{q}})\, l_{{\rm 2}p}^{+\underline{q}}=
2 (l_{{\rm 1}}^{\#}K^{=i}_{{\rm 1}}+l_{{\rm 2}}^{\#} K^{=i}_{{\rm 2}})
\delta_{qp}
 \; ,
 %\\ \nonumber
 \\ \label{ggKK=}
&& \gamma^{i}_{q(\dot{q}} \tilde{\gamma}^{j}_{\dot{p})p} \left(K^{=i}_{{\rm 1}} \, K^{=j}_{{\rm 1}} \, l_{{\rm 1}q}^{+\underline{q}}\, l_{{\rm 1}p}^{+\underline{q}}+ K^{=i}_{{\rm 2}} \, K^{=j}_{{\rm 2}} \, l_{{\rm 2}q}^{+\underline{q}}\, l_{{\rm 2}p}^{+\underline{q}}+ 2K^{=i}_{{\rm 1}} \, K^{=j}_{{\rm 2}} \, l_{{\rm 1}q}^{+\underline{q}}\, l_{{\rm 2}p}^{+\underline{q}}\right)=2 (l_{{\rm 1}}^{\#}(\vec{K}^{=}_{{\rm 1}})^2+l_{{\rm 2}}^{\#} (\vec{K}^{=}_{{\rm 2}})^2)\delta_{qp}\;.
 %\\ \nonumber
 \eea
(To derive these equations we have used the  constraints \eqref{D0:u0s=vv} and  \eqref{u--s=v-v-}).

As $K^{=i}_{{\rm 1}}$ and $K^{=i}_{{\rm 2}} $ are not restricted by momentum conservation condition \eqref{u0=u1+u2},  Eq. \eqref{Kl+l+1+2+12=} implies
\bea\label{l1l1+l1l2=}
&& (l_{{\rm 1}q}^{+\underline{q}}+l_{{\rm 2}q}^{+\underline{q}})\, l_{{\rm 1}p}^{+\underline{q}}=2 l_{{\rm 1}}^{\#}
\delta_{qp} \; ,
 \qquad  \nonumber
 %\\ && \nonumber
 \\ && (l_{{\rm 1}q}^{+\underline{q}}+l_{{\rm 2}q}^{+\underline{q}})\,l_{{\rm 2}p}^{+\underline{q}})=2 l_{{\rm 2}}^{\#}
\delta_{qp} \; . \qquad
%\\ \nonumber
  \eea
The sum of this equation reproduces \eqref{(l1+l2)2=} which is hence dependent.
Furthermore any of these equations can be used to find $l_{{\rm 1}[q}^{+\underline{q}}\,l_{{\rm 2}p]}^{+\underline{q}}=0$. Hence, as the basis of symmetric $8\times 8$ matrices is given by $\delta_{qp}$ and $\gamma^{i_1i_2i_3i_4}_{qp}$,
we conclude that
\bea\label{l1l2=}
l_{{\rm 1}q}^{+\underline{q}}\,l_{{\rm 2}p}^{+\underline{q}}=2 l_{{\rm 12}}^{\#}\delta_{qp}+\frac 1 {4!} l^{i_1i_2i_3i_4\#}\gamma^{i_1i_2i_3i_4}_{qp}
\eea
with some $l_{{\rm 12}}^{\#}$ and $l^{i_1i_2i_3i_4\#} =l^{[i_1i_2i_3i_4]\#}$.
Then Eqs. \eqref{l1l1+l1l2=} can be written in the form of
\bea\label{l1l1=}
&& l_{{\rm 1}q}^{+\underline{q}}\,l_{{\rm 1}p}^{+\underline{q}}=2  (l_{{\rm 1}}^{\#}-l_{{\rm 12}}^{\#})\delta_{qp}-\frac 1 {4!} l^{i_1i_2i_3i_4\#}\gamma^{i_1i_2i_3i_4}_{qp}  \; ,
 \qquad
%\\ \nonumber
\\ \label{l2l2=} && l_{{\rm 2}q}^{+\underline{q}}\,l_{{\rm 2}p}^{+\underline{q}}=2  (l_{{\rm 2}}^{\#}-l_{{\rm 12}}^{\#})\delta_{qp}-\frac 1 {4!} l^{i_1i_2i_3i_4\#}\gamma^{i_1i_2i_3i_4}_{qp}  \; .
 \qquad
%\\ \nonumber
\eea

Thus all $l_{{\rm i}q}^{+\underline{q}}\,l_{{\rm j}p}^{+\underline{q}}$ are symmetric which implies that Eq. \eqref{ggKK=} contains $\gamma^{i}_{(q|(\dot{q}} \tilde{\gamma}^{j}_{\dot{p})|p)}=\gamma^{(i|}_{q(\dot{q}} \tilde{\gamma}^{|j)}_{\dot{p})p}$ and, furthermore, that this equation reduces to the following equation involving as a parameter
the vector $K^{=i}_{{\rm 21}} =K^{=i}_{{\rm 2}} -K^{=i}_{{\rm 1}}$
\bea
\label{ggKK==}
0&=& \gamma^{(i}_{q(\dot{q}} \tilde{\gamma}^{j)}_{\dot{p})p}
K^{=i}_{{\rm 21}} \, K^{=j}_{{\rm 21}}
 \left(l_{{\rm 12}}^{\#}\delta_{qp}+\frac 1 {4!} l^{i_1i_2i_3i_4\#}\gamma^{i_1i_2i_3i_4}_{qp}\right) \equiv
 %\\ \nonumber
 \\ && \equiv
 - 2(\vec{K}^{=}_{{\rm 21}})^2
 l_{{\rm 12}}^{\#}\delta_{\dot{q}\dot{p}} -  K^{=i}_{{\rm 21}} \, K^{=j}_{{\rm 21}} \, \frac 1 {4!} l_{{\rm 12}}^{\# k_1...k_4}
 (\tilde{\gamma}^{(i}\gamma^{k_1...k_4}\tilde{\gamma}^{j)})_{(\dot{q}\dot{p})}
\;.
%\nonumber   \\
 \eea
Decomposing this matrix equation on the irreducible parts, we arrive at
\bea\label{l12xK2=0}
&& l_{{\rm 12}}^{\#}\, (\vec{K}^{=}_{{\rm 21}})^2=0\; \qquad
%\\ \nonumber
\\  \label{lijklK=}
&& 8{K}^{=j}_{{\rm 12}}  l^{\# j [i_1i_2i_3} {K}^{=i_4]}_{{\rm 12}} + l^{\# i_1i_2i_3i_4} (\vec{K}^{=}_{{\rm 12}})^2=0\; .
\eea
As for nontrivial process $\vec{K}^{=}_{{\rm 21}}\not= 0$, we find that \eqref{l12xK2=0} is solved by $l_{{\rm 12}}^{\#}=0$ and \eqref{lijklK=} -- by $l^{\# i_1i_2i_3i_4}=0$ (he simplest way to find this latter solution is to use SO(8) frame in which
${K}^{=i}_{{\rm 12}}=|\vec{K}^{=}_{{\rm 21}}|\,\delta^i_8$), so that
$l_{{\rm 1}q}^{+\underline{q}}\,l_{{\rm 2}p}^{+\underline{q}}=0$.

So far
we have proved that for the nontrivial process involving one massive and two massless (super-)particles
the coefficients in the decomposition \eqref{v=v1l1+v2l2} of massive helicity spinors over two massless helicity spinors
obey the relations
\bea\label{l+1+l1=l++I}
&& l_{{\rm 1}q}^{+\underline{q}}\,l_{{\rm 1}p}^{+\underline{q}}=2  l_{{\rm 1}}^{\#}\delta_{qp} \; ,
 \qquad {}  \qquad  l_{{\rm 1}q}^{+\underline{q}}\,l_{{\rm 2}p}^{+\underline{q}}=0 \; ,
 \qquad {}  \qquad  l_{{\rm 2}q}^{+\underline{q}}\,l_{{\rm 2}p}^{+\underline{q}}=2  l_{{\rm 2}}^{\#}\delta_{qp}  \; .
 \qquad
 %\\ \nonumber
\eea

It is convenient to define $ l_{{\rm i}q}^{\; \underline{q}}:=\, \frac 1 {\sqrt{2l_{{\rm 1}}^{\#}}}\,  l_{{\rm i}q}^{+\underline{q}} $ (for simplicity we restrict our consideration by the case of $l_{{\rm 1}}^{\#}>0$ which corresponds to $\rho_{{\rm 1}}^{\#}<0$, see \eqref{l=-r:m})
which obey
\bea\label{l1l1=I}
&& l_{{\rm 1}q}^{\;\underline{q}}\,l_{{\rm 1}p}^{\;\underline{q}}= \delta_{qp} \; ,
 \qquad {}  \qquad  l_{{\rm 1}q}^{\; \underline{q}}\,l_{{\rm 2}p}^{\;\underline{q}}=0 \; ,
 \qquad {}  \qquad  l_{{\rm 2}q}^{\;\underline{q}}\,l_{{\rm 2}p}^{\; \underline{q}}=\delta_{qp}  \;
 \qquad
 %\\ \nonumber
\eea
and hence form the SO(16) valued matrix
\bea\label{l1l2inSO16}
&&
\left(\, l_{{\rm 1}q}^{\; \underline{q}}\, , \,  l_{{\rm 2}q}^{\; \underline{q}}\right)\quad \in\quad O(16)\; .
%\\ \nonumber
\eea
This also implies
\bea\label{I=l1l1+l2l2}
&& \delta_{\underline{q}\underline{p}}=
l_{{\rm 1}p}^{\;\underline{q}}\,l_{{\rm 1}p}^{\;\underline{p}}+  l_{{\rm 2}p}^{\;\underline{q}}\,l_{{\rm 2}p}^{\; \underline{p}}\; .
 \qquad
 %\\ \nonumber
\eea

In this terms the decomposition \eqref{v=v1l1+v2l2} of the helicity spinor of the massive superparticle in the case of the the three point process becomes
\begin{eqnarray}\label{v==v1l1+v2l2}
{\rm v}_{\alpha}{}^{\underline{q}} &=&
\sqrt{2l_{{\rm 1}}^{\#}} v_{\alpha q{\rm 1}}^{\; -} l_{q {\rm 1} }^{\; \underline{q}} + \sqrt{2l_{{\rm 2}}^{\#}} v_{\alpha  {q}{\rm 2}}^{\; -}
l_{{q}{\rm 2}}^{\; \underline{q}}
\; . \qquad
\end{eqnarray}

Notice that for inverse spinor moving frame matrix  of the massive superparticle in 3 point process we can obtain an expression similar to \eqref{v==v1l1+v2l2}   in the same way as   \eqref{v==v1l1+v2l2} but from momentum conservation \eqref{u0=u1+u2} multiplied by $\tilde{\sigma}_\mu^{\alpha\beta}$,
\begin{eqnarray}\label{v-1=v1l1+v2l2}
{\rm v}_{\underline{q}}{}^{\alpha} &=&
\sqrt{2l_{{\rm 1}}^{\#}} v^{-\alpha}_{\dot{q}{\rm 1}}
l_{\dot{q} {\rm 1} }^{\; \underline{q}} + \sqrt{2l_{{\rm 2}}^{\#}} v^{-\alpha}_{\dot{q}{\rm 2}}
l_{\dot{q}{\rm 2}}^{\; \underline{q}}
\; \qquad
%\\ \nonumber
\end{eqnarray}
with $ l_{\dot{q} {\rm 1,2} }^{\; \underline{q}} $ also forming
$O(16)$ matrix, which is to say, obeying
\bea\label{ld1ld1=I}
&& l_{{\rm 1}\dot{q}}^{\;\underline{q}}\,l_{{\rm 1}\dot{p}}^{\;\underline{q}}= \delta_{\dot{q}\dot{p}} \; ,
 \qquad {}  \qquad  l_{{\rm 1}\dot{q}}^{\; \underline{q}}\,l_{{\rm 2}\dot{p}}^{\;\underline{q}}=0 \; ,
 \qquad {}  \qquad  l_{{\rm 2}\dot{q}}^{\;\underline{q}}\,l_{{\rm 2}\dot{p}}^{\; \underline{q}}=\delta_{\dot{q}\dot{p}}  \; ,
 \qquad
 %\\ \nonumber
 \\ \label{I=ld1ld1+ld2ld2}
&& \delta_{\underline{q}\underline{p}}=
l_{{\rm 1}\dot{p}}^{\;\underline{q}}\,l_{{\rm 1}\dot{p}}^{\;\underline{p}}+  l_{{\rm 2}\dot{p}}^{\;\underline{q}}\,l_{{\rm 2}\dot{p}}^{\; \underline{p}}\; .
 \qquad
 %\\ \nonumber
%\\ \label{ld1ld2inSO16} && \left(\, l_{{\rm 1}\dot{p}}^{\; \underline{q}}\, , \,  l_{{\rm 2}\dot{p}}^{\; \underline{q}}\right)\quad \in\quad O(16)\; .   \\ \nonumber
\eea

The relation between  $ l_{\dot{q} {\rm 1,2} }^{\; \underline{q}}
$ and $ l_{{q} {\rm 1,2} }^{\; \underline{q}} $ can be then obtained from that \eqref{v-1=v1l1+v2l2} is inverse to \eqref{v==v1l1+v2l2}:

\bea
 l_{\dot{q} {\rm 2} }^{\; \underline{q}}&=& \, - \, \frac 1 {2\sqrt{l^{\#}_{{\rm 1} }l^{\#}_{{\rm 2}}}\, (\vec{K}^{=}_{ {\rm 21}})^2 }\, \vec{K}^{=}_{ {\rm 21}}\,  l_{{q} {\rm 1} }^{\; \underline{q}} \vec{\gamma}_{q\dot{q}}  \; , \qquad \\ \nonumber \\
  l_{\dot{q} {\rm 1} }^{\; \underline{q}}&=& \,  \frac 1 {2\sqrt{l^{\#}_{{\rm 1} }l^{\#}_{{\rm 2}}}\, (\vec{K}^{=}_{ {\rm 21}})^2 }\, \vec{K}^{=}_{ {\rm 21}}\,  l_{{q} {\rm 2} }^{\; \underline{q}} \vec{\gamma}_{q\dot{q}} = \, - \, \frac 1 {2\sqrt{l^{\#}_{{\rm 1} }l^{\#}_{{\rm 2}}}\, (\vec{K}^{=}_{ {\rm 21}})^2 }\, \vec{K}^{=}_{ {\rm 12}}\,  l_{{q} {\rm 2} }^{\; \underline{q}} \vec{\gamma}_{q\dot{q}}  \; , \qquad
\eea

\subsection{Three point amplitude with D$0$-brane}

\label{A00m=}

Let us use our results on three particle kinematics to calculate the first multiplier in \eqref{App4p=m1}, namely the second rank SO(9) tensor $p_{{\rm 1}\mu}{\rm u}^{\mu I}_{{\rm 3}} p_{{\rm 2}\nu}{\rm u}^{\nu J}_{{\rm 3}}$:
\bea\label{p1uIp2uJ=} p_{{\rm 1}\mu}{\rm u}^{\mu I}_{{\rm 3}} p_{{\rm 2}\nu}{\rm u}^{\nu J}_{{\rm 3}}=  \frac {(\rho^{\#}_{{\rm 1}}\rho^{\#}_{{\rm 2}})^2}{16 {\rm m}^2}\, <2^-_q|1^-_{\dot{q}}><2^-_p|1^-_{\dot{q}}> \, l^{\underline{q}}_{q{\rm 2}}\gamma^I_{\underline{q}\underline{p}}l^{\underline{p}}_{p{\rm 2}}\, \, <1^-_r|2^-_{\dot{s}}><1^-_s|2^-_{\dot{s}}> \, l^{\underline{r}}_{r{\rm 1}}\gamma^J_{\underline{r}\underline{s}}l^{\underline{s}}_{s{\rm 1}}\, \nonumber
%\\ \nonumber
\\ =  \frac {(\rho^{\#}_{{\rm 1}}\rho^{\#}_{{\rm 2}})^2}{64{\rm m}^2}\, |\vec{K}^=_{{\rm 21}}|^4  \, l^{\underline{q}}_{q{\rm 2}}\gamma^I_{\underline{q}\underline{p}}l^{\underline{p}}_{q{\rm 2}}\, \, \, l^{\underline{r}}_{p{\rm 1}}\gamma^J_{\underline{r}\underline{s}}l^{\underline{s}}_{p{\rm 1}}\, .
%\\ \nonumber
\eea
To obtain the first equality we have used Eqs.
\eqref{D0:uIs=vgv},   \eqref{v===v1l1+v2l2} and $ p_{{\rm i}\mu}=\frac  1 8 \rho^\#_{{\rm i}} v_{{\rm i}\dot{q}}^{-} {\sigma}_\mu v_{{\rm i}\dot{q}}^{-}$, for ${\rm i}=1,2$, which follows from \eqref{pi=ru--i} and ${\rm i}$--th counterpart of \eqref{u--ts=v-v-}.
The second equality uses \eqref{v-iv-j=} and algebra of $SO(8)$ counterparts of Pauli matrices, \eqref{gitgj+=I}.

Then, using
\eqref{bUg=}--\eqref{wgIw=UIcU},  we find
\bea\label{p1Up2bU=}
  p_{{\rm 1}\mu}{\rm u}^{\mu I}_{{\rm 3}} p_{{\rm 2}\mu}{\rm u}^{\mu J}_{{\rm 3}}  U_{(I} \bar{U}_{J)} =
   \frac {(\rho^{\#}_{{\rm 1}}\rho^{\#}_{{\rm 2}})^2}{128 {\rm m}^2}\, |\vec{K}^=_{{\rm 21}}|^4  \,  \left(
  (l_{q{\rm 2}}\bar{{\rm w}}_A){\bar{\cal U}}{}^{AB} (l_{q{\rm 2}}\bar{{\rm w}}_B)\; (l_{p{\rm 1}}{\rm w}^C)\, {\cal U}{}_{CD} (l_{p{\rm 1}}){\rm w}^D+ {\rm 1}\leftrightarrow {\rm 2})\,
  \right)\qquad
 \nonumber
 %\\  \nonumber
 \\  = - \frac {(\rho^{\#}_{{\rm 1}}\rho^{\#}_{{\rm 2}})^2}{64{\rm m}^2}\, |\vec{K}^=_{{\rm 21}}|^4  \,  \left|
  (l_{q{\rm 2}}\bar{{\rm w}}_A){\bar{\cal U}}{}^{AB} (l_{q{\rm 2}}\bar{{\rm w}}_B)\;
  \right|^2 = - \frac {(\rho^{\#}_{{\rm 1}}\rho^{\#}_{{\rm 2}})^2}{64{\rm m}^2}\, |\vec{K}^=_{{\rm 21}}|^4  \,  \left|
  (l_{q{\rm 1}}\bar{{\rm w}}_A){\bar{\cal U}}{}^{AB} (l_{q{\rm 1}}\bar{{\rm w}}_B)\;
  \right|^2 \; , \qquad
  %\\ \nonumber
  \eea
where $(l_{q{\rm i}}\bar{{\rm w}}_A)= l_{q{\rm i}}^{\; \underline{q}}\bar{{\rm w}}_{\underline{q}A}=
(l_{q{\rm 2}}{\rm w}^A)^*$ and, to arrive at the final expression, we have used
 \be (l_{q{\rm 2}}{\rm w}^A)\, (l_{q{\rm 2}}{\rm w}^B)=- (l_{q{\rm 1}}{\rm w}^A)\,  (l_{q{\rm 1}}{\rm w}^B) \ee and its c.c. These as well as
 \be \label{l1wl1bwscUI=} (l_{q{\rm 2}}{\rm w}^A)\,  (l_{q{\rm 2}}\bar{{\rm w}}_B)\,
  (\sigma^{(\check{I}|}  {\bar{\cal U}})_A{}^{B} =-(l_{q{\rm 1}}{\rm w}^A)\, (l_{q{\rm 1}}\bar{{\rm w}}_B)\,
  (\sigma^{(\check{I}|}  {\bar{\cal U}})_A{}^{B}\ee
are  correct, in their turn,  due to \eqref{D0:wbw=I}
%(${\rm w}^B_{\underline{q}}\, {\rm w}^A_{\underline{q}}=0$  and ${\rm w}^B_{\underline{q}}\, \bar{{\rm w}}_{\underline{q}B}=\delta^A_B$),
as well as to Eq.  \eqref{I=l1l1+l2l2} and tracelessness of  $ (\sigma^{\check{J}}  {\bar{\cal U}})_C{}^{D}$.

  Similarly, using \eqref{l1wl1bwscUI=}, we obtain
  \bea
  %\nonumber \\
  \label{p1UcIp2UcJ=}
 p_{{\rm 1}\mu}{\rm u}^{\mu I}_{{\rm 3}} p_{{\rm 2}\mu}{\rm u}^{\mu J}_{{\rm 3}} U_{(I}^{\check{I}} U_{J)}^{\check{J}}  = -
  \frac {(\rho^{\#}_{{\rm 1}}\rho^{\#}_{{\rm 2}})^2}{16{\rm m}^2}\, |\vec{K}^=_{{\rm 21}}|^4  \, (l_{q{\rm 2}}{\rm w}^A)\, (l_{q{\rm 2}}\bar{{\rm w}}_B)\,
  (\sigma^{(\check{I}|}  {\bar{\cal U}})_A{}^{B} \;
    (l_{q{\rm 1}}{\rm w}^C)\, (l_{q{\rm 1}}\bar{{\rm w}}_D)\,
  (\sigma^{|\check{J})}  {\bar{\cal U}})_C{}^{D} \qquad
 \nonumber
 %\\ \nonumber
 \\  = +
 \frac {(\rho^{\#}_{{\rm 1}}\rho^{\#}_{{\rm 2}})^2}{16{\rm m}^2}\, |\vec{K}^=_{{\rm 21}}|^4  \, (l_{q{\rm 1}}{\rm w}^A)\, (l_{q{\rm 1}}\bar{{\rm w}}_B)\,
  (\sigma^{\check{I}}  {\bar{\cal U}})_A{}^{B} \;
    (l_{q{\rm 1}}{\rm w}^C)\, (l_{q{\rm 1}}\bar{{\rm w}}_D)\,
  (\sigma^{\check{J}}  {\bar{\cal U}})_C{}^{D} \qquad
  %\\ \nonumber
\eea

In such a way we have found the expression \eqref{App4p==m2} for the amplitude \eqref{App4p=m1}.

\end{widetext}

%\end{widetext}

\end{document}